\documentclass[proof]{WileyASNA-v1}

\articletype{Original article}%

\received{xx xxxxx xxxx}
\revised{xx xxxxx xxxx}
\accepted{xx xxxxx xxxx}

\begin{document}

\title{Review of light curves of novae in the modified scales. II. Classical novae.}

\author[1]{Rosenbush A.*}




\address[1]{\orgname{Main Astronomical Observatory of the National Academy of Sciences of Ukraine}, \orgaddress{\state{27 Akademika Zabolotnoho St., 03143 Kyiv}, \country{Ukraine}}}



\corres{*\email{aeros@mao.kiev.ua}}


\abstract{The presentation of the light curves of the novae on the logarithmic scale of the radius of the shell ejected during the outburst and in the scale of the amplitude of the outburst simplified the review of the light curves of all the known, about 500, classical novae of our Galaxy and the Large and Small Magellanic Clouds. As a result, the modified light curves about of 250 classical novae were grouped into 7 typical groups with subgroups defined by the light curves of prototypes.   
	
The largest group is the DQ Her group with V1280 Sco as the main prototype and the key to grouping. Novae of this group have three or four variants of a dust condensation. Novae of the small GQ Mus group may be a bright X-ray source during an outburst. 
	
CP Pup, CP Lac, V1974 Cyg, and V1493 Aql are prototypes for 4 groups with close tilts of the initial brightness decline phase of the light curves. The group with the prototype RR Pic, including the HR Del subgroup, has a prolonged state of maximal brightness with the presence of several brightness variations before the final decline phase. 

The relationship of groups of novae, or form of the light curve, with processes in the circumstellar and circum-binary system environment is discussed. The material ejected during an outburst forms expanding shells of a generally ellipsoidal shape. In the Lac, Pup, and Cyg groups, the shells do not show a pronounced regular structure, except for their ellipsoidal shape. Shells of novae with a dust condensation (the DQ Her group) have an ellipsoidal shape with an equatorial belt. The unique shell of CK Vul, Nova Vul 1670, gives an idea about the possible shape of the shells of other members in the V1493 Sql group. The prototypes RR Pic and HR Del have a very bright belt. 

Each of groups has own position along the well-known "absolute maximal magnitude, rate of brightness decline" relation.}


\jnlcitation{\cname{%
\author{Rosenbush A.}} (\cyear{2020}), 
\ctitle{Review of light curves of novae in the modified scales. II. Classical novae.}, \cjournal{Astron. Nachr.}, \cvol{20--;00:1--6}.}


\maketitle


\section{Introduction}\label{sec1}

In certain binary systems, in specific conditions, a short-term surface source of thermonuclear energy may arise \citep{Starrfield2016}. As a result, the phenomenon of “new” stars or a nova is observed. With the outburst amplitude of more than 9-11$^{m}$, the nova is classified as a classical nova (CN). The energy of such an outburst is enough to eject a large mass of a matter that can form a expanding shell existing for a long time. The duration of outburst is several or tens years, unlike recurrent novae outbursts with the duration of 70-700 days. 

The ejected shell becomes second radiation source, but due to the re-radiation of flux from a central high-temperature source. We fix the interaction of these two variable sources, in particular, in the form of a visual light curve. The spectrum of an expanding source transforms over time from an absorption spectrum to purely an emission one. If the energy distribution in the spectrum of a nova at outburst maximum has a maximum in the optical spectral region and the absolute visual brightness is almost equal to the absolute bolometric magnitude of the nova, then as the shell expands, the maximum of the energy distribution in the spectrum of a nova shifts to the ultraviolet. It is believed that sometimes the bolometric luminosity of a nova can a constant for a long time \citep{Gallagher1974, Wu1977}. By the end of a outburst, the contribution of the central source radiation to the observed brightness in the optical region of the spectrum is greatly reduced in comparison with the line spectra of the shell radiation \citep{Vorontsov-Velyaminov1953, Kaler1986}. 

One of the results of such set of radiation sources is a prolonged light curve. And if at the beginning of the outburst a light curve shows clearly noticeable fast and large changes, then as you move away from the outburst maximum, all changes slow down in time, decrease in amplitude and are harder to notice. Therefore, sometimes a logarithmic time scale is used to represent the light curve \citep{Vorontsov-Velyaminov1953}, since in this case the initial and some other parts of the light curve of the nova are displayed by straight lines, and the prolonged second half of the light curve is compressed and the light curve becomes compact, convenient for her full review.  

Each year, the list of novae stars is replenished with new objects. To date, the number of known classical novae in our Galaxy reaches 400. 

The most extensive and accessible information on novae is provided by photometry, which has a 150-year history, and in some cases centuries-old. A great amount of photometric information has always served as the basis for the systematization of objects. Spectral observations mainly served for the confirmation of the identification of the star as nova. It has long been noticed that both the photometric and spectral behaviour of the novae have common features. This consistency in behaviour found itself expression in the definition of phases in the development of brightness and spectrum of a nova during an outburst. For light curves, the result is expressed in the phases of a schematic light curve \citep{Payne-Gaposchkin1957}. Spectrum development went through 9 general phases \citep{McLaughlin1942}. The presence of common details of light curves in different combinations served as the basis for novae classification schemes. An increase in the sample size of novae and a more detailed idea of the light curves revealed shortcomings in the existing classifications of novae, starting with the classical scheme based on the rate of brightness decline after a maximum of 2$^{m}$ or 3$^{m}$: t$_{2}$ or t$_{3}$ \citep{Payne-Gaposchkin1957}. A lot of efforts were aimed at creating classification schemes for light curves of novae based on the grouping of parts on light curves: by the presence of a plateau, temporary decline of brightness, or the absence of details on light curves \citep{Duerbeck1981, Duerbeck1987a, Strope2010}.

But \cite{Strope2010} concluded that "the primary reason why none of these classification schemes has been usefully adopted is that the classes are so broad that greatly different physical settings are lumped together and they have not been found to correlate with anything else". The reason for this state of affairs may also be the wrong approach to solving the problem. Existing schemes rely mainly on the shape of the first half of the light curve and pay less attention to the completion of the outburst. Our review of the light curves of recurrent novae (\citep{Rosenbush2020}, hereafter Paper I) showed that for a recurrent nova by any part of an incomplete light curve in any outburst, we can be restored the complete light curve in a given outburst if there is a template. This is the classification of light curves of classical novae should also strive for such a state. Then the combination of some physical characteristics of the novae in the same group will most likely be the same. It is possible that in the other group a slight difference in one characteristic can lead to a striking difference in the shape of the light curve. Such a scheme is proposed in this study. 

We have changed the basic provisions. The basis of the new approach was the transition to the construction of light curves in modified scales: the logarithmic scale of time, or radius of the shell ejected during an outburst \citep{Rozenbush1996a}, and the amplitude of outburst \citep{Rosenbush1999a}. The impetus for using the shell radius as an argument was the work of \citep{Clayton1976} on the study of dust condensation in expanding shells of novae, one of the results of which can be interpreted as follows: when the luminosity and temperature of two novae are equal, dust condensation occurs at equal distances. Such novae, with a dust condensation, have always been combined into a novae group of the DQ Her types, in which the transition phase of the outburst, a temporary decline in brightness, is identified with dust condensation, which did not violate the general trend of the light curve. \cite{Rozenbush1996a} showed that the same phases of the light curve and the phases of spectra development in the shell radius scale for different novae, including DQ Her, occur at a smaller relative radius range than the ratio of the corresponding time intervals after the maximum brightness when the shell is reset. As a consequence of this circumstance, the studies of \cite{Rosenbush1999a, Rosenbush1999b, Rosenbush1999c, Rosenbush1999e} showed the possibility of grouping novae according to a form of light curves in the scale of the radius of the ejected shell. Applying the results to specific objects made it possible to suspect the identification mistake of the V605 Aql infrared (IR) source with its optical counterpart \citep{Rosenbush1999d}. There was made the suggestion that V2487 Oph, the nova of 1998, could be assigned to the recurrent novae \citep{Rosenbush2002}, which was confirmed by archival searches of \cite{Pagnotta2009}. The weak point of the proposed method was the assumption of the possibility of shifting the light curves along the abscissa, motivated by the desire to compensate for the difference in the expansion velocities of novae shells with similar light curves. The solution of this situation was helped the outburst in 2007 of the Nova Scorpii, which showed a very unique light curve, for unusual details of which we offer below a very simple explanation. As a result, the rule arose: to compare as it is; i.e. do not make any additional conversions. In view of the need to avoid zero values of the argument of the logarithmic function, a rule was developed for determining the instant of outburst maximum brightness when the expanding shell is separating from the star system. But a definite need exists for a shift along the outburst amplitude scale, which is related to both the frequent impossibility to know the brightness of the binary system in a quiet state and the change in the brightness of the binary system due to spatial orientation with respect to the line of sight, the amplitudes of which reach 2-3$^{m}$ \citep{Warner1987}.

Application of the proposed approach in the presentation of light curves to recurrent novae allowed us to unify the methodology for constructing modified light curves and to test it when reviewing recurrent novae (Paper I). The result was a statement of the fact that the light curves of the recurrent nova in the sequence of outbursts are almost completely identical. It also turned out that 10 known recurrent novae can be combined into groups. Candidates for recurrent novae among the novae of our Galaxy and the the Large (LMC) and Small (SMC) Magellanic Clouds were proposed. In this publication, that technique will be applied to classical novae. 

The aim of given review of modified light curves of classical novae is to highlight groups with similar light curves. 

The basis of this study is the observations contained in numerous publications and extensive publicly available databases of international associations of variable stars observers (AAVSO, AFOEV, BAA VSS). The main source, the AAVSO data \citep{Kafka2019}, was supplemented by data from sky surveys: OGLE \citep{Wyrzykowski2014}\footnote{http://ogle.astrouw.edu.pl/ogle4/transients/}, in particular, in the region of the galactic bulge \citep{Mroz2015} and the LMC and the SMC \citep{Mroz2016b}; ASAS-SN \citep{Shappee2014, Kochanek2017}; SMARTS specialized review \citep{Walter2012}, data of mission Gaia  and others. In addition to the SIMBAD database, sources of bibliography and catalogue data on objects served as \cite{Payne-Gaposchkin1957, Samus2017, Duerbeck1987a, Strope2010},  \cite{Pietsch2010}\footnote{http://www.mpe.mpg.de/\_m31novae/opt/index.php} and especially the VSX AAVSO \citep{Kafka2019}\footnote{https://www.aavso.org}. 

Classification on this phase occurs through visual comparison with a set of summarized templates of light curves. 

The name of the groups of novae according to \cite{Rosenbush1999a, Rosenbush1999b} was retained, but the main prototype can be offered another in view of a more detailed light curve or its particular features. The result proposed below is a consequence of several iterations: at first we were guided by grouping according to \cite{Rosenbush1999a, Rosenbush1999b}, then other novae took up the positions of prototypes, the criteria were simultaneously refined and tightened, and finally there became clear that one unique nova (V1280 Sco) joined properties of prototypes of almost all groups. This last version is presented here. It should be noted that our set of novae groups is not exhaustive, since there are examples of single novae with unique light curves. 

The structure of the paper is following. In section 2, we consider the methodological rules and situations that are arose when constructing and analysing modified light curves of novae. Separately, the modified light curve of Nova Scorpii 2007, V1280 Sco, will be presented which is very unique in its shape and had one of the decisive influences on the determination of distinctive features for the separation of novae into groups based. Section 3 is devoted to introducing the main groups of novae with comments for some novae. Separate attention will be paid to the novae - members of LMC and SMC, two nearest galaxies. Discussion into Section 4, and conclusions will complete the paper.
 
\section{The procedure of constructing novae light curves in modified scales}\label{sec2}

For a graphical display of long-term physical processes, such as outbursts of novae, the logarithmic representation of the argument is compact and more convenient for the review and analyse of simultaneously the entire light curve. \cite{Rozenbush1996a} suggested moving from the argument of time to the radius of shell, which the classical novae eject during a outburst. This made it possible to bring close together the same name phases of the light curve in a number of novae. Therefore, light curves of novae were proposed to be constructed on a logarithmic scale of time (Paper I, \cite{Rosenbush1999a, Rosenbush1999b, Rosenbush1999c, Rosenbush1999d, Rosenbush1999e}). It was proposed to normalize the shell radius on r$_{0}$=1.4$\times$10$^{13}$ cm (log(r$_{0}$)=13.1), where r$_{0}$ is the radius of a star at the instant t$_{0}$ of maximum brightness for fast novae or for slow novae at pre-maximum halt or the beginning of the final rise. In reality, this value of r$_{0}$ can vary within one or two orders of magnitude; formally, this radius in the logarithmic representation of the abscissa axis is achieved on the first day after the maximum brightness. The final rise in brightness can last a dozen or more days for slow novae. At the maximum of brightness, there occurs the ejection of the matter from an active star of the binary system that forms the shell of a diverse structure. The expansion velocity is assumed to be constant. After a series of transformations and simplifications, an expression was obtained for the argument (Paper I)

log(r)=log(t-t${_0}$)+C,    ${    }$       (1)

where C=13.1 is the constant to which the light curve is normalized in the logarithmic scale of a time t, and which is equal to the logarithm of radius r${_0}$. Time t and shell radius r in expression (1) are in days and cm, respectively. The value of the parameter t${_0}$ is found in successive iterations, starting from the zero approximation equal to the integer value of date of the first estimate of the nova brightness. The observation moments are recorded with an accuracy no worse than hundredth or thousandths fraction of a day, therefore, for the first observation, the value of log(t-t${_0}$) has a negative value. This “formalism" will be used by us later (see below). 

Since the radius of the shell is often unknown, time becomes the working argument. [It would be more correct to talk about the effective radius of the shell.] Therefore, we transform the light curve from the time using expression (1), which as a result gives an idea of the geometric dimensions of the active system. The light curves in the modified abscissa axis have linear and non-linear sections, the transitions between which are in certain ranges of the argument. For the convenience of further exposition, we supplement the concepts of the phases of the schematic light curve of a nova \citep{Payne-Gaposchkin1957} with the concept of the "state of maximal brightness", which combines the final rise to the principal maximum, this maximum itself and the transition to a linear portion of the early decline; in this state, the brightness of a slow nova may experience fluctuations up to 1-2${^m}$. In the logarithmic time scale, the state of maximum brightness passes into a linear section of the initial (early) decline in brightness, continuing up to the transition phase. The transition phase of the light curve can be both linear and non-linear before the start of another linear section - the final decline in brightness. A linear final decline ends by a level of quiescent brightness or by 2-2.5${^m}$ to a level of quiescence brightness a decelerating monotonous decrease in a brightness begins, which can be regarded as the contribution of nearby field stars or sources directly in the binary system. As it became clear after the outburst of V1280 Sco, the transition phase in different novae even within the same group develops at different values of the effective radius of the shell. We can say that the result of \cite{Rozenbush1996a} was obtained thanks to a happy set of classical novae that formed the basis of that study. This result made it possible to see the direction in which is need to move to create a classification scheme of the light curves of novae and at the nodal point of which is located V1280 Sco. In describing the modified light curve, we will also use the term "moment" as applied to the logarithmic radius scale. The brightness decline after the maximum state maintains its pace until the beginning of the final decline therefore instead of an early brightness decline we will use the initial (linear) brightness decline to emphasize the difference from the final (linear) decline. In fact, there are two phases of a brightness decline: initial and final. Often the initial decline is divided into two parts by the transitional phase.

The concept of the state of maximal brightness can be also described as follows. If we draw a horizontal line at a level 1-2${^m}$ below the maximum on the schematic light curve, then the upper part of the light curve will be the state of maximum brightness, the start time of this section will be t${_0}$, length of the horizontal line is the duration of maximal brightness state. 

It is clear that the moment t${_0}$ has an error, both by definition (which we examined earlier \citep{Rosenbush1999e}) and because of the frequent absence of observations at the maximum brightness. The most significant gaps of the order of hundred days are associated with the invisibility seasons due to the conjunction of the object with the Sun. The use of a modified light curves convinced us that light curves with characteristic details within the same group can be shifted relative to each other in the range $\Delta$log(r)=0.1$\div$0.2. This is due to both the inaccuracy of the determination of t${_0}$ and the peculiarities of the novae themselves, as could been seen from the example of recurrent novae groups of T Pyx (Paper I). The upper limit of the displacement range may already mean a possible belonging to another or even unknown group. In the linear scale, $\Delta$log(r)=0.2 means an inaccuracy of about 50\%, i.e. at a distance of 10 days from the maximum, this will be an inaccuracy of 5 days, for 100 days - 50 days, etc. It is clear that each nova has its own individualities, which as a result will lead to a certain scattering of points on the summarized light curve both prototypes and all novae of a separate group. Characteristic details of the modified light curves are the transition from the state of maximum brightness to the linear portion of the initial decline, the transition phase and the transition to the final decline; there may be other specific details of the light curves. In the light curves of one group, the displacement to the final phase may increase and will no longer be a sign of belonging to another group, but may be a consequence of a change in the shell expansion speed (see, for example, \citep{Rosenbush1999e}). The shape of the light curve stabilizes 10-15 days after the outburst maximum, which confirms the hypothesis of \citep{Buscombe1955}, and if only this initial part of the light curve is absent, then the rest of the details will allow classification with a good degree of confidence using a small sequence of iterations moment t${_0}$. In practice, this looks like the following sequence of our actions. First, for the time t${_0}$, the integer of the date of the first observation is taken. Using the shape of the light curve for log(r)>14, the most probable belonging to any group of novae is estimated. Using the trial-and-error method to the parameter t${_0}$, we obtain the best agreement between the light curves of the studied nova and prototype(s).  

Near the brightness maximum of a fast nova, the light curve can be reconstructed by averaging several fast novae of one group, varying the moment t${_0}$ for each nova. For recurrent novae problems with this did not arise. For classical novae, it needs to be sure that they are novae from the same group. During the iteration, the numerical value of t${_0}$ can decrease or increase. The restriction on decreasing, i.e. (t-t${_0}$) will increase, is determined by the appearance of an increasing deviation of the initial portion of the light curve to the right upwards from a straight line approximating the subsequent initial brightness decline, i.e. here occurs  an atypical upward bend in the light curve. An example of such a deformation is the V4160 Sgr light curve in Fig. 4 of \cite{Strope2010}. Accordingly, an increase in t${_0}$, i.e. (t-t${_0}$) will decrease, lead to a decrease in the slope of the initial part of the light curve and an increase in its length to the left. Thus, the procedure for choosing the moment t${_0}$ for fast novae can lead to a formal shift of the maximum of the light curve by log(t t${_0}$)<13.1. In the course of our research, this took place in several rare cases (about a dozen to three or four hundred novae that passed our qualifications). This circumstance was noted as a drawback of \cite{Bode2016} when using a light curve with a logarithmic representation of the time scale in the study of the LMC nova of LMCN 2009-02a. [By this means our method of constructing a light curve in a logarithmic scale reduces this drawback.] For slower novae, the task is simplified. Here the requirement for the accuracy of the maximum moment is relaxed. For slow novae, we propose defining t${_0}$ as the moment of transition from a rapid increase in brightness to a slower change in a prolonged state of maximum brightness, or the beginning of the final rise in brightness. It is important that within a specific, fast or slow, group of novae, the determination of the parameter t${_0}$ proceeds alike. It is also desirable to know the date of the last observation of the pre-nova that also indicates the permissible limits of possible changes in t${_0}$. 

The second modification in constructing the light curve of novae concerns the transition to a scale of the amplitude of the outburst. After reviews of recurrent novae, \cite{Schaefer2010} and (Paper I), we can confidently speak of the constancy of amplitude in the sequence of outbursts. Within one group, the outburst amplitudes were also very close. Therefore, it can be assumed that the classical novae of the same group will also have equal amplitudes of outbursts. There is also a range of values for this parameter. A nova is an outburst in the system of a close binary star with several bright extended radiation sources. Their spatial orientation relative to the line of sight results in a brightness range of 2-3${^m}$ \cite{Warner1987}; secular changes in the orientation of the binary system itself and other possible causes will bring to the brightness changes of a quiescence state. The shape of the elongated ellipsoidal shell of DQ Her \citep{Williams1978} allows us to estimate that the influence of the spatial orientation of the shell on the amplitude of the brightness variations at the maximum of the outburst can be 0.5${^m}$ when viewed from the poles or along the equator. During the maximum, many novae are characterized by brightness fluctuations with amplitudes of up to 1-2${^m}$. Therefore, the strip width for a summary light curve of up to 2${^m}$ can be regarded as acceptable provided that the overall trend of the light curve is maintained, or by shift along the amplitude axis two light curves are coinciding. As prototypes of the groups, we will choose stars with known initial parameters, and with respect to them we will already evaluate the brightness of the pre- or post-nova. 

As the result of our classification of the light curve of a nova, we will give for it belonging to a certain group; the brightness of the quiescent state mq and the accepted instant of maximum brightness t${_0}$. The pre- or/and post-nova of prototypes are known, this means the amplitudes of the outbursts are also known. If there are several prototypes, then the average amplitude is accepted for them and, accordingly, the reduced brightness of a quiescence state, which may thus differ from the known one. With this data, we can unambiguously reproduce any modified curve of a known nova and selecting these parameters, classify a nova that has just erupted. Therefore, we will not present the averaged tabular light curves: the averaging procedure can introduce some subjectivity into the details of light curves. It should also be understood that the peculiarities of each binary pre-nova system are added by their own characteristics to the light curve. In the future, when the database is accumulated for novae, it will be possible to obtain averaged modified light curves from the beginning of the outburst to its completion, as well as automate the entire classification process. 

Our study focused mainly on visual and photographic observations. The data for old novae are often presented mainly only by photographic observations. Visual observation of amateurs always significantly complemented the observations of professionals and now it is actively developing CCD photometry among amateur-astronomers. The large number and temporal density of such observations make it possible to describe the light curve in sufficient details. At the final phases of an outburst, when the stars were significantly weaker and inaccessible to amateurs, we used data in the photometric V band from journal publications and databases of modern photometric sky surveys (see paragraphs below). If necessary, based on multi-colour photometry data, colour index estimates were performed and the resulting visual light curve was obtained after appropriate corrections if the data belong to other photometric bands (R or R${_C}$, I or I${_C}$). 

Historical outbursts have a lot of data on the results of photographic surveys of the sky, which is equivalent to photometric B band. A comparison of the light curves in the photographic and visual photometric bands did not show significant differences in the shape of the light curve (see, for example, Fig. 20 of \cite{Schaefer2010}). Modern surveys, for example, the OGLE project \citep{Mroz2016b}, use observations in the I band, in which the light curve can noticeably differ from the visual one: for example, V1141 Sco according to visual data from AAVSO \citep{Kafka2019} and according to the OGLE photometry OGLE \citep{Mroz2015}. Extensive multi-colour photometry of\cite{ Walter2012} indicates that the V-I colour indices have different behaviour at the maximum brightness and at later phases of an outburst, the amplitude of the changes can reach 1.5${^m}$. Therefore, some caution should be exercised here. 

When comparing light curves in different photometric bands, it must be understood that the brightness in a given band is the sum of the radiation in spectral lines and in the continuum. If the photometric strip is wide enough (several hundred angstroms), then two photometric bands adjacent in spectrum will give light curves that differ slightly from each other. It is known that the emission lines observed in the spectrum of a nova, especially at the nebular phase of the outburst, increase the brightness of a star by several magnitudes \citep{Vorontsov-Velyaminov1953}. In this case, the light curves for the pure spectral continuum and for the spectral region with strong emission lines will differ significantly: the brightness in a continuum declines much faster than the brightness in spectral region with strong emission lines \citep{Kaler1986}. The intensity of emission lines increases sharply after the transition phase of the outburst, by an order of magnitude or more, exceeding the radiation in a continuum.

A critical condition for the successful classification of a nova is the presence of a light curve throughout the outburst from the maximum brightness to a return into the quiet state. But there are situations with sufficient data when only a few observations (including the last pre-outburst observation) are distributed over key parts of the light curve. 

Another problem arose when constructing the light curve using CCD photometry: when measuring images, it can be used aperture photometry with annular diaphragms, within which, under a dense stellar field, faint background stars, including variables, can fall. This is especially true for photometric surveys of the Magellanic Clouds. Authors themselves warn of this \citep{Walter2012}. A contrasting example is given by light curves in the I band of LMCN 2005-11a according to \cite{Mroz2016b} and \cite{Walter2012}. In order to reconcile data, the brightness should be reduced by 3${^m}$ from the SMARTS. Further, until JD 2453900, the curves coincide, then the SMARTS data are located above  on 1${^m}$ and gradually this difference increased; after JD 2454100, a light curve according to \cite{Walter2012} shows an almost constant level. In this situation, we preferred the data of \cite{Mroz2016b}, since this specialized review was originally focused on a dense stellar field, the original measurement technique was used and the results have already passed a long period of testing in other studies. \cite{Walter2012} are working on a solution to this problem, so we will be critical with similar data in our further research. 

\subsection{V1280 Sco light curve in modified scales}

The outburst of Nova Scorpii 2007, V1280 Sco, can be considered one of the main bases for any classification scheme, as the properties of this unique nova are very multifaceted. Just the light curve is enough to understand the reasons for some differences in the light curves of some novae on the transition phase of the outburst.

The V1280 Sco light curve is unique and the only such of the novae known light curves, which increases its value in terms of methodology. The brightness of the nova in the maximum and the uniqueness of the light curve attracted much attention from the community of observers, which made it possible to obtain a well-presented light curve at all phases of the outburst. As of end-2019, the visual brightness of the nova has remained at the same level of about 10.3${^m}$ since November 2007 \citep{Kafka2019}. Perhaps due to the incompleteness of the outburst, there are no data for the IR monitoring, which would simplify and make the only possible interpretation of some details of the nova behaviour that will be presented below.

A temporary brightness decline during the transition phase of the outburst of some novae is a contrast detail. The reason for the temporary decline in brightness is unambiguously associated with the condensation and scattering of dust in the substance ejected during the outburst \citep{Gehrz1988}. The light curve during a dust condensation has a characteristic shape: a rapid monotonous decline in brightness and its slower recovery to the general trend of the light curve. The theory of carbon dust condensation indicated a certain radius of the dust condensation zone, depending on the parameters of the nova \citep{Clayton1976}. 

But for the V1280 Sco, the light curve in the transition phase had several temporal declines of a typical shape and their follow successive with overlap led to a complex and unusual shape of this part of the light curve. When comparing the modified V1280 Sco light curve with the light curves of some other novae with temporary light dip, it turned out that each of the three light dips in the V1280 Sco coincides with a single dip in the other nova. Occasionally, there was a coincidence of two such temporary dips (mainly, provided that there were observations at the corresponding phase of the outburst). By comparing the light curves of these novae, it was found that there are three values of the radius of the shell at which the maximum dust density is achieved. The difference between these values exceeds the error of the calibration of the radius scale for a separate nova ${\Delta}$log(r)=0.1-0.2 (see above in this section). Figure 1 shows examples of the light curves of novae with temporary light dip that coincide with the corresponding three light minima of the V1280 Sco for log(r)${\approx}$14.6, 14.9, and 15.3 (in our shell radius scale). The moment t${_0}$ of the V1280 Sco corresponds to the beginning of the final brightness rise near JD 2454146.8 from a level of about 5.2${^m}$ to the maximum brightness state (for other stars see Section 3). All light curves are presented without additional displacements along the abscissa; V2361 Cyg might also have a second episode with dust condensation, but the light curve is practically absent at this phase. It is very possible that there is one more value of the shell radius at which the 4th episode of dust condensation can very rarely occur: log(r)${\approx}$15.7. Of the entire novae collection there were isolated examples of such objects: for example, V5579 Sgr with three episodes of dust formation, but not with such detailed light curve. There is only one conclusion about the chemical composition of dust: the available data for V1280 Sco are interpreted as the formation of two types of dust: carbon and silicate \citep{Sakon2016}. The combination of the novae in Fig. 1 on the amplitude scale was normalized to the known amplitudes of the DQ Her and V842 Cen outbursts, which required a correction of the possible amplitude of the V1280 Sco by -2.5${^m}$; the brightness of the quiescence state of V2361 Cyg is unknown. A common feature of all these light curves is a fairly smooth plateau in the state of maximum light, which lasts until the phase of temporary light dip. The brightness recovery after dip in these stars is poorly represented by observations, but, apparently, the final brightness decline in all novae occurs after the shell reaches an effective radius of about log(r)${\propto}$16.0. 

For determining the cause of the temporary dip of brightness at the transitional phase of the outburst, the decisive role was played by detailed IR photometry, which unambiguously related the temporary dip of the visual brightness with a temporary increase in the IR brightness at wavelengths of more 2 microns (see the review by \cite{Gehrz1988}). This research method is not so common and requires a lot of time in comparison with photometry in the visual range of the spectrum. Therefore, few groups work in this direction, and observations with a wide variety of research objects are often fragmentary. Compensating for the arising shortcomings can be reached using the results of studies of other objects where similar external manifestations are encountered. For example, stars with the R Coronae Borealis type variability also show temporary dip of visual brightness due to the formation of a dense layer of dust on the line of sight \citep{Rosenbush1996b}. It is possible the successive formation of several such layers with different parameters and they forms a light curve of complex shape, which can be approximated using a simple empirical formula \citep{Rozenbush1988a, Rozenbush1988b, Rozenbush1992, Rosenbush1996b, Rosenbush2016}. Figure 2 shows the total result of such a presentation of successive independent episodes of the formation of four dust layers for the V1280 Sco (we did not pursue the goal of accurate approximation of the light curve, the principal possibility of such an approximation was shown). Each of the 4 temporal light dips was represented by a simple empirical two-parameter formula with gamma function, where the parameters are the depth of light dip and the time during which the brightness declines in this dip. The 4th episode with dust condensation was the longest and shallowest. It is less expressive on the modified light curve than on the traditional light curve in Fig. 2. It is very possible that a sequential increase in the duration of dips from first to fourth is real. 

Dust condensation has one more effect on the parameters of the ejected material. Thanks to the radiant pressure, the dust particles acquire an additional impulse and transmit it to the gaseous medium. In stars with R CrB type variability, the gas medium thus acquires a velocity of up to 200-300 and more km s${^-1}$. Novae also demonstrates a similar effect. \cite{Payne-Gaposchkin1957} in Fig. 4.9 compares the DQ Her visual light curve with the hydrogen line data of \cite{McLaughlin1954}, which allow us to suspect this effect: 10 days before the start of the visual brightness decline, when the dust condensation actually begins (for details, see the review of \cite{Rosenbush1996b}), the displacement of hydrogen lines increased sharply by 200-300 km s${^-1}$. 

The appearence of dust is indicated by the presence and strengthening of a graphite band at wavelengths of about 2200 A and only for ten days we have possibility to see the visual brightness decline \citep{Rosenbush1996c}). This occurs simultaneously with changes in the IR range. For this time radius of the dust particles increases up to values to they begin to affect the optical and IR ranges of the spectrum.

\begin{figure}[t]
	\centerline{\includegraphics[width=78mm]{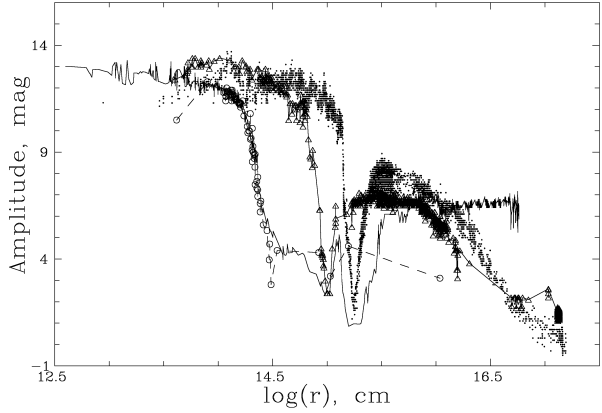}}
	\caption{Modified light curve of V1280 Sco (solid broken line) compared to V2361 Cyg (dashed line with circles), V842 Cen (solid line with triangles) and DQ Her (dots). The zero-point of the ordinate scale is hereafter shifted by -1 in such figures to display the brightness variability of objects in a quienscence state.\label{Fig1}}
\end{figure}

\begin{figure}[t]
	\centerline{\includegraphics[width=78mm]{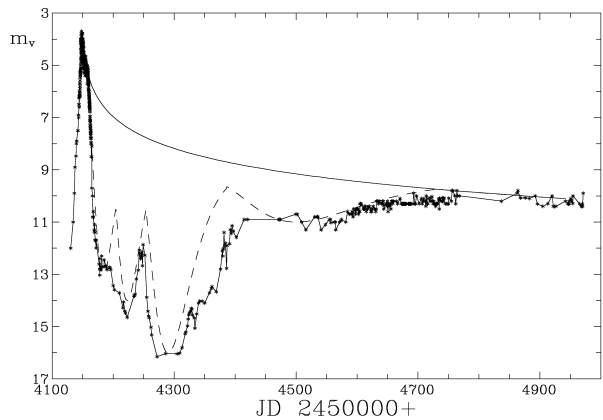}}
	\caption{The visual light curve of V1280 Sco (dotted line, AAVSO data) and an example of its possible approximation during the transition phase of the outburst, as a result of the sequential independent formation of 4 dust layers in the line of sight (dashed line). The upper solid line is a typical light curve without temporarily weakening the brightness.\label{Fig2}}
\end{figure}

After the transition phase of an outburst, the V1280 Sco has a unique relative brightness stabilization at 10${^m}$, which continues to the present (end-2019). Since the pre-nova was not identified among stars up to 20${^m}$ and, therefore, the V1280 Sco may still have a significant evolution of a luminosity, further surprises can be expected from this unique nova. 

On the V1280 Sco modified light curve, several points can be distinguished (Table 1), which are key for the novae groups that we have identified. This is, first of all, the moment of maximum brightness t${_0}$. The above-mentioned 4 temporary light declines will be called dust condensation episodes of the D1, D2, D3, and D4 type of V1280 Sco. The characteristics of these episodes will be the moments D1, D2, D3 or D4 of the beginning of the brightness decline in the corresponding dip; they are fixed more confidently than the moments of maximum brightness weakening. We add to this the moments of the starts of the initial (E) and final (F) brightness declines, the position of which varies in a certain range of values of log(r) or a time from group to group. The contents of Table 1 will simplify us the description of the characteristics of light curves. 

\begin{center}
	\begin{table}[t]%
		\centering
		\caption{Key points of modified light curves of novae.\label{tab1}}%
		\tabcolsep=0pt%
		\begin{tabular*}{240pt}{@{\extracolsep\fill}rccccccc@{\extracolsep\fill}}
			\toprule
			\textbf{Point}& \textbf{t${_0}$} & \textbf{E} & \textbf{D1} & \textbf{D2} & \textbf{D3}& \textbf{D4}& \textbf{F } \\
			\midrule
			log(r)    & 13.1 & (13.3-13.4) & 14.3 & 14.8 & 15.1 & 15.5 & 15.5-15.8 \\
			    &  & $\div$ & &  &  &  &  \\
			    &  & (13.8-14.3) &  &  &  &  &  \\
			\midrule
			t-t${_0}$, d  & 1   & 2-15   & 15 & 50 & 100  & 250  & 250-500  \\
			\bottomrule
		\end{tabular*}
	\end{table}
\end{center}					

The main conclusion that the V1280 Sco light curve offers at this phase and important for purposes of our review of light curves in modified scales that is that it confirms the rule that light curves must be compared "as is", i.e. it need to compare actual observed light curves without any shift along the abscissa (Paper I). And the logarithmic modification of the abscissa scale makes the transition between the phases of the outburst more expressed. 

\section{Groups of classical novae}\label{sec3}

The first group with which we will begin our review includes classical novae with the most recognizing and long-known detail of the light curve and associated with the condensation of dust in the envelope ejected during the outburst. This is an important group of novae, because it allows us in the future to put at least one more point for calibration of the radius scale in addition to the radius at the moment of maximum. 

Dust condensation is possible under certain conditions \citep{Clayton1976}. Mostly it is the carbon dust, sometimes the silicate dust condenses \citep{Gehrz1988}, or both possible in the same shell \citep{Sakon2016}. 

\subsection{DQ Her group}

The name of the group according to \cite{Rosenbush1999a} was left unchanged, since the characteristic detail of the DQ Her light curve — the temporal light minimum (or the dust dip, or the light minimum) in transition phase — has always been the main argument to separate out such stars into the group of novae (see, for example, the classification scheme of \cite{Duerbeck1981, Strope2010}). 

The main prototype is V1280 Sco with the unique light curve and with its three, but possibly and four, episodes of dust condensation. The chemical composition of the dust of all these three or four layers is unknown, the carbon composition is unambiguous and silicate dust is allowed to form \citep{Sakon2016}. In addition to the V1280 Sco, the prototypes of the subgroups are V2361 Cyg (subgroup D(ust)1), V842 Cen (subgroup D2) and DQ Her together with T Aur (subgroup D3) (Fig. 1). The amplitude of the V1280 Sco outburst was reduced to novae with well-known amplitudes: DQ Her, V842 Cen (Table 2, 3, 4, and 5). For each of these novae, the dust condensation occurred once. While some doubts may arise regarding the V2361 Cyg, since the light curve in this part is very fragmented, the light curves of the other two novae exclude such possibilities. (Here we can see the advantage of prolonged wide-angle sky surveys, since monitoring of an individual object can ceased due to an unexpected, prolonged temporary minimum below the sensitivity limit of the equipment.) 

IR observations play an important role in detecting or/and confirming the fact of dust condensation in the ejected matter from a nova. The value of IR and UV observations increases in the case of a smooth light curve, without temporal minimum, since the latter occurs only in the case of dust condensation on the line of sight to a central source (see, for example, \citep{Rozenbush1988b}). \cite{Rozenbush1988b} suggested that the dust condensation in DQ Her occurred out in the equatorial and "tropical" belts and polar caps of its ellipsoidal shell \citep{Williams1978}; the equatorial belt is in the plane of the orbit of the binary system. In the case of dust condensation, but in the absence of a temporal minimum, the light curve of a nova exhibits an almost linear initial decline of brightness, which will be replaced by the final decline near log(r)>15.4. The light curves of the novae in this group have the state of maximal brightness of different lengths; with rare exceptions, these are slow objects in terms of brightness decline in the first day after the outburst maximum.  

In the designation of the group, we will focus on "DQ Her", as the traditional designation of similar novae, and in the future we will use the nomenclature: the DQ Her group, the subgroups D1 (DQ Her), D2 (DQ Her), D3 (DQ Her), D4 (DQ Her) (which is mutually equivalent to the above designation "D1 V1280 Sco", etc.). 

After a visual comparison of the light curves of almost all known novae with prototypes (asterisked in following tables), Tables 2, 3, 4, and 5 were formed by members of the DQ Her group. Candidates into the group are noted by italics. This and subsequent similar tables contain the main two parameters for reproducing light curve in the modified scales: m${_q}$ - accepted magnitude of a quiescence state, visual or V-band, if nothing else is indicated (p - photographic or in the photometric B bands, R (R${_C}$), I (I${_C}$)), and t${_0}$=[nnnnn]=mJD$\equiv$t-JD 2400000. Column 3 contents the known pre/post-nova brightness and the data source, column 4 shows the sources of observational data (especially the first dates of outburst detection) that complement the observational data of the AAVSO members \citep{Kafka2019}, which are present in almost all light curves. The ":" icon after the magnitude of a post-nova is the brightness of its possible candidate. Similar tables are accompanied by comments for some novae, if when determining membership in a particular group it was necessary to justify our conclusion or the nova showed a own peculiarity. For all novae of the DQ Her group, the light curves at the maximal brightness state have the same shape: basically a convex light curve with slight fluctuations and a monotonous trend toward a decrease in brightness, all within about 1.5${^m}$. The duration of the maximal brightness state depends on the moment of onset of dust condensation according to the type D1, D2 or D3. A trend of the initial decline only has time to become apparent and is interrupted by a rapid drop in brightness of a typical shape with the subsequent recovery of the trend of the initial decline. 

A temporal brightness drop of type D1 starts at log(r)$\propto$14.2-14.4 (Fig.3, Table 1). For some members of the D1 subgroup, the observations could was ended at the initial phase of brightness drop (V720 Sco), which allows for the possible presence of other dust condensation episodes. In the absence of further observations, there is a danger of including dwarf novae in the DQ Her group, especially those defined as "tremendous outburst amplitude dwarf novae" or TOADs (see review of \cite{Kato2015}). The brightness drop of dwarf novae coincides in time with the beginning of dust condensation of the D1 DQ Her type and is equally fast. At the maximum state, the light curve of dwarf novae also has a trend towards a decrease in brightness, but it is monotonous and without local fluctuations (see the previous paragraph, as well as our comment on V476 Sct below). Spectral observations can help to avoid this erroneous conclusion. 

A once time, only for the subgroup D1, in Table 1 is presented the shift $\Delta$ of the light curve  which are necessary for a more complete coincidence of the light curve with the prototype. The average displacement of all novae subgroups D1 is practically zero and this illustrates both the admissible inaccuracy of determining the maximum moment t${_0}$ and the individuality of novae. For the remaining subgroups D2 and D3, the light curves did not be shift. 

The light curves of the members of the D2 subgroup are presented in Fig.4, 5. The main distinguishing feature of this subgroup is the start of a temporary weakening of brightness at log(r)$\approx$14.6-14.8 that is later in comparison with the first subgroup. In the absence of a brightness minimum, i.e. smooth light curve, the conclusion of membership to the DQ Her group is based on the detection of dust formation process by the IR or UV observations (see comments below for specific novae). The light curves of some members of the subgroup D2 are intermediate with the D3 subgroup. In Fig.5, light curves are plotted, which, with an equivalent shift, can be included in the D3 subgroup. For the purpose of our study - a review of the light curves of novae - this is not critically, therefore, we acted in this way. As a kind of boundary for including the intermediate novae to the D2 group, the criterion was accepted that the onset of minimum occurs at log(r)<14.9.

Figure 6 presents light curves of novae of the D3 subgroup; the V1280 Sco light curve is not shown so as not to overload the drawing. The outbursts amplitudes was equated with DQ Her and T Aur, the prototypes of the subgroup. In Fig.4, 5, and 6, there is no identification of light curves, since this does not carry additional semantic load, except for a general idea of the range of variations of light curves. The onset of a brightness minimum or the transition phase of a outburst refers to log(r)$\approx$15.1.

The ambiguity with some novae of the D2 subgroup was reflected on the population of this subgroup in comparison with the other two subgroups. The first subgroup confidently assigned 18 galactic novae, the second - 31, the third - 17. For the LMC and the SMC: 0, 9 and 1, respectively. This result rather qualitatively reflects the trend in the distribution of novae among subgroups, since there are no clear boundaries and some novae occupy an intermediate position, especially from subgroups D2 and D3. 

\begin{figure}[t]
	\centerline{\includegraphics[width=78mm]{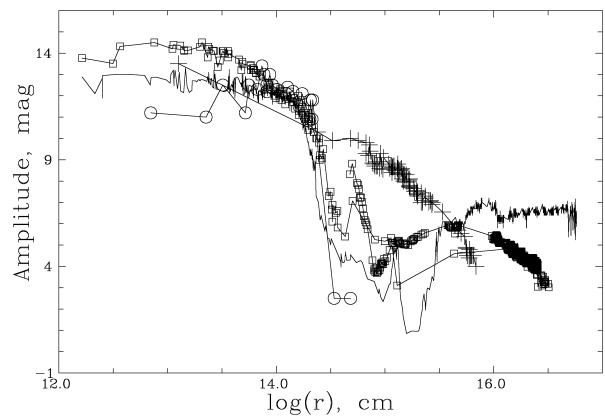}}
	\caption{The light curves of novae of the D1 subgroups: V1280 Sco - solid broken line, V697 Sco - solid broken line with pluses, V720 Sco – broken line with circles. V5579 Sgr is represented by two light curves: a line with squares - visual data from the AAVSO database \citep{Kafka2019} and SMARTS V photometry \citep{Walter2012}, a dashed line with squares with an abscissa beginning at about 14.7 - the OGLE photometry \citep{Mroz2015}.\label{Fig3}}
\end{figure}

\begin{figure}[t]
	\centerline{\includegraphics[width=78mm]{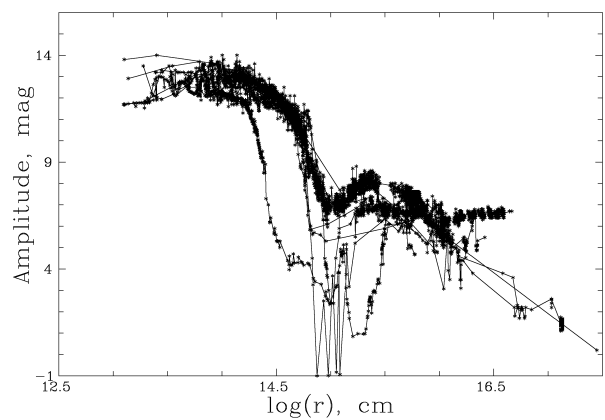}}
	\caption{Light curves of novae of the D2 subgroup (see text).\label{Fig4}}
\end{figure}

\begin{figure}[t]
	\centerline{\includegraphics[width=78mm]{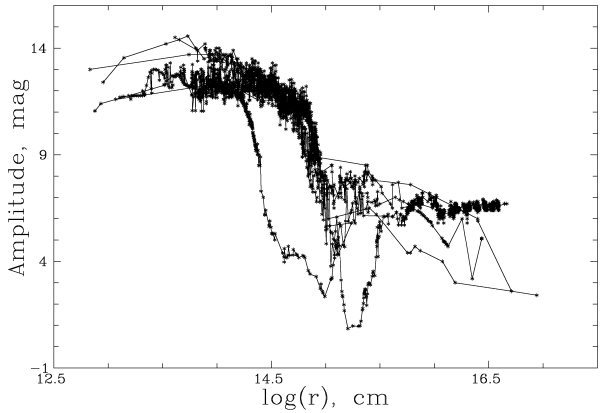}}
	\caption{Light curves of novae of the D2 subgroups, which occupy an intermediate position with the D3 subgroup (see text). \label{Fig5}}
\end{figure}

\begin{figure}[t]
	\centerline{\includegraphics[width=78mm]{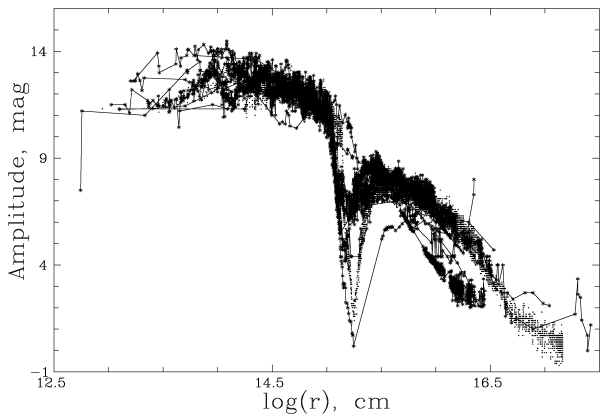}}
	\caption{Light curves of novae the D3 subgroup. DQ Her - small dots; T Aur – broken line with dots (the lowest and widest line during the transition phase).\label{Fig6}}
\end{figure}

V1280 Sco is the only nova, which, after the light minimum reached the stable high level of brightness (this emphasizes the need for long-term monitoring of novae brightness). The V5579 Sgr can be considered the second nova with three episodes of dust condensation: the first two are visible on the visual light curve according to AAVSO and SMARTS data, the second and third ones is presented in the OGLE photometry (Fig.2), but its light curve from the I photometry does not finished by the stable high level of brightness similarly to V1280 Sco.

\begin{center}
	\begin{table}[t]%
		\centering
		\caption{List of the D1 DQ Her subgroup.\label{tab2}}%
		\tabcolsep=0pt%
		\begin{tabular*}{240pt}{@{\extracolsep\fill}rccc@{\extracolsep\fill}}
			\toprule
			\textbf{Nova} & \textbf{m${_q}$/t${_0}$/$\Delta$} & \textbf{m1/m2, [ref]} & \textbf{Ref}  \\
			\midrule
			V1280 Sco* &	17.0/54146.8 &	>19.3R/ [1]	& [2] \\
			V2361 Cyg* &	22/53405/0 &	>19 &  \\
			OS And &	20.5/46767/-0.2 & 18.44B/ [3] &	[4]   \\
			V1301 Aql &	23.5/42567/0 &	>20.0 &	[5] \\
			\textit{V1831 Aql} &	\textit{28.5/57290/-0.25}	& &	\textit{[6]} \\
			DD Cir &	22/51414/0 &	/20.1 [7]&	[8] \\
			DZ Cru&	22.5/52871/0.05&	>20/&	[9]\\
			DI Lac&	17p/18997/0&	/14.8B [10]&	[11]\\
			V2295 Oph&	22/49091/0&	>21.0&	[12]\\
			V2574 Oph&	22.5/53109/0& />18.20I [13]&	[14]\\
			V3889 Sgr&	22/42606/0&	/20.86 [13]&	[15]\\
			V4740 Sgr&	19.5/52157/0&	/>18 [16]&	[17]\\
			V5579 Sgr&	21/54579/0&	>23.0&	[18]\\
			V5855 Sgr&	20.2/57684/0&	>22&	[19]\\
			V5857 Sgr&	24/58217/0&	21.9&	[20]\\
			V697 Sco&	20.5p/30044/0&	/19.8 [21]&	[22]\\
			V719 Sco&	23p/30479/-0.1&	/20.5j [25]&	\\
			V720 Sco&	20.5/33492/-0.1&	/>18 [24]&	[24]\\
			V721 Sco&	21p/43514/0&	/>18j [25]	&[24]\\
			V1662 Sco&	23.5/58157/0&	>18	&[26]\\
			LMCN 2000-07a&	23/51726&	&	\\
			\bottomrule
		\end{tabular*}
    	\begin{tablenotes}
	    	\item Note: The main data source \citep{Kafka2019} is supplemented by data from publications: 1 - \citep{Das2008}, 2 - \citep{Hounsell2016},  3 - \citep{Collazzi2009}, 4 - \citep{Suzuki1986}, 5 - \citep{Wild1975a}, 6 - \citep{Shappee2015}, 7 - \citep{Woudt2003}, 8 - \citep{Liller1999}, 9 - \citep{Tabur2003}, 10 - \citep{Pavlenko2002}, 11 - \citep{Walker1933}, 12 - \citep{Camilleri1993}, 13 - \citep{Mroz2015}, 14 - \citep{Takao2004a}, 15 - \citep{Kuwano1975}, 16 - \citep{Strope2010}, 17 - \citep{Pereira2001}, 18 - \citep{Nishiyama2008a}, 19 - \citep{Itagaki2016}, 20 – \citep{Kojima2018}, 21 - \citep{Warner2002}, 22 - \citep{Mayall1946}, 23 - {Szkody1994}, 24 - \citep{Herzog1951}, 25 - \citep{Duerbeck1987a}, 26 - \citep{Nishimura2018b}.
    	\end{tablenotes}
	\end{table}
\end{center}

\begin{center}
	\begin{table}[t]%
		\centering
		\caption{List of the D2 DQ Her subgroup.\label{tab3}}%
		\tabcolsep=0pt%
		\begin{tabular*}{240pt}{@{\extracolsep\fill}rccc@{\extracolsep\fill}}
			\toprule
			\textbf{Nova} & \textbf{m${_q}$/t${_0}$} & \textbf{m1/m2, [ref]} & \textbf{Ref}  \\
			\midrule
			V842 Cen*&	18/46754&	18.3/ [27]&	[27] \\
			V606 Aql&	20p/14759&	/20.8B [28]&	\\
			V1229 Aql&	21/40686&	/18.59 [23]&	[29]\\
			V1419 Aql&	20.8/49119&	21/ [16]&	[30]\\
			V1723 Aql&	29/55435&	<18Rc&	[31]\\
			V1212 Cen&	21.5/54703&	>19.0&	[33]\\
			V1368 Cen&	23.5/56007&	>17.8/&	[34]\\
			V809 Cep&	25/56321&	>20	&[35]\\
			V357 Mus&	19.5/58120&	>18&	[36]\\
			V3662 Oph&	26.5/57864&	>22&	[37]\\
			V3698 Oph&	25I/55848	& &	[13]\\
			HS Pup&	21/38378&	/20.5p [25]&	[38,39]\\
			FM Sgr&	22p/24722	&/17.792:I& [13]	\\
			\textit{V1175 Sgr} &\textit{21.5p/34060}&\textit{/>21I  [13]}&\textit{[40]}\\
			V2572 Sgr &	20.2p/40407 &	/17.8R [90] &	91 \\
			V5584 Sgr&	22.5/55129&	>21	&[41]\\
			V5592 Sgr&	21.5/56115.9&	>20	&[42]\\
			V5669 Sgr&	22.3/57291&	20.9&	[43]\\
			V382 Sco&	22p/15620&	>16.5p&	[44]\\
			V1178 Sco&	22.5/52041&	$\geq$20.5B/ [45]&	[46]\\
			V1324 Sco&	23/56086&	>19&	[47]\\
			V1655 Sco&	25.5/57548&	>19:&	[48]\\
			V1656 Sco&	26.3/57638&	<22/&	[49,50]\\
			V1661 Sco&	25.5/58135&	>18	&[51]\\
			V1663 Sco&	25.5/58179&	>17.7&	[52]\\
			\bottomrule
		\end{tabular*}
    	\begin{tablenotes}
        	\item Note: Continued references: 27 - \citep{McNaught1986}, 28 - \citep{Tappert2016}, 29 - \citep{Honda1970b}, 30 - \citep{Yamamoto1993}, 31 - \citep{Nishiyama2010b}, 33 - \citep{Pojmanski2008}, 34 - \citep{Seach2012}, 35 - \citep{Nishiyama2013}, 36 - \citep{Kaufman2018b}, 37 - \citep{Itagaki2017}, 38 - \citep{Strohmeier1964}, 39 - \citep{Huth1964}, 40 - \citep{Taboada1952}, 41 - \citep{Nishiyama2009a}, 42 - \citep{Nishiyama2012}, 43 - \citep{Itagaki2015}, 44 - \citep{Swope1936}, 45 - \citep{Andersen2001}, 46 - \citep{Haseda2001}, 47 - \citep{Wagner2012}, 48 - \citep{Nishimura2016}, 49 - \citep{Fujikawa2016}, 50 - \citep{Stanek2016a}, 51 - \citep{Nishimura2018a}, 52 - \citep{Stanek2018a}, 90 - \citep{Bateson1969}, 91 - \citep{Tappert2012}. 
	    \end{tablenotes}
		\end{table}
\end{center}

\begin{center}
	\begin{table}[t]%
		\centering
		\caption{List of the D2 (continued Table 3) and D3 DQ Her subgroups.\label{tab4}}%
		\tabcolsep=0pt%
		\begin{tabular*}{240pt}{@{\extracolsep\fill}rccc@{\extracolsep\fill}}
			\toprule
			\textbf{Nova} & \textbf{m${_q}$/t${_0}$} & \textbf{m1/m2, [ref]} & \textbf{Ref}  \\
			\midrule
			EU Sct	&22/33129&	/19.1: [7]&	[53]\\
			V475 Sct&	21/52877&	>16	&[54]\\
			LW Ser&	22/43568&	/19.4 [16]&	[55] \\
			XX Tau	&20/25152	&/19.8 [23]&	[56]\\
			CQ Vel&	22.5/29731&	/21.1 [57,58]&	[59]\\
			QV Vul&	20/47112&	/18 [16]&	[60]\\
			LMCN 1988-03a&	24/47241	&&	[61]\\
			LMCN 1999-09a&	24.5I/51410	& &	[62]\\
			LMCN 2002-02a&	25/52335	& &	[63] \\
			LMCN 2003-06a&	25/52808	& &	[64] \\
			LMCN 2011-08a&	25I/55760	& &	[62]\\
			LMCN 2018-07a&	24.5/58280	& &	[65]\\
			SMCN 1897-10a&	24.5p/14209	& &	[66]\\
			SMCN 1994-06a&	27B/49522	& &	[67]\\
			\textit{SMCN 2016-10a}&	\textit{24.5/57671}&	 &	\textit{[68]}\\
			D3 subgroup & & &	\\		
			T Aur* &	16.6p/12076 &	/14.9 [16] &	[11]  \\
			DQ Her*	 &14.7/27784 &	/14.3 [16] &	\\
			\textit{V604 Aq}l &	\textit{23/17074} &	\textit{18p} &	\textit{[11]}\\
			OY Ara &	20p/18761 &	/18.55 [25] &	[11]\\
			V705 Cas &	19.2/49327 &	/16.4 [16] &	[69]\\
			V435 CMa &	23/58202 &	21.3 &	[70]\\ 
			V450 Cyg &	20.5p/30515 &	/20.06B [71] &	[72]\\
			QY Mus &	21/54704 &	20B	 &[2]\\
			V2615 Oph &	23/54178 &	20 &	[73]\\
			\bottomrule
		\end{tabular*}
		\begin{tablenotes}
			\item Note: Continued references: 53 - \citep{Bertaud1953}, 54 - \citep{Nakano2003}, 55 - \citep{Honda1978}, 56 - \citep{Beyer1929}, 57 - \citep{Woudt2001}, 58 - \citep{Schmidtobreick2005}, 59 - \citep{Hoffleit1950}, 60 - \citep{Beckmann1987}, 61 - \citep{Garradd1988}, 62 - \citep{Mroz2016b}, 63 - \citep{Liller2002}, 64 - \citep{Liller2003}, 65 - \citep{Stanek2018c}, 66 - \citep{McKibbenNail1951}, 67 - \citep{deLaverny1998}, 68 - \citep{Shumkov2016}, 69 - \citep{Kanatsu1993}, 70 - \citep{Nakamura2018}, 71 - Rosenbush, unpublished, 72 - \citep{Steavenson1950}, 73 - \citep{Nishimura2007}.
		\end{tablenotes}
	\end{table}
\end{center}				
	
\begin{center}
	\begin{table}[t]%
		\centering
		\caption{List the D3 (continued Table 4) and V838 Her subgroups\label{tab5}}%
		\tabcolsep=0pt%
		\begin{tabular*}{240pt}{@{\extracolsep\fill}rccc@{\extracolsep\fill}}
			\toprule
			\textbf{Nova} & \textbf{m${_q}$/t${_0}$} & \textbf{m1/m2 [ref]} & \textbf{Ref}  \\
			\midrule
			V2676 Oph &	23.5/56010 &	>21 &	[74]\\
			V3665 Oph &	23.5/598187 &	21r:/ &	[75]\\
			V726 Sgr&	24.5p/28300&	/19.4&	[76]\\
			V732 Sgr&	18.8p/28260&	/16 [16]&	[77]\\
			V1014 Sgr&	23.5p/2415520&	/19.598I [13]&	[11]\\
			V5668 Sgr&	17.5/2457096&	16.2&	[78]	\\
			FH Ser&	18/40630&	/16.8 [16]&	[79]\\
			NQ Vul&	20/43070&	/17.2 [16]&	[80]\\
			LMCN 2009-05a&	25/54935&	&	[81]\\
			LMCN 2018-05a&	26.5/58242&	&	[82]\\
			V838 Her subgroup	& & &	\\
			V838 Her*&	19.1/48340&	/19.1 [83]&	[84, 85]\\
			V3661 Oph&	22.5/57459&	/>21I [86]&	[87]\\
			V630 Sgr&	17.6/28445.65&	/17.6 [57]&	[88]\\
			LMCN 1991-04a&	21.5/48368&	&	[89]		\\	
			\bottomrule
		\end{tabular*}
		\begin{tablenotes}
			\item Note: Continued references: 74 - \citep{Nishimura2012}, 75 - \citep{Nishimura2018c}, 76 - \citep{Mayall1938}, 77 - \citep{Swope1940}, 78 - \citep{Seach2015}, 79 - \citep{Honda1970a}, 80 - \citep{Alcock1976}, 81 - \citep{Liller2009}, 82 – \citep{Chomiuk2018}, 83 - \citep{Szkody1994a}, 84 – \citep{Sugano1991}, 85 - \citep{Ingram1992}, 86- \citep{Mroz2016d}, 87 - \citep{Yamamoto2016}, 88 - \citep{Gaposchkin1955}, 89 - \citep{Liller1991b}.
		\end{tablenotes}
	\end{table}
\end{center}

\textbf{Subgroup D1}. The fact, that the light curve of V1301 Aql with the parameters of Table 2 belongs to the DQ Her group, suggests that the most likely candidate for a post-nova we can consider the star ‘F’ (R=22.18${^m}$ from the list of 4 candidates of \cite{Tappert2015}. 

To classify of Nova Aquilae 2015, V1831 Aql, reddened by extremely high interstellar absorption, was significantly helped the study of \cite{Banerjee2018}, in which records the fact of dust condensation.

The light curve of DD Cir has indications for a dust condensation by  of the 1st and 3rd episodes types; there are no enough observations at the moments of the second episode. Both temporary brightness minima had the depth of about 1${^m}$, which can be interpreted as a consequence of the small inclination of the orbit plane of the binary system with respect to the line of sight. IR observations of DD Cir are absent. 

Visual observations of the DZ Cru ended at the very beginning of the temporary weakening, which does not give an unambiguous indication of the formation of a dust envelope, since a similar decline in brightness may take place in the TAODs. But for the latter, the outburst amplitude should be lower by 3${^m}$, also the drop in brightness after the state of maximal brightness should be faster and smoother. The presence of dust in the vicinity of the post-nova is confirmed by IR data of \cite{Evans2014}. 

The V2361 Cyg outburst can be compared with V5579 Sgr: a short and peak state of maximum, a deep and prolonged temporary weakening of brightness. A few initial days of the state of maximal brightness, about 6 days, were missed. Dust condensation is confirmed by infrared observations of \cite{Russell2005}. 

DI Lac dust condensation occurred outside the line of sight of the main radiation source. 

We supplemented the AAVSO data for V2574 Oph in several steps with several points averaged over the data of \cite{Kang2006} and \cite{Walter2012}. We, as well as the authors, shifted the data from the first source by +2${^m}$ to compensate for differences in instrumental systems. Since the photometry was not calibrated for the data from the second source, we estimated that the shift necessary for combining with photometry in the I band of the OGLE survey \cite{Mroz2015} is to be 11.1${^m}$. And the OGLE data supplemented in this way was shifted by +1${^m}$ to compensate for differences in instrumental systems: visual and in the I band; here we assumed that the colour index V-I decreased to this value (+1${^m}$) compared to +2${^m}$ at the phase of the initial brightness drop. As a result, we received a more complete picture of the behaviour of brightness of a nova at the final phase of the outburst (Fig.6), and some conclusions can be drawn. V2574 Oph is an analogue of V1280 Sco, but without obvious signs of dust condensation on the visual light curve: a typical temporary light weakening. Dust condensation can be suspected based on the increasing deviation of the visual light curve of the nova from the smooth light curves of the Her1 subgroup (see below) in terms of dust condensation by types D1 and D2. An interesting detail of the V2574 Oph light curve that should be noted is that the light curve reaches a very possible plateau of the V1280 Sco type at the final phase of the outburst. But the brightness level on this plateau is lower than that of V1280 Sco. The plateau length is definitely shorter: according to the data of \cite{Mroz2015}, the I brightness began to decline rapidly for log(r)>16.4 (Fig.7). Other novae did not display the similar brightness behaviour, i.e. the presence of a plateau at the final phase of the outburst is a rare phenomenon and to record it prolonged observations are needed. 

\begin{figure}[t]
	\centerline{\includegraphics[width=78mm]{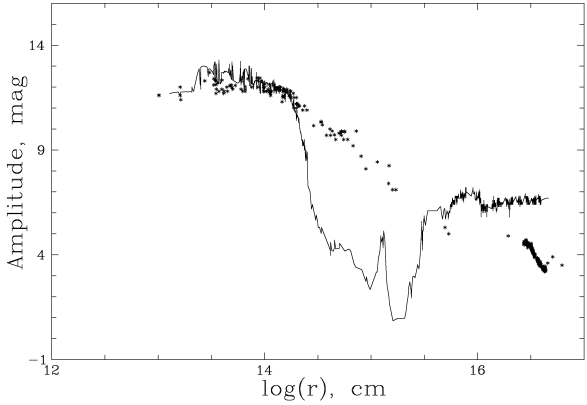}}
	\caption{Light curves of V1280 Sco (line) and V2574 Oph (points).\label{Fig7}}
\end{figure}

V697 Sco classification is based on graphic material from a publication of \cite{Mayall1946}, covering the entire period of visibility of a given area of the sky. But the presentation of the light curve in the modified scales in the presence of (1) a sufficiently clear indication on temporary light weakening, (2) the subsequent light curve and (3) the outburst start date limit allowed us to dwell on the option of parameters that more successfully satisfy the D1 prototype (Table 3). A confidence is strengthened by one of the results of the study of \cite{Warner2002} about the belonging of the post-nova to the intermediate polars, like DQ Her. Such an interpretation allows us to say that V697 Sco had a temporary brightness weakening of shallow depth and the outburst was detected just at the time when it was achieved  the minimal brightness  in this minimum. By extrapolation it can be estimated that the maximum near m${_pg}$$\approx$7${^m}$ was reached near the date of JD 2430044. That is, the maximum brightness of nova was occurred almost 20 days before the first observations of \cite{Mayall1946}. 

V719 Sco light curve \citep{Herzog1951, Bok1951} has a typical shape with a shallow minimum of the D1 DQ Her type (Table 3) and ends before reaching the phase of a final decline. 

LMCN 2000-07a has a limited photometry. Outburst detection by \cite{Liller2000} occurred in JD 2451737.91. \cite{Duerbeck2000} based on the state of the spectrum for JD 2451740.9 concluded that the nova was already almost a week after the maximum. In the same short message of\cite{Duerbeck2000} is added: the CCD image on JD 2451724.85 has not a nova, but it appeared after 0.33 day and already at JD 2451727.18 it was in its brightest state. Unfortunately, the authors did not fulfilled their promise to calibrate the images. \cite{Greiner2003} with a link to a private message of I.Bond noted: all images except the first one on JD 2451725.18 show the nova to be saturated. Based on these boundary values, we assume that at JD 2451726 the nova was already in a state of maximum brightness, and this is the parameter t${_0}$. With this parameter, the modified light curve of LMCN 2000-07a corresponds well with the prototype V1280 Sco at the very beginning of the D1-D3 light minimum, and it can be seen that minima of the D2 and D3 types were with same slight weakening of 1-1.5${^m}$. Latest Photometry by \cite{Greiner2003} refers to JD 2451998: the brightness in the U band was 15.70$\pm$0.02${^m}$. Assuming that the colour index U-V=-1.03${^m}$ of \cite{Kilmartin2000} has not changed over a time, we find V=14.7${^m}$. In such a manner we are convinced that the nova began to restore brightness after a series of three minima and this is really a nova of the DQ Her group with dust condensation of the D1, D2, and D3 types of a small depth of light minima. Since the LMCN 2000-07a, like the V1280 Sco, had three light minima and both light curves are similar, we make assuming the physical parameters of stellar systems are equal. We can transfer the numerical values of these parameters from one system to another, including the outburst amplitude (different the depth of light attenuation at minima can arise due to the difference in the orientation of stellar systems with respect to the observer’s line of sight). The average outburst amplitudes are A$\approx$12.8$\pm$0.5${^m}$, from here we find the average maximum brightness LMCN 2000-07a V=10.0$\pm$0.5${^m}$. This value is almost equal to the brightness extrapolation of \cite{Greiner2003}: m${_V}$V=10.5${^m}$. 

\textbf{D2 subgroup}. For V1419 Aql, there are two AFOEV member observations into the AAVSO database: JD 2449107 and 2449108, but they are not in the AFOEV database. But with them, better agreement is reached with the D2 prototype (Table 3), so we left them in our data array.

The classification of V1723 Aql is a very illustrative example to demonstrate the possibilities of the proposed application of modified light curves. V1723 Aql was very weak nova: the visual brightness at the maximum was about 16.5${^m}$, which led to a very limited amount of photometry \citep{Kafka2019}, especially since the outburst was detected 15 days after the extrapolated moment t${_0}$. To obtain a more complete picture for the light curve, it was assumed that the colour indices V-R and V-I did not change during the observations period. This allowed us to transform the light curves in the bands R and I into the light curve in V. The parameters of Table 3 were chosen to better match this light curve of V1723 Aql with the typical light curve with the D2 type condensation. This visual minimum coincided with the early-time flare of the radio flux of V1723 Aql according to \cite{Weston2016}. So, all three episodes \citep{Weston2016} when the radio flux densities reached local maximum: "the early-time flare around 45 days, ... another local maximum at all frequencies between days 200 and 400, ... and an additional local maximum between these two bright peaks at the higher frequencies"  are coincided in time with episodes D2, D3 and D4. Such a set of coincidences, in addition to the meagre light curve, leaves us no other choice but to include V1723 Aql in the DQ Her group with dust condensation by the D2, D3 and D4 types.

The belonging of the light curve of V809 Cep to the D2 subgroup is not in doubt. It becomes clear that the data for JD 2456384.33 and 2456385.33 with magnitudes of about 16.5${^m}$ from the AAVSO database are most likely erroneous: at this time there was a temporary deep weakening of the visual brightness below 18-19${^m}$. Photometry after the completion of a temporary weakening is presented by \cite{Munari2014}.

Photometry of the ASAS-SN survey \citep{Shappee2014, Kochanek2017} made it possible to trace the development of the N Mus 2018, V357 Mus, 10 days before the outburst was discovery: on 3 days, the nova rise from 15${^m}$ to 8.6${^m}$ and weakened to 11.6${^m}$. These data eliminated the shift in the visual light curve relative to the prototypes light curve of the D2 subgroup. The shape of the light curve is completely typical in the state of maximal brightness with the parameters of Table 3.

The light curve of Nova Oph 2017, V3662 Oph, as a member of the DQ Her group, according to the AAVSO with the addition of \cite{Walter2012, Itagaki2017}, is well represented by the parameters of Table 3. The possible dust condensation in the ejected envelope has been reported of \cite{Raj2017} and others. The data of \cite{Walter2012} make it possible to estimate the moment of the maximum optical thickness of the dust layer on the line of sight near JD 2457980: at this date, there is a minimum of R magnitude and the most red colour index H-K.

The HS Pup light curve \citep{Strohmeier1964, Huth1964} ends with a typical D2 light minimum. A shell is observed near the post-nova \citep{Gill1998}. 

The outburst of V1175 Sgr was recorded after the end of the period of the connection of this region of the sky with the Sun. The modified light curve has a characteristic detail: a slightly variable state of maximal brightness is replaced by a rapid decline of 2${^m}$. In the OGLE survey \citep{Mroz2015}, it was not possible to propose a variable candidate for a post-nova within a radius of 10" from inaccurate coordinates of \cite{Taboada1952}. Based on the limiting magnitude of the OGLE I>21${^m}$ survey, we can assume that the photographic magnitude of the post-nova is not brighter. In this case, it is possible that V1175 Sgr belongs to the group of novae with dust condensation according to the D2 type (Table 3). 

The light curve of Nova Sagittarii 2015, V5669 Sgr, had the light curve form during the initial decline similar to novae with a dust condensation by the D2 type but without the sign of dust drop. It is followed by the season of object invisibility, after which the brightness followed according to the typical phase of the final decline (Table 3). 

The old nova of V382 Sco has 5 photographic magnitudes and several limiting values. But such a small set allows us to admit that this nova belongs to the D2 subgroup with the parameters of Table 3. 

The light curve of V1178 Sco was classified as the D2 DQ Her (Table 3). But, like \cite{Kato2001}, we would like also to note the presence of a brightness fluctuation with an amplitude of about (-2${^m}$ ), there may have been also a second similar fluctuation, but less noticeable due to the large spread of the brightness estimates, before the subsequent light minimum due to dust condensation by the D2 DQ Her type according to the OGLE photometry data \cite{Mroz2015}. These brightness fluctuations were not related to dust condensation, since the latter gives a light curve of a characteristic shape: a rapid decline and a slow recovery (see also above V1280 Sco). In this regard, one should pay attention to a similar fluctuation in NQ Vul (see below), which it coincides with the onset of dust condensation by the D1 type (Fig.8) but it not due to a dust. Another feature is associated with a rather "flat" shape of the light curve in the state of maximal brightness, usually it is convex. \cite{Andersen2001} identified a possible candidate for the pre-nova: a faint red star with B-R$\geq$2.3${^m}$ . 

The V1656 Sco light curve was assigned to D2 subgroup: IR photometry of \cite{Walter2012} showed the development of IR excess in the corresponding time. But the visual light curve did not demonstrate a noticeable light minimum due to dust condensation, therefore, the inclination of the binary system orbit is enough less than 90${^o}$. 

The post-nova EU Sct has not yet been identified: Table 3 shows the brightness of the candidate according to \cite{Woudt2003}. Therefore, modern photometry in the AAVSO database refers to a field star. 

The visual light curve of LMCN 1988-03A was without a temporary light weakening during the transition phase, but we classified it as a nova of the D2 DQ Her group, since \cite{Schwarz1998} from IR and UV observations found that between 50 and 100 days after the principal visual maximum there was a secondary maximum of IR brightness and a local minimum of UV brightness, and these are signs of dust condensation \cite{Rosenbush1996c}. The time interval of 50-100${^d}$ in our scale corresponds to the second episode of dust generation in the V1280 Sco. The absence of a temporary weakening of the visual brightness indicates the condensation of dust outside the line of sight, possibly the dust structure had a disk shape \cite{Rozenbush1988b}. 

For LMCN 1999-09a, there is only OGLE photometry in the I band \citep{Mroz2016b} and an estimate of the maximum brightness V=12.5${^m}$ by \cite{Shida2004}. But this data is enough to confidently classify as the nova of the DQ Her groups (Table 4). The shape of the light curve allows us confidently to refer the moment of maximum brightness 22${^d}$ earlier than the first observation and the maximum brightness is I$\approx$12.2${^m}$. It should be noted that the outburst amplitude in the I band may be less than in the V band. 

Visual observations of \cite{Liller2002, Mason2005} and the CCD data of \cite{Mroz2016b} in the I band complement well  each other to give us an indication on the onset of temporary light decline in LMCN 2002-02a by the D2 type (Table 4). 

Visual observations of \cite{Liller2003} and OGLE data \citep{Mroz2016b} complement well each other to have an idea of the light curve of LMCN 2003-06a (Table 2 )4)). One can speak of dust condensation of the D2 DQ Her type with the orientation of dust structures outside the line of sight. Note the stabilization of I brightness at the level of 19-20${^m}$, which is higher than the threshold sensitivity of the OGLE equipment. This stabilization can be attributed to the contribution of a weak field star, but this level almost coincides with the level of long-term brightness stabilization of V1280 Sco. It is possible that the latter fact reflects the rarely observed property of the novae associated with a stable high level of brightness at the final phase of the outburst of the V1280 Sco type: the final phases of the outburst are often beyond the sensitivity thresholds of the telescope equipment and here we have deal the effect of the observations selection due to the cessation of the interest to the non-remarkable variability of the object. We add that the brightness in a quiescence state in the I band can be at the level of 26${^m}$. 

LMCN 2011-08a was discovered from archival materials of the OGLE-IV review \citep{Mroz2016b}. The light minimum is well represented by observations, but the state of maximum is represented by a single brightness estimate. As a result, we attached the maximum brightness to the moment JD 2455750 and, thus, the duration of the maximum brightness state was about 38${^d}$, which is close to the typical values for the novae in this group. 

Two Gaia mission brightness measurements for LMCN 2018-07a (=Gaia18boy) \cite{Stanek2018c} substantially complement the ASAS-SN and SMARTS survey data, which allows us to include this nova in the D2 subgroup. [The light curve with these data is similar to the light curves of dwarf novae, for example, GW Lib in 2007, but with an shift along the abscissa axis of more than 0.2, that we described above as a sign of belonging to another group.] 

For SMC 1994-06a, the only data source \citep{deLaverny1998} presented the light curve only in graphical form, but for our purpose this was enough: a light minimum was unambiguously recorded. Due to the lack of typical light recovery after temporary weakening, it is possible also to classify it as a dwarf nova. But we exclude this possibility on the basis of a longer and more horizontal state of maximum brightness, in contrast to dwarf novae with their short maximum state and a steady trend towards a weakening of brightness.  

When classifying SMCN 2016-10a, it was necessary to solve the problem of non-system photometry during the state of the maximum brightness. \cite{Munari2017} and \cite{Aydi2016} discussed it in detail and adjusted the data of \cite{Lipunov2016} by +1.0${^m}$. Our construction of the light curve in the modified abscissa scale required an additional correction of +0.5${^m}$, i.e. +1.5${^m}$ to the data of \cite{Lipunov2016}. \cite{Aydi2016} in their study noted plateau in the UV emission between at least days 90 and 170, which coincides with the main interval of bright, high temperature, super-soft X-rays (see below our discussion). This plateau is clearly visible in Fig.12 of \cite{Aydi2016}. Such depressions are similar to the ultraviolet sign of a temporary visual light weakening at the transition phase of an outburst \citep{Rosenbush1996c}. But IR photometry by \cite{Walter2012} did not show a local maximum of the corresponding dust condensation. Despite the lack of dust condensation in the dropped shell, we included this unique nova as a possible candidate for the D2 subgroup, since the 2016-10a SMCN visual light curve is more consistent with the V842 Cen prototype than with any other. The uniqueness of SMCN 2016-10a was also manifested in the conclusion of \cite{Hachisu2018} about its belonging to our Galaxy. 

\textbf{D3 subgroup.} The V604 Aql light curve \citep{Walker1933} was included in the D3 DQ Her subgroup, since in this case it has fewer differences from the typical light curve in contrast to other options. The next to last date of the nova productive observations falls precisely at the beginning of a possible light minimum according to the D3 type, then only the limits are estimated, and the last productive observation corresponds to a typical brightness level after the light minimum. 

QY Mus has a fairly well-presented brightness maximum and a quick decline into a shallow typical dusty minimum with subsequent recovery of the general trend. The data of \cite{Hounsell2016} substantially clarify the time when the nova reached the near-maximum brightness state, but the correction on -1${^m}$ was made for these data to match the brightness level with visual photometry. 

The onset of light weakening after a prolonged state of maximum brightness for the V726 Sgr \citep{Mayall1938} exactly coincided with a typical light weakening of the D3 DQ Her type. \cite{Mroz2015} give the likely brightness of post-nova V=19.395${^m}$, which, taking into account the colour index of V-I=1.141${^m}$, means that our estimate of the brightness in a quiescence state can differ by almost +4${^m}$, and this is the highest difference from all that we made estimations. But we do not find other options for classification.

The NQ Vul light curve has one interesting expressive detail (Fig.8), which is found in other novae, but not so clearly expressed. The onset of the first temporary weakening of D1 in all novae of the DQ Her group coincides with the onset of brightness growth in NQ Vul during an anomalous fluctuation near log(r)=14.4. 

\begin{figure}[t]
	\centerline{\includegraphics[width=78mm]{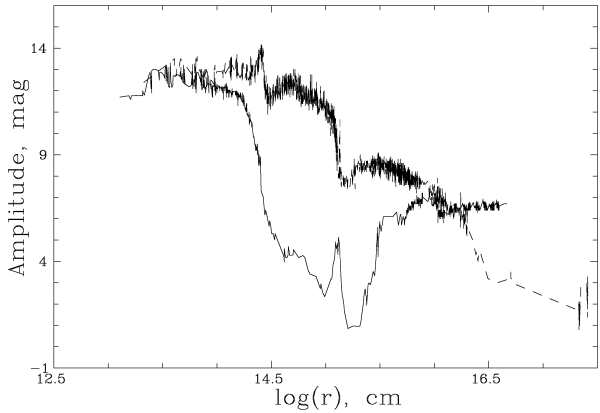}}
	\caption{Light curves of NQ Vul (upper broken dashed line) and V1280 Sco (lower broken solid line).\label{Fig8}}
\end{figure}

The parameter t${_0}$ for LMCN 2009-05a (Table 5) was obtained due to the similarity with V2615 Oph, the nova of this subgroup: by the shape of the light curve and light variations with the amplitude of about 2${^m}$ at the maximum state. According to our estimation, LMCN 2009-05a reached a state of maximum brightness almost 20 days before the detection of a outburst. 

Table 5 includes also a small \textbf{unique subgroup of V838 Her}. Already in our early research \citep{Rosenbush1999a}, we included the very fast nova of V838 Her in the DQ Her group according to the IR sign of dust formation in the ejected matter, although this was not displayed on the visual light curve. In the course of presented study, relying on the unique V1280 Sco light curve, we can confirm that the V838 Her belongs to this group along with some peculiarities (Fig.9). 

The first peculiarity concerns the maximum of light curve. Before detecting the outburst, the nova increased its brightness by more than 4${^m}$ to 5.4${^m}$ during a day \citep{Sugano1991}. And then, also during a day, it first increased its brightness to 5${^m}$ and then weakened to the level of 7${^m}$ \citep{James1991}. Of course, this could be attributed to the low accuracy of photometry, but the amplitude of the variability exceeds the possible errors and this is the data of several observers. The amplitude of the V838 Her outburst is 1${^m}$ higher than the amplitude of the V1280 Sco, but the subsequent light curve passes 1.5-3${^m}$ below the prototype. In other classical novae, we have not seen such uniqueness in the brightness change near the maximum.

\begin{figure}[t]
	\centerline{\includegraphics[width=78mm]{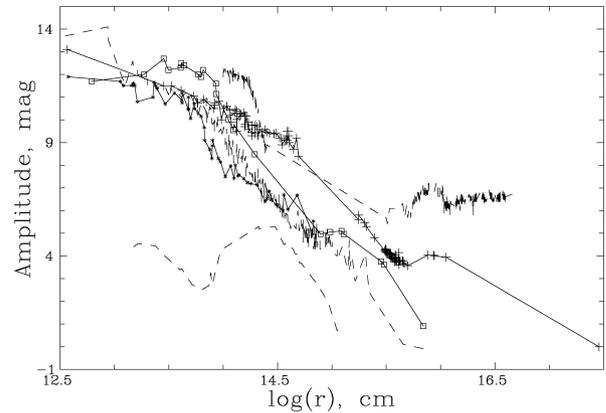}}
	\caption{Light curves of novae of V838 Her subgroup: V838 Her itself (visual and IR (L-magnitude) - upper and lower dashed lines, respectively, (the latter is shifted arbitrary down for ease of display), V3661 Oph (dotted line), V630 Sgr (line with pluses), LMCN 1991-04a (line with squares) and the prototype of V1280 Sco (rightmost dashed line from log(r)=14.0 to the level of about 9${^m}$ and continued after log(r)=15.5, in a truncated form without light minima for the convenience of displaying the other light curves).\label{Fig9}}
\end{figure}

Another feature is the outburst duration of not more than 350 days, which is atypically fast for a classical nova, but slightly longer than fast recurrent novae, for example, U Sco or V745 Sco. \cite{Pagnotta2014} consider V838 Her as a candidate for a recurrent nova with a high degree of probability: "category B contains strong RN candidates, for which many of our indicators strongly point to the system being recurrent". 

The second local maximum of the IR light curve \cite{Harrison1994} captures the moment of maximum bolometric luminosity of the dust, which corresponds to the D1 episode of V1280 Sco. This has not been displayed on the visual light curve, which can mean that dust condensation occurred outside the line of sight. 

The V838 Her subgroup contents V3661 Oph with a light curve almost a textbook example for a Fe II nova \citep{Munari2017} but which repeats well the light curve of the prototype (Table 5). The latter circumstance allowed us to include this fast nova in the V838 Her subgroup without data of dust condensation. The only difference from V838 Her in the brightness behaviour is the absence of a very short peak maximum with an amplitude of about 2${^m}$ relative to the recorded brightness (this is possibly due to the lack of corresponding observations). 

To construct the light curve of the fast nova of V630 Sgr (Fig.9), we used data from five sources: \cite{Gaposchkin1955, Kafka2019, Okabayasi1936, Parenago1949, Ponsen1957}. The parameters of Table 5 allow us to include V630 Sgr in this unique group due to a very similar shape of the light curve, but shifted to the right along the abscissa axis (Fig.9). At log(r)$\geq$14.2, there is a slight brightness depression on the light curve, which coincides with the moment of dust condensation D1 for V1280 Sco. In the final part of the light curve, which corresponds to the resumption of observations after the season of invisibility of the sky region with a nova, there is again a depression that coincides with the episode of dust condensation D4 in V1280 Sco. Therefore, it is possible that the V630 Sgr also experienced dust condensation. It is also possible that the V630 Sgr light curve after the last depression had a consistently high level of brightness like V1280 Sco. 

Another interesting case is the outburst of Nova LMC 1991, LMCN 1991-04a \citep{Liller1991b, Schwarz2001}. The rise of brightness from a quiescence state to a maximum took 8 days. Slow final rise for 3 days gave way to the same slow initial decline for 3 days. In the modified scales with the parameters of Table 5 in Fig.9, it looks like a plateau and a quick drop of 2${^m}$, followed by a light curve parallel to V838 Her with a plateau near log(r)$\approx$16.0 before the final brightness decline. Fortunately, the observations recorded the beginning of the outburst and a return to a state close to the initial one; therefore, the recorded outburst amplitude may be close to the actual. Thus, we can say that for the two novae, LMCN 1991-04a and V838 Her, the light curves differ at a maximum state, but are similar at the phases of early and final brightness decline. 

We also note some similarity of the light curves of V630 Sgr and SMCN 2016-10a, if the outburst amplitude of the latter is reduced by 2${^m}$. 

The light curves of the V838 Her subgroup can be called a kind of "exception to the rule": the light curve in the state of maximum brightness before the start of the initial decline is almost the same as the fast novae, for example, the Lacerta group (see below), and then they decrease their brightness more quickly, which is more typical for recurrent novae. But they differ from recurrent ones in the larger outburst amplitude. This combination of characteristics provides a basis for concluding that the status of V838 Her is intermediate between classical novae with a ejection of a substance and recurrent novae without a ejection of a substance during an outburst.

\subsubsection{Her1 subgroup – novae with smooth light curves}

At the initial phase of research, when reviewing novae, groups with smooth light curves inevitably arose. The light curves of some novae of the DQ Her groups were smooth, but at the transition phase of the outburst, a strong IR excess was appearing as a result of condensation of dust in the ejected shell. Since the dust component in the envelope may be distributed non-uniformed (see, for example, \cite{Rozenbush1988b}, then the weakening of the visual brightness does not occur if the line of sight does not pass through the dust medium. SMCN 2016-10a could also be assigned to a subgroup with smooth light curves, as the visual light curve was smooth, but the registration of IR excess (dust condensation) became the basis for inclusion in the D2 DQ Her subgroup (Table 3). 

The absence of a sign of temporary weakening of visual light and other observations to identify dust condensation and, on this basis, the inclusion of such novae in the DQ Her group, can be compensated by a sign of the start of an early decline. But at the same time, the slope of the light curve at the phase of an early decline is close to the slope of light curves in novae of the CP Lac group and some others.

The subgroup of novae with a shape of light curve at the beginning of the phase of early light decline, close to novae with dust condensation of the D1 type, was noted as the subgroup with the prototype of V827 Her. Denote the subgroup as Her1. The outburst amplitude was normalized by combining with the linear portion of the V1280 Sco light curve within log(r)$\approx$13.75-14.3 (Fig.10), i.e. from the beginning of the brightness decline after the maximum state to the beginning of the brightness decline as a result of the first episode of dust condensation in the V1280 Sco. 

\begin{figure}[t]
	\centerline{\includegraphics[width=78mm]{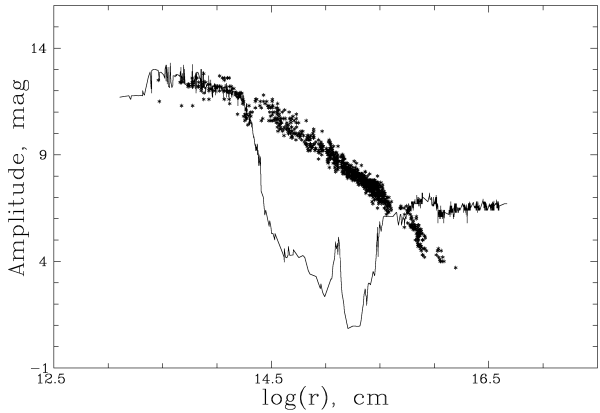}}
	\caption{Nova with a typical smooth light curve of the V827 Her (dots) compared to the V1280 Sco (solid line). \label{Fig10}}
\end{figure}

\begin{center}
	\begin{table}[t]%
		\centering
		\caption{Novae of Her1 subgroups.\label{tab6}}%
		\tabcolsep=0pt%
		\begin{tabular*}{240pt}{@{\extracolsep\fill}rccc@{\extracolsep\fill}}
			\toprule
			\textbf{Nova} & \textbf{m${_q}$/t${_0}$} & \textbf{m1/m2 [ref]} & \textbf{Ref}  \\
			\midrule
			V827 Her* &	20/46822&	/18.1 [1] &	2 \\
			V1663 Aql&	23/53530&	>18&	3\\
			V679 Car&	22/54799&	>20&	4\\
			V834 Car&	23.5/55987&	>19&	\\
			V962 Cep&	24/56725&	>21.0&	\\
			V693 CrA&	21/44696.7&	/>21 [1]&	5\\
			V2274 Cyg&	24.8/52108&	>20/ [1]&	6\\
			GI Mon	&18.7p/21593&	/16.2 [7]&	8\\
			KT Mon&	23.5/30724&	>20.0&	9\\
			V2264 Oph&	23/48356&	>21/ [1]&	10\\
			V2575 Oph&	23.5/53777&	>23.5:&	11\\
			V2576 Oph&	22.5/53831&	>17&	13\\
			\textit{V2671 Oph}&	\textit{25/54615}&	\textit{>19}&	\textit{13}\\
			V2674 Oph&	22/55256&	>20	&14\\
			V363 Sgr&	21/25100&	/19 [15]&	16\\
			V1583 Sgr&	22/25421&	/20.5 [17]&	18\\
			V4169 Sgr&	20.5/48812 &	>17 [1] &	21\\
			V4171 Sgr&	20.5/48910&	20.5j/ [22]&	23\\
			V5114 Sgr&	21.5/53080&	>18	&24\\
			V5117 Sgr&	21.5/53778&	/20.238 [25]&	26\\
			V5853 Sgr&	24/57609&	>22	&27\\
			V5926 Sgr&	26I/56387	& &	25\\
			V723 Sco&	23.5p/34236	&/18.978: [25]&	28\\
			V1141 Sco&	23/50604.5&	>20.0&	29\\					
			\bottomrule
		\end{tabular*}
		\begin{tablenotes}
			\item Note: Possible members of the subgroup are in italics. References: 1 - \citep{Strope2010}, 2 - \citep{Sugano1987}, 3 - \citep{Pojmanski2005a}, 4 - \citep{Malek2008}, 5 - \citep{Honda1981}, 6 - \citep{Nakamura2001a}, 7 - \citep{Szkody1994a}, 8 - \citep{Walker1933}, 9 - \citep{Gaposchkin1954}, 10 - \citep{Camilleri1991a}, 11 - \citep{Pojmanski2006}, 12 - \citep{Williams2006}, 13 - \citep{Nishiyama2008b}, 14 - \citep{Nishimura2010b}, 15 - \citep{Tappert2014}, 16 - \citep{Walton1930}), 17 - \citep{Tappert2016}, 18 - \citep{Dishong1955}, 21 - \citep{Liller1992a}, 22 - \citep{McNaught1992}, 23 - \citep{Camilleri1992b}, 24 - \citep{Nishimura2004}, 25 - \citep{Mroz2015}, 26 - \citep{Liller2006}, 27 - \citep{Nishiyama2016}, 28 - \citep{Taboada1952}, 29 - \citep{Liller1997}.
		\end{tablenotes}
	\end{table}
\end{center}

\begin{center}
	\begin{table}[t]%
		\centering
		\caption{Novae of Her1 subgroups (continued Table 6). \label{tab7}}%
		\tabcolsep=0pt%
		\begin{tabular*}{250pt}{@{\extracolsep\fill}rccc@{\extracolsep\fill}}
			\toprule
			\textbf{Nova} & \textbf{m${_q}$/t${_0}$} & \textbf{m1/m2 [ref]} & \textbf{Ref}  \\
			\midrule
			V1142 Sco&	22/51107&	>20.0	30\\
			V1188 Sco&	22.5/53575&	>19/ [1]&	31\\
			V1281 Sco&	23/54151&	>20	&32\\
			\textit{V373 Sct}&	\textit{19/425539}&	\textit{/18.8}&	\textit{33,34}\\
			LV Vul&	17.5/39963&	/16.9p [35]&	36\\
			V459 Vul&	20.5/54459&	20/ [37]&37\\	
			LMCN 1970-11a&	26/2440894	& &	38\\
			SMCN 2012-06a&	24.5/2456083&	&	39, 40\\
			\bottomrule
		\end{tabular*}
		\begin{tablenotes}
			\item Note: Possible members of the subgroup are in italics. References: 30 - \citep{Liller1998b}, 31 - \citep{Pojmanski2005b}, 32 - \citep{Nakamura2007}, 33 - \citep{Wild1975b}, 34 - \citep{Mattei1975}, 35 - \citep{Duerbeck1987a}, 36 - \citep{Alcock1976}, 37 - \citep{Kaneda2007}, 38 - \citep{Graham1971b}, 39 - \citep{Wyrzykowski2012}, 40 - \citep{Levato2012}.
		\end{tablenotes}
	\end{table}
\end{center}

The light curve of KT Mon \citep{Gaposchkin1954} in modified scales with the parameters of Table 6 is in good agreement with the prototype. \cite{Kato2002c} assume possible belonging of KT Mon to recurrent nova. But we can see that the fixed portion of the KT Mon light curve is typical for this group (Table 6) and the outburst amplitude is higher by 3-5${^m}$m than in the recurrent novae: \citep{Kato2002c} estimated the post-nova brightness as weaker than 20${^m}$. 

After a report by \cite{Camilleri1991a} on the discovery of a outburst of Nova Ophiuchi 1991, V2264 Oph, photometric data were supplemented by earlier observation of \cite{Yamamoto1991}. 

The few observations of V363 Sgr are successfully distributed over the large amplitude and allow us to admit belonging to the Her1 subgroup with the parameters of Table 6. This corresponds to a maximum near the photographic brightness of 8.5${^m}$ and a moment t${_0}$ in JD 2425100. 

The short light curve of V1583 Sgr, when the maximum and the beginning of the light decline on 4${^m}$ covered the time interval 76 days, allows us to include this nova according to parameters of Table 6 to this subgroup. 

The smooth light curve of V5117 Sgr was included by us to the Her1 subgroup. In this case, spectrophotometry of \cite{Lynch2006} and photometry of \cite{Walter2012} indicate the dust condensation of type D3. 

Both the visual light curve of V1188 Sco and from the I photometry of the OGLE survey have a change in the rate of light decline at the log(r) values corresponding to the beginning of the second dust condensation episode of V1280 Sco. Unfortunately, IR photometry of this nova was not carried out.  

Some difficulty arose in the classification of V373 Sct.  10–20-day brightness variations of this nova with the initial amplitude of up to 2${^m}$ were superimposed on the general trend of the initial brightness decline, that is not characteristic of this subgroup. A outburst was detected in JD 2442578.528 with a photovisual brightness of 7.9${^m}$ \cite{Wild1975b}. The state of maximal brightness has only fragmentary brightness estimates based on archival materials of observers: a dozen estimates within 50 days before the start of regular observations, and also indicate the presence of such variations. For JD 2442542, a brightness of 6.1${^m}$ is given by \cite{Mallama1975}. And the earliest date is displayed on the graphic material in the study \cite{vanGenderen1979}: near the date JD 2442512, the photovisual magnitude of the nova was 9.6${^m}$ (in the publication of \cite{Mattei1975} this corresponds to the limiting estimate), but, unfortunately, without specifying links to the source. At the same time, \cite{Rosino1978} indicated a limiting value of 12${^m}$ near JD 2442539 on the light curve graph, also without specifying an additional source. \cite{Mattei1975} also gives a graphical light curve, where for a date near JD 2442539 the nova brightness of about 9.2${^m}$ is indicated. This date was accepted by us as the start date t${_0}$ of the final rise of brightness (Table 7). The  light curve of V373 Sct with parameters of Table 7 was included to this subgroup, but with one peculiarity: this is the only nova of subgroup in which the brightness fluctuations noted above were observed at the initial phase of brightness decline. 

The light curve of LMCN 1970-11a can be included in this subgroup with the parameters of Table 7, but it should be borne in mind that it may been have a dust of the D3 type (there is a single and last point that sharply bounced down from general trend of the light curve) \citep{Graham1971b}. 

\subsubsection{Novae of Magellanic Clouds}

10 novae in the LMC and 4 in the SMC were classified by us as members of the DQ Her group (Table 2-7). In view of the limited capabilities of the telescopes equipments and the limited duration of the observational time, the final part of the light curve is absent in all cases: the magnitude of the brightest nova in a quiescence state, according to our estimate, is V$\approx$24${^m}$, that is 3${^m}$ weaker than the detection limit (I$\geq$21${^m}$) of the OGLE survey \citep{Mroz2016b}. 

\begin{center}
	\begin{table}[t]%
		\centering
		\caption{The absolute magnitude of novae of the Large and Small Magellanic Clouds of the members of the DQ Her group. \label{tab8}}%
		\tabcolsep=0pt%
		\begin{tabular*}{250pt}{@{\extracolsep\fill}rccc@{\extracolsep\fill}}
			\toprule
			\textbf{Subgroup} & \textbf{Nova} & \textbf{m$_{peak}$} & \textbf{M}  \\
			\midrule
			D2 DQ Her &	LMCN 1988-03a&	11.2V&	-7.5V\\
			&LMCN 1999-09a&	12.5V&	-6.2V \\
			&LMCN 2000-07a&	10.0V&	-8.7V\\
			&LMCN 2002-02a&	10.5v&	-8.2v\\
			&LMCN 2003-06a&	11r&	-7.6r\\
			&LMCN 2011-08a&	13.1I&	-5.4I\\
			&LMCN 2018-07a&	10.7g&	-8.0g\\
			&SMCN 1897-10a&	11.4p&	-7.8p\\
			&SMCN 1994-06a&	14.5B&	-4.7B\\
			&SMCN 2016-10a&	10.0V&	-8.9V*\\
			D3 DQ Her&	LMCN 2009-05a&	12.1V&	-6.6V\\
			V838 Her&	LMCN 1991-04a&	8.8v&	-9.9v\\
			Her1&	LMCN 1970-11a&	12.5V&	-6.2V\\
			&SMCN 2012-06a&	11.4I&	-8.7I\\
			\bottomrule
		\end{tabular*}
		\begin{tablenotes}
			\item Note: M - absolute magnitude in the corresponding photometric bands (B, p, v(is), V, g(reen), r(ed), R, I). * - \cite{Hachisu2018} express opinion on the belonging SMCN 2016-10a to our Galaxy.
		\end{tablenotes}
	\end{table}
\end{center}

The presence of the MCs novae in this group makes it possible to estimate their absolute magnitudes with small errors. 

The question of interstellar absorption for the MCs was solved by studying field stars \cite{Haschke2011}. Fortunately, the absorption in the V band in the vicinity of the novae in the LMC was low: all values were in the range A$_{V}$=0.05-0.36$^{m}$. Since these values relate to field stars, for the extinction correction for our objects, we used the average interstellar absorption A$_{I}$$\approx$0.1$^{m}$ and A$_{V}$$\approx$0.2$^{m}$. In the SMC, the average absorption is even lower: A$_{V}$$\approx$0.1$^{m}$ and A$_{I}$$\approx$0.05$^{m}$. When here was only for 1-2 stars near nova than was used values average for the MCs, or average by a set of field stars if the number of field stars was more. At this phase of the study, these values can be applied to our objects, since the possible differences will not be very high, and the scatter of the absolute brightness values for novae will be significantly higher due to its individual differences.

The distance modules of the LMC and SMC by \cite{Graczyk2014} are 18.493$^{m}$ and 18.951$^{m}$, respectively. 

The absolute magnitudes of the MOs novae are presented in Table 8. The greatest absolute magnitude is for LMCN 1991-04a, as it were, underlines the uniqueness of this nova, on that we turned to when comparing it with the V838 Her. The absolute magnitude of SMCN 2016-10a can be aggravated by calibration errors of non-system photometry in the maximum (see also the opinion of \cite{Hachisu2018} that it is a member of our Galaxy, but this reduces the absolute magnitude of the nova to 2.4$^{m}$). But the exclusion of these novae from consideration does not greatly reduce the scatter of the remaining values, which may mean that there are actual differences in the absolute magnitudes of the novae within the same group. 

The similarity of the light curves of V2615 Oph and LMCN 2009-05a, which we noted above, can also mean the similarity of their fundamental characteristics, including absolute magnitudes. \cite{Das2008} for V2615 Oph give M$_{V}$=-7.16$\pm$0.12$^{m}$. For the LMC nova, we obtained M$_{V}$=-6.6$^{m}$ (Table 8). 

The assumption made and used above about the equality of the parameters of the stellar systems LMCN 2000-07a and V1280 Sco also means the equality of absolute brightness. Consequently, the V1280 Sco can have an absolute magnitude of M$_{V}$=-8.7$^{m}$. 

\subsubsection{Preliminary discussion}

The DQ Her group novae with a characteristic property of light curves: a light minimum, is the largest and most easily recognizable group, and therefore is always recognized among the entire array of novae \citep{Duerbeck1981, Strope2010}. The light curve of nova of the DQ Her group can be characterized by a sequence from the state of maximum brightness lasting up to 5 days (moment "E" of Table 1). This is followed by an initial, linear, decline of brightness, which can be interrupted on the 15th-100th day by a light minimum of type D1-D3 V1280 Sco and ends with a final, linear, decline, which begins at 250-500 days after the maximum brightness state (moment "F" Table 1). In addition to the temporal dusty light minimum, a distinctive sign of the group is the noticeable duration of the state of maximum brightness or the value of moment E (as the Her1 group). 

The presence of 3-4 variants of the sequential condensation of dust or any combination of them in one nova is not in doubt. The population of the subgroups D1:D2:D3:Her1=18:30:16:28$\approx$1:2:1:2 can be represented as the ratio of the number of novae with dust condensation on the line of sight and outside it 4:2. And this ratio is a consequence of the non-uniform distribution of dust condensation zones in the expanding shell of a nova, i.e. dust condensation zones cover up to 66\% of the shell surface. Dust density in these zones varies from the maximum value to almost zero value: from the maximum depth of the light minima (T Aur) to 1$_{m}$ or less (DD Cir, DI Lac). The structure of the dust component in the ejected shell most likely corresponds to the bright luminescence regions in the H$_{\alpha}$ line of DQ Her \citep{Williams1978}. The position of these belts, equatorial belt and two "tropical" belts \citep{Slavin1995}, and polar caps is most likely given by the orbit of the binary system – a pre-nova system. Accordingly, if the inclination of the orbital plane of a binary is as favourable as that of DQ Her, then we will observe a temporary weakening of the central radiation source \citep{Rozenbush1988a, Rozenbush1988b}. A strong magnetic field can play an important role: DQ Her itself is an intermediate polar (see the catalogue of Mukai (2014) \footnote{//asd.gsfc.nasa.gov/Koji.Mukai/iphome/catalog/alpha.html}). In connection with the issue of dust distribution over the nova shell we draw attention to another very important assumption: in a series of light minima of two or all three variants of the D1, D2, D3 type, the depth of light weakening is never follow in a sequence - one is shallow and the other is very deep or conversely, they are always almost equal. 

The ratio of the population of the subgroups D1:D2:D3$\approx$1:2:1 can be interpreted as an almost equally likely possibility of dust condensation at appropriate distances from the star. The fourth option is a much rarer phenomenon. The population of the Her1 subgroup with the ratio of the population of the subgroups D1:D2:D3$\approx$1:2:1 can be distributed among these three subgroups.

Outbursts of a different nature may have some similar details or parts of light curves. For example, the visual light curve of the V1309 Sco outburst, interpreted as a result of the merger of two stars \citep{Tylenda2011}, can be considered as a possible unique candidate for the DQ Her group. Therefore, additional information is needed, especially spectral, starting with the velocity of motion away of the substance. It should also be borne in mind that some dwarf novae, the so-called TOADs \citep{Kato2015}, have the first half of the light curve with the initial portion of a sharp and significant drop in brightness after a maximum state, which is very similar to novae of the DQ Her group with dust condensation of type D1. 

Light minima occur in other groups, for example, RR Pic (see below).

Many novae of the DQ Her group in the classification of \citep{Strope2010} also formed a separate group - class D. But the sign of the presence of temporary brightness decline is sufficient for class D, but it is not sufficient to include a nova in the DQ Her group: other details of the light curve must also be coinciding. V476 Cyg with depression at the transition phase of the light curve near log(r)$\approx$14.5 is included in class D in the classification scheme \citep{Strope2010} only on a basis of this unique feature. In our scheme, similar specific depressions, together with the whole shape of the light curve before and after it, are a sign of novae the CP Pup group (see Section 3.3 below).  

\subsection{Lacertae group}

A group with CP Las as a prototype combines novae with smooth, light curves with out any considerable details, in which the phase of the initial decline begins immediately at the time of maximum brightness t$_{0}$. I.e. these are only fast novae in the classification by the parameter t$_{m}$. The modified light curve of nova of this group has a maximum at log(r)$\le$13.1 (point t$_{0}$ in Table 1). In what follows, we will use the such notations: the Lacertae, Lac, or CP Lac group. 

Typical light curves of novae of the CP Lac group: CP Lac and V1500 Cyg, are presented in Fig.11 with the parameters of Table 9. A change in the dispersion of points in the local parts of the light curves, for example, when nova enters in the phase of the final brightness decline near log(r)$\approx$15.6, can reflect spectral lines variations: the appearance/disappearance of bright emission lines changes the instrumental colour indices during photometry. The amplitudes of the objects in Table 5 are normalized to the amplitude of CP Lac. The light curve of the V1500 Cyg is unique in its high amplitude and shape in the brightness peak, in other respects the differences are negligible. The difference in outbursts amplitudes may reflect the structure and spatial orientation of the binary system with respect to the line of sight \citep{Warner1987} and also the spectral energy distribution of very hot white dwarf. To demonstrate the differences between fast classical and recurrent novae, Fig.11 shows also the light curve of U Sco for the outburst in 2010: the recurrent nova has a significantly larger slope of the initial brightness decline and the light curve is located in the more inner region of the Fig.1, than classical novae (see general discussion).

\begin{figure}[t]
	\centerline{\includegraphics[width=78mm]{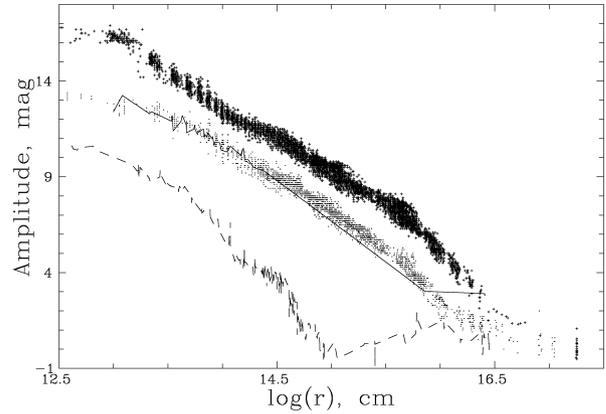}}
	\caption{The light curves of the prototype CP Lac (dots), V1500 Cyg (pluses), CP Cru (solid line) and the recurrent nova of U Sco (dashed line).\label{Fig11}}
\end{figure}

\begin{center}
	\begin{table}[t]%
		\centering
		\caption{Novae of the CP Lac group.\label{tab9}}%
		\tabcolsep=0pt%
		\begin{tabular*}{250pt}{@{\extracolsep\fill}rccc@{\extracolsep\fill}}
			\toprule
			\textbf{Nova} & \textbf{m${_q}$/t${_0}$} & \textbf{m1/m2 [ref]} & \textbf{Ref}  \\
			\midrule
			CP Lac* &	15.2/29339.7&	/15.0 [1] &\\	
			V1425 Aql&	20/49745&	20B/ [2]&	3 \\
			V1722 Aql	&22.5/55181&	>21&	4\\
			IV Cep&	18/41135&	19.3B&	5\\
			\textit{CP Cru}& \textit{22.5/50321.5}& \textit{/19.48 [6]}& \textit{7}\\
			Q Cyg&	15/06583.5&	/15.6 [8]&	\\
			V1500 Cyg&	18.5/42655.6&	/17.9 [1]	\\
			V2275 Cyg&	20.5/52141&	/>19R [9]&	10\\
			DM Gem&	17.5p/16178&	/17.54B [11]&	12\\
			V533 Her&	14/38058&	15/ [1]	&\\
			V407 Lup&	19.5/57656.5&	<17.5/ [13]&	13\\
			V959 Mon&	17.5/56095&	18.0r/ [14]&	15\\
			V2673 Oph&	20.5/55211.5&	/20.93I [16]&	17\\
			V2677 Oph&	23.5/56067&	>21&	18\\
			V597 Pup&	20.5/54419&	20.0&	19\\
			V598 Pup&	16.2/54256&	15.7B/ [20]&	21\\
			V1059 Sgr&	16/14353&	/18.1B [8]&	12\\
			V4157 Sgr&	20/48665&	/20.820: [16]&	22\\
			V4444 Sgr&	21/51295.5&	>21&	23\\
			V4741 Sgr&	22/52379&	>18&	24\\
			V5115 Sgr&	20.5/53458.5&	>21/ [25]&	26\\
			\bottomrule
		\end{tabular*}
		\begin{tablenotes}
			\item Note: In italics - possible member. References: 1 - \citep{Strope2010}, 2 - \citep{Skiff1995}, 3 - \citep{Takamizawa1995}, 4 - \citep{Nishiyama2009b}, 5 - \citep{Bahng1972}, 6 - \citep{Downes2000}, 7 - \citep{Liller1996}, 8 - \citep{Duerbeck1987a}, 9 - \citep{Balman2005}, 10 - \citep{Nakamura2001b}, 11 - \citep{Szkody1994a}, 12 - \citep{Walker1933}, 13 - \citep{Starder2016}, 14 - \citep{Greimel2012}, 15 - \citep{Cheung2012}, 16 - \citep{Mroz2015}, 17 - \citep{Nishimura2010a}, 18 - \citep{Seach2012}, 19 - \citep{Pereira2007}, 20 - \citep{Read2007}, 21 - \citep{Hounsell2016}, 22 - \citep{Liller1992c}, 23 - \citep{Yamamoto1999}, 24 - \citep{Liller2002}, 25 - \citep{Henden2007a}, 26 - \citep{Nishimura2005}.
		\end{tablenotes}
	\end{table}
\end{center}

\begin{center}
	\begin{table}[t]%
		\centering
		\caption{Novae of the CP Lac group. (Contined Table 9) \label{tab10}}%
		\tabcolsep=0pt%
		\begin{tabular*}{250pt}{@{\extracolsep\fill}rccc@{\extracolsep\fill}}
			\toprule
			\textbf{Nova} & \textbf{m${_q}$/t${_0}$} & \textbf{m1/m2 [ref]} & \textbf{Ref}  \\
			\midrule
			\textit{V5589 Sgr}&	\textit{22.5/56038.8}&	\textit{19.49/20.05 [16]}&	\textit{27}\\
			\textit{V5858 Sgr}&	\textit{22.5I/50505}	&&	\textit{16}\\
			\textit{T Sco}	&\textit{19/00551}&	\textit{>12}	&\textit{28}\\
			\textit{V977 Sco}&	\textit{22.5/47755.5}&	\textit{>18.4}&	\textit{29}\\
			V1187 Sco&	20.2/53222.5&	>18.8&	21, 30\\
			LMCN 1987-09a&	24/27056	&&	31\\
			\textit{LMCN 1992-11a}&	\textit{22.5/48938}	&	&\textit{32} \\
			\bottomrule
		\end{tabular*}
		\begin{tablenotes}
			\item Note: In italics - possible member. References: 27 - \citep{Korotkiy2012}, 28 - \citep{Sawyer1938}, 29 - \citep{Liller1989b}, 30 - \citep{Takao2004b}, 31 - \citep{Garradd1987}, 32 - \citep{Liller1992b}.
		\end{tablenotes}
	\end{table}
\end{center}

V1425 Aql is interesting due to the episode of the formation of a small amount of dust: the ratio of gas mass to dust mass in the emission is about 400-1000, and intense coronal emissions in the IR range \citep{Mason1996}. But, despite this, we classified the nova according to the shape of the light curve as a member of the Lacertae group. We proceeded from the magnitude of the brightest star nearly the pre-nova, which has B$\approx$20$^{m}$ \citep{Skiff1995}. In this case, the V1425 Aql light curve with the parameters of Table 9 is in good agreement with the V1500 Cyg light curve according to the linear initial decline, the moment of transition to the linear final decline, and the slope of this part of light curve. The nova reached its maximum for 5 days before detection by \citep{Takamizawa1995} or 2.5$^{d}$ later the date estimating by \cite{Mason1996}. But the outburst amplitude is most likely less than V1500 Cyg, otherwise in the maximum it would have been as a star with a brightness of about 4$^{m}$ and would have easy been noticed by naked eye after the end of the conjunction season with the Sun. If comparison with the prototype CP Lac, then the maximum brightness would be about 5$^{m}$ and the visual brightness of a quiescence state would be about 18$^{m}$. 

The parameters of Table 9 for the IV Cep light curve as a nova of this group mean that a possible maximum with a brightness of about 5$^{m}$ took place 5 days before the detection of outburst \citep{Kuwano1971}; the recorded maximum visual magnitude was 7.5$^{m}$. This corresponds to the gap for 12$^{d}$ in the observations of the discoverer (see references \citep{Bahng1972} after the previous image of the region with the limiting magnitude of the equipment of 10.5$^{m}$. \citep{Bahng1972} studied the spectrum of the nova and, according to the degree of its development, concluded that the possible maximum brightness of the nova was 6-7$^{m}$. 

CP Cru (Fig. 11, Table 9) was included in this group as a possible member, since the light curve is presented only in the maximum, the first half of the initial decline and only single measurement at the final phase of the light decline \citep{Downes2000}. On the same level of brightness, the nova was after 4 years \citep{Woudt2003}. Another equivalent opportunity may be to belong to the Puppis group (see below). An increase of maximum brightness up to 5.2$^{m}$, as propose \cite{Woudt2003}, will lead to unacceptable deformation of the light curve (deviation of the light curve in the maximum state from the main linear trend of the brightness initial decline). I.e. parameters of Table 9 is close to true ones. 

The Q Cyg light curve exists on the 19th century photometric system, which manifests itself in a systematic underestimation of the star’s brightness in the range of 4-7.5$^{m}$ and is clearly visible in the unusual deviation to -1.5$^{m}$ down this part of the light curve in modified scales according to \cite{Schmidt1877} or \cite{Lockyer1891}. If we exclude these data from consideration and will rely on data near the maximum and the second half of the light curve according to the data of \cite{Lockyer1891}, we can conclude that Q Cyg belongs to the Lac group (Table 9) and assume a maximum brightness was about 2$^{m}$. 

The detection $\gamma$-rays of V959 Mon \cite{Cheung2012}, long before this nova was detected in optics after the end of the invisibility season, facilitated us the classification of the visual light curve, which begins from the second half of the transition phase of the outburst. The current state of knowledge of the development of $\gamma$-ray radiation in classical novae allows one to tie the maximum moment of the optical outburst to the appearance of $\gamma$-ray source radiation. On this basis, extrapolation of a smooth visual light curve to the beginning of the outburst allows us to estimate the maximum brightness of about 5.5$^{m}$ near the date of detection of $\gamma$-ray source (Table 9). Other groups, for example, the Her1 group, is excluded by the inclination of the final part of the initial brightness decline. Radio flux from the nova peaked near log(r)$\approx$15.20-15.35 \cite{Healy2017}), i.e. before the transition of a nova from the initial brightness decline to the final decline phase for log(r)$\approx$15.4. 

When classifying the light curve of V2677 Oph we encountered a very contrasting difference between the light curves according to the visual data of AAVSO and the V photometry of SMATRS, and according to the I photometry of OGLE. In the initial phase of the brightness decline, the slope of the light curve in the I band was much larger than for the visual brightness; after the transition to the final drop in brightness, the difference in the tilts changed places. The reason for this uniqueness is, apparently, in the development of the emission lines spectrum. The classification is based on the data of AAVSO \citep{Kafka2019} and SMARTS \citep{Walter2012}. 

Nova Pup 2007, V598 Pup, was first discovered as a new bright X-ray source in October 2007 and possibly identical with the star USNO-A2.0 0450-03360039 \citep{Read2007}. Intensive observations by AAVSO members, starting with \cite{Dvorak2007} and others \citep{Kasliwal2007}, continued almost until the end of the final decline in brightness. Observations in the SMEI project \citep{Hounsell2016} provided a significant addition to the shape of the light curve at the maximum state. But the latest data went through our cleaning procedure: data with a brightness weaker of 8.5$^{m}$ and more were discarded,  and also as clearly erroneous estimates with a brightness of 7$^{m}$ and brighter were discarded for this period (the general trend of the light curve was determined by previous and subsequent data by AAVSO and \cite{Walter2012}). With the parameters of Table 9, the light curve exactly fits the CP Lac prototype.

V1059 Sgr with parameters of Table 9 could been have a brightness of about 3$^{m}$ in the peak of maximum. 

V4444 Sgr did not have been our doubt, like also of \cite{Mroz2015}, about belonging its to classical novae, while there was a discussion about its belonging to repeated novae \cite{Kato2004}. The light curve have an almost complete coincidence with CP Lac. 

The light curve of V4741 Sgr \cite{Liller2002, Kafka2019, Mroz2015} fits well with the CP Lac prototype. 

V5589 Sgr is included as a candidate in this group. This nova is interesting by the brightness depression at log(r)=14.9-15.7, which is typical for dust condensation of the D3 V1280 Sco type with a small optical thickness on the line of sight. But the data of \cite{Walter2012} do not support this assumption: the brightness in the photometric K band did not show a typical temporary increase associated with dust condensation, that should be observed independently to the space orientation of a dust structure. The denser light curve of the OGLE sky survey demonstrates in this time interval the clear brightness variations simultaneously with the depression of the type of the Puppis group. Depression or fluctuations, but not together, are characteristic novae of the Puppis group, and they occur somewhat earlier: in the interval log(r)=14.6-15.2. Perhaps the V5589 Sgr is a unique case of a nova given group or the Puppis group. 

Detection of outburst of OGLE-1997-NOVA-01, V5858 Sgr, became a by-product of OGLE sky surveys \citep{Mroz2015} (in the publication of the OGLE-II sky survey \citep{Cieslinski2003}, this nova has a different designation Bul\_sc5\_2859 with the same coordinates and light curve in the graph). The light curve for this nova allows us the possibility to include it in the Lac group. The light curve covers the final part of the initial, linear decline and possible is come into the final decline, but the brightness of the object is already near the limit of the equipment or the contribution of a close faint additional radiation source appears and the light curve smoothly transitions to a stable level. At the outburst maximum near JD 2450505, the brightness of the object could reach I=10$^{m}$. 

T Sco is the first nova in the constellation Scorpii and has only visual observations \citep{Sawyer1938}. For the T Sco light curve, nova in 1860 in the region of the M80 cluster, we admitted a possible belonging to the Lac group (Table 10), which may be interesting from the point of view of searching for a nova remnant in a dense stellar field. 

V977 Sco is represented by a few observations and, mainly, they belong to the first half of the light curve, which leaves room for the assumption about its to also belong the Puppis group. But regardless of this interpretation, it can be concluded that the observations after JD 2447782 in the AAVSO database refer to field stars, including possible identification in the OGLE sky survey \citep{Mroz2015} (difference with the catalogue coordinates 0.11"). 

The V1187 Sco light curve with the parameters of Table 10 was unambiguously assigned to this group of novae with a possible brightness in a maximum of about m$_{vis}$=8.8$^{m}$. 

LMCN 1992-11a was discovered at the final rise phase \citep{Liller1992b}. The light curve is incompleteness there are observations only in the maximum and in the initial phase of brightness decline, and they are finished by the data of \cite{Gilmore1992}. This allows us only to assume possible belonging to this group.

\subsubsection{Novae of the LMC}

Of the nearly 50 novae of the LMC, including recurrent novae, and similarly, of the 27 novae of the SMC, only 2 novae of the LMC were classified as members of the CP Lac group (Table 10). 

\begin{center}
	\begin{table}[t]%
		\centering
		\caption{Novae of the LMC - members of the Lacerti group\label{tab11}}%
		\tabcolsep=0pt%
		\begin{tabular*}{210pt}{@{\extracolsep\fill}lccc@{\extracolsep\fill}}
			\toprule
			\textbf{$ $ $ $ $ $Nova} & \textbf{m$_{peak}$, mag} & \textbf{M, mag}  \\
			\midrule
			LMCN 1987-09a &	9.5\textit{v}&	-9.2\textit{v}\\
			LMCN 1992-11a&	10.2\textit{v}&	-8.5\textit{v}\\
			\bottomrule
		\end{tabular*}
	\end{table}
\end{center}

The issue of interstellar extinction has been resolved by studying field stars \citep{Haschke2011}. The absorption in the V band in the field of the novae in the LMC was adopted equal AV$\approx$0.2$^{m}$. 

Absolute maximal visual magnitude of two novae are presented in Table 11.  

\subsubsection{Preliminary discussion}

It is noteworthy that the bright, well-studied Nova Cygni 1975, V1500 Cyg, retained its uniqueness: exceptionally high outburst amplitude. The light curve in the maximum state also has a unique - dome-shaped shape, which in the modified scales has the prolonged horizontal part, and it is difficult to imagine the beginning of the monotonous light decline in the modified scales (Fig.11). 

The light curves have the simplest form of all groups: a fast short maximum (excluding V1500 Cyg, see above), the extended linear initial decline and the short final decline. The slopes of these two linear sections differ within small limits. 

Novae of this group have the smoothest light curves, and here engages our attention the obvious difference in the amplitudes of the outbursts of novae. If in other groups we had practically no problems with normalizing to the amplitudes of prototypes, then here the V1500 Cyg stands out sharply with its highest amplitude, and the light curve of CP Cru with the reduced amplitude locates on the opposite side from the prototype CP Lac. This will need to consider in the future. It is possible that the very high amplitude, as in V1500 Cyg, leads into the unique shape of light curve in the maximum. 

This group of novae is also one of the many: 25 novae of our Galaxy and 2 novae in the LMC. But the erroneous classification and thus the belonging of some novae to other groups is not ruled out. Example V5589 Sgr points to possible individualities of outbursts: the "late" depression by a Puppis type. 

\subsection{CP Pup group}

The basis for this group was 4 fast novae: V603 Aql, V476 Cyg, GK Per and CP Pup (Fig.12, Table 7). The reason for this number of prototypes is the diversity of the brightness behaviour in the transition phase of an outburst with almost identical behaviour in the remaining phases. A composite light curve was obtained by shifting to a common average level of the brightness in the area of light curves about of log(r)$\approx$14.1 (in Fig.12 this looks like the area with the smallest dispersion of points). This corresponds to our remark in Section 2, that the shape of the light curve stabilizes 10-15 days after the outburst maximum. 15 days after the outburst maximum is defined by \cite{Buscombe1955, Shara2018} as a critical date. As a result of this procedure, the average outburst amplitude of this set of 4 novae was decreased by 0.4$^{m}$ with a full range of differences of the observed amplitudes of 3.7$^{m}$. A characteristic feature of the CP Pup group is the presence of oscillations with an amplitude of 0-2$^{m}$ (V603 Aql, GK Per) or a depression of variable amplitude (V476 Cyg, CP Pup) on the transition phase for log(r)=14.4-15.5. [In the classification of \cite{Strope2010} such depressions were identified as a plateau, which became the group's symbol "P".] The initial decline phase begins immediately after the maximum brightness, as also in the Lacertae group. After the maximum brightness, starting from log(r)$\approx$13.6, a short linear section follows up to 14.28 (the marker for 15 days is a vertical dashed line in Fig.12), followed by the transition phase. In the interval 15.4-15.7, the second short linear section begins - the final decline, which, at a brightness level of about 2.5$^{m}$ above the quiescence state, can slow down the rate of brightness decline until the outburst completes whole. Outbursts duration is at least 9-10 years. V476 Cyg just now, 100 years after the outburst, has approached the quiescence level. 

These 4 novae were distributed over 4 different classes of light curves in the catalogue of \cite{Strope2010}, but we have no reason to separate them. 

To include a nova in this group, a classical light curve is enough to note the presence of light depression at the transition phase of a outburst (the shape of this part of the light curve is somewhat different from the light minimum in the DQ Her group) or a series of light fluctuations with an amplitude of 1-2$^{m}$ and a characteristic time of about ten days. It remains to estimate the brightness of a pre-/post-nova or to assume a certain value by calibrating it by prototypes if it is lower than the limits of observation complexes. 

\begin{figure}[t]
	\centerline{\includegraphics[width=78mm]{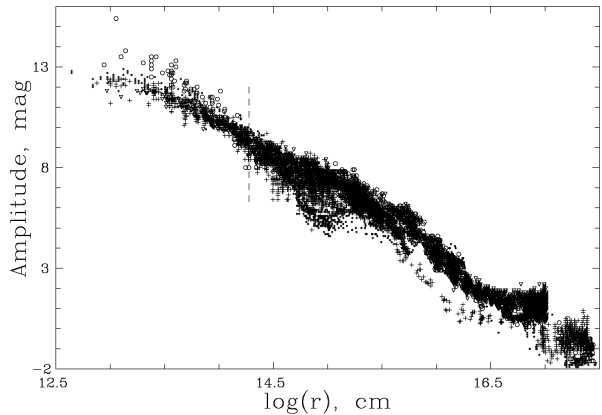}}
	\caption{Light curves of prototypes of the CP Pup group. CP Pup - circles, GK Per - pluses, V603 Aql - triangles, V476 Cyg - points. All observations from the AAVSO database. The vertical dashed line corresponds to 15 days after the maximum brightness.\label{Fig12}}
\end{figure}

\begin{center}
	\begin{table}[t]%
		\centering
		\caption{Novae of the CP Pup group.\label{tab12}}%
		\tabcolsep=0pt%
		\begin{tabular*}{250pt}{@{\extracolsep\fill}rccc@{\extracolsep\fill}}
			\toprule
			\textbf{Nova} & \textbf{m${_q}$/t${_0}$} & \textbf{m1/m2 [ref]} & \textbf{Ref}  \\
			\midrule
			V603 Aql* & 	12.2/21755 & 	/12.0  [1]&	\\
			V476 Cyg* & 	14.5/22561 & 	/17.2B    [1]	&\\
			GK Per*	 & 12.4/15438 & 	/12     [1] &	\\
			CP Pup*	 & 14/30675 & 	/15   [1]	&\\
			V368 Aql & 	17.0p/28435 & 	/16.9B [2]&	3,4\\
			V528 Aql & 	18.1/31694 & 	/18.1p  [1]	&\\
			V1370 Aql & 	19/44995.5 & 	/19.5p  [1]	&5\\
			V1494 Aql & 	16.5/51515 & 	/16.5  [1]&	6\\
			V1065 Cen & 	21/54120/0 & 	>18.3&	7\\
			V1213 Cen & 	20.5/54959 & 	24.4/ [8]  &	9\\
			X Cir & 	16p/24748	/19.3 &   [10]&	11\\
			V1668 Cyg & 	17.8/43763 & 	/20B  [12]&	13\\
			V2467 Cyg & 	19.5/54174 & 	18.46r'/ [14]&	15\\
			V2468 Cyg & 	19/54533 & 	18R/ [16]&	17\\
			V339 Del & 	16.5/56521 & 	17.5/ [18]&	19\\
			V446 Her & 	15.5/36997.5 & 	16.1/ [10a]&	64a\\
			DK Lac & 	16.5/33304 & 	15.5p/  [1] &	\\
			LZ Mus	 & 21/51174 & 	>18 &	20\\
			V382 Nor & 	20/53443 & 	>19.2 &	21\\
			\textit{V2104 Oph} & 	\textit{21/43044} & 	\textit{/20.94 [10]}&	\textit{22}\\
			V2313 Oph	 & 19.5/49505 & 	>12.5 &	23\\
			V3663 Oph & 	22/58068.5 & 	>20.3G:&	24\\
			HS Sge	 & 18/43145 & 	/20.0 [25]&	26\\
			\bottomrule
		\end{tabular*}
		\begin{tablenotes}
			\item Note: In italics - possible group member. References: 1 - \citep{Duerbeck1987a}, 2 - \citep{Szkody1994a}, 3 – \citep{Hoffmeister1936}, 4 - \citep{Beyer1939}, 5 - \citep{Honda1982a}, 6 - \citep{Pereira1999}, 7 - \citep{Liller2007a}, 8 - \citep{Mroz2016c}, 9 - \citep{Pojmanski2009}, 10 - \citep{Tappert2014}, 10a – \citep{Strope2010}, 64a - \citep{Hassel1960}, 11 - \citep{Cannon1930}, 12 - \citep{Kaluzny1990}, 13 - \citep{Collins1978}, 14 - \citep{Steeghs2007}, 15 - \citep{Tago2007}), 16 - \citep{Chochol2012}, 17 - \citep{Kaneda2008}, 18 - \citep{Deacon2014}, 19 - \citep{Itagaki2013a}, 20 - \citep{Liller1998c}, 21 - \citep{Liller2005a}, 22 - \citep{Kuwano1976}, 23 - \citep{Tago1994}, 24 - \citep{Kaneko2017}, 25 - \citep{Tappert2015}, 26 - \citep{Hosty1977}.
		\end{tablenotes}
	\end{table}
\end{center}

\begin{center}
	\begin{table}[t]%
		\centering 
		\caption{Novae of the CP Pup group. (Continued Table 12)\label{tab13}}%
		\tabcolsep=0pt%
		\begin{tabular*}{250pt}{@{\extracolsep\fill}rccc@{\extracolsep\fill}}
			\toprule
			\textbf{Nova} & \textbf{m${_q}$/t${_0}$} & \textbf{m1/m2 [ref]} & \textbf{Ref}  \\
			\midrule			
			V909 Sgr & 	21.5p/30171.8	 & /20.71B [27]&	28\\
			V927 Sgr & 	21.5p/31196.5	 & /19.391 [29]&	30\\
			\textit{V1012 Sgr} & 	\textit{19.8p/20355} & 	\textit{/17.398 [29]}&	\textit{31}\\
			V1016 Sgr & 	17.5p/14877 & 	/17p [1]&	32\\
			\textit{V4021 Sgr}	 & \textit{17.5/43210} & 	\textit{/18.2 [33}]&\textit{	34}\\
			V4027 Sgr & 	21p/39993 & 	<21/ [1]&	35,36\\
			V4160 Sgr & 	21.5/48467 & 	>18I/ [37]&	38\\
			\textit{V4742 Sgr} & 	\textit{19.5/52532} & 	\textit{<20/ [39]}&	\textit{40}\\
			V4743 Sgr & 	17.5/52536 & 	16.8:&	41\\
			V4745 Sgr & 	18/52742.5 & 	17.9R/ [42]&	42\\
			V5583 Sgr & 	20/55050 & 	20.73/20.31 [29]&	43\\
			V5585 Sgr & 	20/55212 & 	<21I/18.31 [29]&	44\\
			V5591 Sgr & 	24.8/56105.4 & 	>16.5V&	45\\
			V5856 Sgr & 	17.5/57699 & 	>22I/ [46]&	47\\
			V1312 Sco & 	22/55712 & 	>21	&48\\
			V1313 Sco & 	22.5/55810.5 & 	>15.5&	49\\
			V1535 Sco & 	21/57064 & 	16.6: &	50\\
			V368 Sct & 	18/40796 & 	/17.7 [2] &	51\\
			V444 Sct & 	23/48498 & 	>20 &	52\\
			V476 Sct & 	22.5/53639 & 	>17	&53\\
			V477 Sct & 	23.5/53656 & 	>20B/     [54]&	54\\
			CT Ser & 	16/32589 & 	/16.2 [2]&	1\\
			V382 Vel & 	15.2/51321.5 & 	16.6&	55\\
			LMCN 1977-03a & 	22.5/43214&	&	56,57\\
			\bottomrule
		\end{tabular*}
		\begin{tablenotes}
			\item Note: In italics - possible group member. References: 27 - \citep{Tappert2012}, 28 - \citep{Mayall1946}, 29 - \citep{Mroz2015}, 30 - \citep{Campbell1947}, 31 - \citep{Bailey1920}, 32 - \citep{Walker1933}, 33 - \citep{Ringwald1996}, 34 - \citep{Kuwano1977}, 35 - \citep{MacConnell1977}, 36 - \citep{Sarajedini1984}, 37 - \citep{McNaught1991b}, 38 - \citep{Camilleri1991b}, 39 - \citep{Yamaoka2002}, 40 - \citep{Liller2002}, 41 - \citep{Haseda2002}, 42 - \citep{Brown2003}, 43 - \citep{Nishiyama2009c}, 44 - \citep{Seach2010}, 45 - \citep{Itagaki2012}, 46 - \citep{Mroz2016b}, 47 - \citep{Stanek2016b}, 48 - \citep{Seach2011a}, 49 - \citep{Seach2011b}, 50 - \citep{Kojima2015}, 51 - \citep{Alcock1970}, 52 - \citep{Camilleri1991c}, 53 - \citep{Takao2005}, 54 - \citep{Pojmanski2005c}, 55 - \citep{Williams1999}, 56 - \citep{Graham1977}, 57 - \citep{vanGenderen1979}.
		\end{tablenotes}
	\end{table}
\end{center}

The initial portion of the light curve with depression at 1370 Aql is typical for this group, but further unique deviation follows. Depression at the transition phase, like V1065 Cen (see the next paragraph), corresponds to a shallow episode of dust condensation by the D2 DQ Her type:  IR and UV photometry of \cite{Snijders1987} confirmed this. After the depression, the general trend of the light curve was restored, but then again there was a decline in brightness on 3.3$^{m}$ at an unusually high rate, forming a plane up to a “recovery” of common trend. Unfortunately, UV or IR observations are not available for this part of the outburst. But it can be suggested that such a rapid decline could be due to the re-condensation of dust by the rare  D4 DQ Her type. The start of the initial brightness decline immediately at the moment of maximum brightness is the argument against the inclusion in the DQ Her group, as they have a prolonged state of maximum brightness. 

V1065 Cen light curve reproduces the GK Per light curve very well: monotonous decrease in brightness immediately after the maximum, depression at the transitional phase, and the final phase. But depression is very short compared to prototype. Spectra in the IR range indicated the presence of dust in the nova shell \cite{Helton2010}. The depression position on the abscissa axis is intermediate between dust condensation of types D1 and D2. By now, the post-nova has almost returned to a quiet state. 

Several observations of Nova Circus 1926, X Cir, are available before and at the beginning of the outburst and they were continued after 1.5 years. Even such a very limited data set made it possible to obtain a modified light curve typical of this group with the parameters of Table 12. According to our extrapolation the outburst maximum with m$_{p}$$\approx$$\approx$4$^{m}$ was about of 14 days early the first fixation of outburst. The observations ended at the beginning of the final brightness decline. 

IR photometry of V1668 Cyg \cite{Gehrz1980} definitely showed dust formation of type D2 DQ Her, which in the visual region showed depression typical for the Pup group (see also Section 3.3.3 for a discussion). 

Depression at the transition phase of the outburst of V339 Del was accompanied by dust condensation, starting from log(r)$\approx$14.7 and with a maximum mass of dust near log(r)$\approx$15.1 \citep{Evans2017}. 

The DK Lac outburst has of the lowest amplitude among the novae of this group. Interestingly, the brightness fluctuations occurred in her later than the usual position of the transition phase on the abscissa scale. 

In V382 Nor, the maximum visual magnitude according to our estimation reached 8.5$^{m}$ near the date t$_{0}$ in Table 12. 

V2104 Oph has a few photometric data, which with some certainty can be interpreted as the light curve of the given group; it was not possible to pick up other options. 

The final observations of V3663 Oph \citep{Shappee2014, Kochanek2017}, which coincided with the position of depression at the transition phase of the outburst, supplemented the AAVSO data and it became possible to include this fast Pup group with the parameters of Table 12. 

HS Sge at a maximum near JD 2443145 could have a visual brightness of 5.5-6.0$^{m}$. 

V909 Sgr was a very fast nova, which gave \cite{Duerbeck1988} a reason to include it in a small list of candidates for recurrent nova. But the light curve in graphical form by \cite{Mayall1946} and with the parameters of Table 13 allows us to reliably relate it to the CP Pup group. 

V927 Sgr provides us with a unique opportunity for the light curve classification by 4 observations \cite{Campbell1947}. The nature of the object as a nova was established by the spectrum. It seems to us possible to classify the light curve as a member of the Puppis group with the parameters of Table 13. The first, negative observation refers to a pre-nova state 11 days before the maximum state or the second brightness estimate, the third brightness estimate was obtained immediately before the onset of depression and the last, below the limiting magnitude for telescope, refers to the most active part of depression. \cite{Mroz2015} presented photometry of a possible candidate for an old nova with a brightness of V=19.391$^{m}$; the colour index V-I=1.490$^{m}$ indicates the possibility of magnitude B$\approx$20$^{m}$, which is close enough to our assessment of the brightness of a quiescent. 

V1012 Sgr has only 8 observational dates, but the range of their location covers a significant part of the total possible light curve of a outburst and allows us to include it in the Puppis group with the parameters of Table 13 in the absence of depression on the light curve, but there are were possible light fluctuations as in V603 Aql. Observations ended before the phase of the final light decline. 

V4021 Sgr was detected after the maximum brightness, which coincided with the season of conjunction with the Sun. The light curve with the parameters in Table 13 is closer to the Puppis group. In this case, we estimate the possible maximum of 6-7$^{m}$. Single IR observation of \cite{Hatfield1977} indicated possible dust formation (their result cites \cite{Gehrz1988}: "..dust formation evidently began before JD 2443285.."). But the dust was not recorded by the WISE project \cite{Evans2014}. 

The maximum of the V4027 Sgr outburst was missed \citep{MacConnell1977} and the only archival series of observations was presented by \cite{Sarajedini1984} and refers to the middle part of the light curve, critical for our classification. In total, photometry data of \cite{MacConnell1977, Sarajedini1984} make it possible to attribute the V4027 Sgr to the CP Pup group with the parameters of Table 13. It can also be noted that a possible candidate for a post-nova \cite{Mroz2015} with a brightness of V=19.939$^{m}$ and a colour index of V-I=1.462$^{m}$ is really post-nova, since for such a red object the photographic magnitude can be about 21$^{m}$ (Table 13). 

\cite{Pagnotta2014} consider V4160 Sgr a good candidate for a recurrent nova. They assigned it the candidate category B contains strong RN candidates, for which many of their indicators strongly point to the system being recurrent. But the modified light curve better corresponds to the Puppis group (Table 13): until the observations were completed when the transition phase of the outburst was reached, the light curve closely followed the typical curve of this group. It was a fast nova: within a day it rose to a maximum and weakened by 1.5$^{m}$. The outburst amplitude ($\approx$14$^{m}$) exceeds the amplitudes of the recurrent novae; the slope of the light curve at the phase of the initial decline is typical of a classical nova.

V4742 Sgr, unlike other novae of the Puppis group, has light variations of low amplitude during a state of the light maximum (see the next subsection). 

V4745 Sgr, Nova 2003 Sgr, had a unique feature: the transition phase oscillations had rapid ups and downs and almost flat maxima. Nova also had one of the lowest outburst amplitudes in the group. 

V5591 Sgr has several series of observations: AAVSO, SMARTS, OGLE, and all of them show the onset of depression. Together with a smooth light curve at the initial brightness decline phase, we consider this as a sign of belonging to CP Pup group (Table 13). A further decrease in brightness according to OGLE data was faster than according to SMARTS data, which with a wide scatter followed the general trend of light curve typical of the CP Pup group. Unfortunately, the final observations of both reviews are fixing apparently of the background sources. 

The bright nova V5856 Sgr is an interesting example of the classification according to the shape of the light curve in addition to the unique detailed observations of $\gamma$-ray \citep{Li2017} and photometry in the optical region of the spectrum \citep{Kochanek2017, Kafka2019} from the completion of the phase the initial increase in brightness and before the transition phase of the outburst, the maximum development of which occurred in the 65-day season of invisibility of the nova. Visual observations of the next season of visibility resumed at the end of the transition phase and stopped at the beginning of the final decline. The early start of the transition phase with a very rapid decline compared to the V476 Cyg, one of the prototypes of this group, but very slow for the D1 DQ Her type dust condensation made it difficult to choose between these novae groups. This was also associated with a rather horizontal portion of the light curve in the state of maximum brightness, which is more characteristic of the DQ Her group. The choice in favour of the latter group was made on the basis of spectral IR observations of \cite{Rudy2016}, which did not record thermal radiation from dust about 3 days after the onset of depression, whereas it should be the other way around: thermal radiation first appears, and then it starts a decline of optical radiation. According to the data of \cite{Mroz2016b}, the pre-nova V5856 Sgr had a brightness of I>22$^{m}$, which at a low interstellar absorption for a given galactic direction \citep{Munari2017} means that the amplitude of the outburst is A$\geq$16.5$^{m}$. 

The brightness decline in V476 Sct, including the first half of the depression at the transition phase of the outburst, occurred typical the CP Pup group. It is possible that V476 Sct had dust condensation, as concluded from spectrophotometry \cite{Perry2005} (JD 2453689.608), when a temporary light decline has only just begun to be reflected in the light curve. \cite{Munari2006} estimate the brightness of a quiescence state of 22-25$^{m}$, which coincides with our estimate (Table 13). 

The CT Ser outburst was detected post-factum \cite{Bartay1948}, which led to regular observations only at the transition phase. But \cite{Duerbeck1987a} notes her presence on the Harvard plate on date JD 2432588 with a magnitude of 8$^{m}$. This data is sufficient to estimate the brightness at the outburst maximum of 3.5-4$^{m}$ in JD 2432589 and to determine CT Ser as a nova of the Puppis group with the parameters of Table 13. 

\subsubsection{V458 Vul subgroup of the novae with flashes in a state of maximum}

Some novae (Table 14) have light curves very similar to V603 Aql (Fig.12), but light fluctuations occur mainly at the maximum state or before the transition phase and may be absent in the transition phase. These brightness fluctuations are flare-like: a rapid increase in brightness by 1-2$^{m}$ and a return to the previous trend of the light curve within 2-3 days, single flash lasting up to 50 days occur at other stage of outburst (V458 Vul: log(r)$\approx$15.6). 

V458 Vul has excellent detailed observations at the state of the maximum brightness \citep{Hounsell2016}, which indicate pulsations with a width at half the amplitude of about a day and a change of brightness on 1.2$^{m}$ in 0.5$^{d}$. It is very likely that such fluctuations in other novae may be skipped by discoverers and single brightness assessments are perceived as irregular brightness fluctuations or even as errors. Subsequently, V458 Vul shown isolate bursts of brightness. The light curve of V458 Vul (Fig.13, Table 14) substantially refines the idea of the brightness behaviour at the beginning of the outburst, which, according to the data for other novae in this small group, is presented very fragmentarily and ambiguously.

\begin{figure}[t]
	\centerline{\includegraphics[width=78mm]{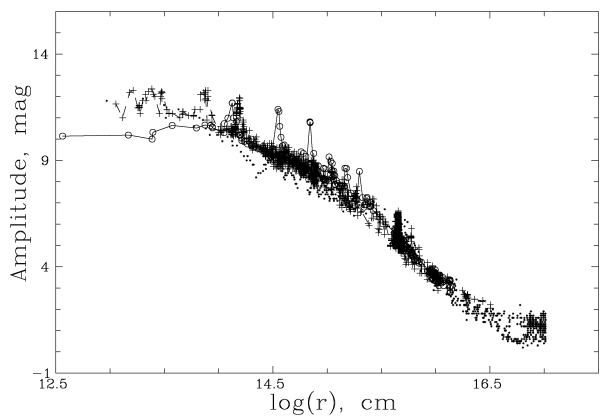}}
	\caption{Light curves of novae with flashes: V458 Vul (dashed line with pluses) and V5588 Sgr (solid line with circles), in comparison with V603 Aql - points.\label{Fig13}}
\end{figure}

\begin{center}
	\begin{table}[t]%
		\centering
		\caption{List of novae of the V458 Vul subgroup.\label{tab6}}%
		\tabcolsep=0pt%
		\begin{tabular*}{240pt}{@{\extracolsep\fill}rccc@{\extracolsep\fill}}
			\toprule
			\textbf{Nova} & \textbf{m${_q}$/t${_0}$} & \textbf{m1/m2 [ref]} & \textbf{Ref}  \\
			\midrule
			V458 Vul*&	20.5/54320&	18.3&	1\\
			PR Lup&	20.3/55768&	>17/&	2\\
			\textit{HR Lyr}&	\textit{17.5p/22299.5}&	\textit{/15-16.5 [3]}&	\\
			V840 Oph&	20p/21342&	/19.1B [4]&	5\\
			\textit{FL Sgr}	&\textit{20.5/23956}&	\textit{/19.456 [6]}&	\textit{7}\\
			\textit{V4135 Sgr}&	\textit{20/46928}&	\textit{>21B/  [8]}&	\textit{8}\\
			\textit{V4327 Sgr}	&\textit{21/49245}&	\textit{/21.069: [6]} &\textit{	9}\\
			V4361 Sgr&	20.5/50265&	>15.5p&	10\\
			V5113 Sgr&	20.5/52898&	20.60I/20.02I [6]&	11\\
			V5588 Sgr&	23/55647&	/17.74 [6]&	12,13\\
			\textit{V902 Sco}&	\textit{21.5/33058}	&\textit{/20j [14]}&	\textit{15}\\
			FS Sct&	21.5p/34186&	/17.4 [16]&	17\\
			V443 Sct& 20/47777& <20/ [18]& 19\\								
			\bottomrule
		\end{tabular*}
		\begin{tablenotes}
			\item Note: Possible members of the subgroup are in italics. References - the main source of catalogue data and photometric data \citep{Kafka2019} is supplemented with data from publications: 1 - \citep{Abe2007}, 2 - \citep{Brown2011}, 3 - \citep{Honeycutt2014}, 4 - \citep{Schmidtobreick2003}, 5 - \citep{Shapley1921}, 6 - \citep{Mroz2015}, 7 - \citep{Gill1927}, 8 - \citep{McNaught1987}, 9 - \citep{Sugano1993}, 10 - \citep{Sakurai1996}, 11 - \citep{Brown2003}, 12 - \citep{Nishiyama2011}, 13 - \citep{Hounsell2016}, 14 - \citep{Duerbeck1987a}, 15 - \citep{Henize1961}, 16 - \citep{Ringwald1996}, 17 - \citep{Harwood1960}, 18 - \citep{Anupama1992}, 19 - \citep{Wild1989}.
		\end{tablenotes}
	\end{table}
\end{center}

The state of maximum brightness for the PR Lup outburst is poorly represented by observations, which raises some doubts in our classification due to the lower outburst amplitude than a prototype by 1-2$^{m}$. But the general trend of the light curve of PR Lup is similar to the prototype (Table 14). It is unusual that in the region of the transition phase there is depression more typical of the main group CP Pup, but shifted in abscissa by +0.3; this can also be interpreted as a consequence of the condensation of dust on the line of sight according to the D2 DQ Her type. 

The beginning of the HR Lyr outburst was recorded by two observations, before the outburst and near the maximum brightness \cite{Bailey1920}, noticeable fluctuations in brightness are also indicated here. This verbal comment was the basis for inclusion in this group. Observations of \cite{Nijland1925}, which began 80 days later, captured the end of the transition phase and the beginning of the final brightness decline. With these data, we with some doubt attributed the HR Lyr light curve to this novae group with flashes (Table 14). 

For FL Sgr \cite{Mroz2015} dive the potential candidate post-nova brightness at a distance of 1.53” from the catalogue coordinates, which for the accuracy of 1924 data may mean correct identification. 

V5113 Sgr, as well as V458 Vul, with their activity at the maximum state, was noticed in the study of novae with re-brightenings of \cite{Tanaka2011}, in a group of which were included several novae of different groups of our review. V5113 Sgr is the only nova from this small subgroup, in which, 9 months after the outburst, the presence of dust was recorded by \cite{Rudy2004} that did not affect the visual light curve.

The V902 Sco outburst was detected as a result of observations with the objective-prism by \cite{Henize1961} and the light curve with 7 very rough light estimates allows inclusion in the V458 Vul group almost as the only possibility (Table 14). Our estimate of the brightness of a quiescence state is close to the catalogue data of \cite{Duerbeck1987a}.

FS Sct has few observations, but we can distinguish the general trend of the light curve, on which increased brightness is superimposed at moments close to the light flashes of the prototype V458 Vul. This gives reason to include it in this group.

\subsubsection{Novae of the LMC}

Only the light curve of LMCN 1977-03a was assigned to the CP Pup group (Table 13): the beginning of the transitional phase of the outburst led to a typical deformation of the linear trend of the initial brightness decline. 

Correction for interstellar extinction and the distance modulus was carried out similarly to the procedure described above. 

As a result, the absolute magnitude of LMCN 1977-03a was estimated as M$_{V}$$\approx$-8.7$^{m}$. 

\subsubsection{Preliminary discussion}

The dust condensation examples of the V1668 Cyg, V1065 Cen and other novae of the Puppis group convince us of the influence of the dust condensation process on the temporal visual brightness depression in the Puppis group novae during the transition phase. Depression occurs at the time of dust condensation of type D2 V1280 Sco. But other options are possible: V1370 Aql at the moment corresponding to the moment D4 there was a sharp drop in visual brightness, unfortunately, there are no IR observations. Novae of the CP Pup groups are distinguished from the DQ Her group by the earlier beginning of the brightness decline: on the first day of the outburst (Fig.12). Also apparently, the outburst amplitudes for the CP Pup group novae are 1-2$^{m}$ lower than for the DQ Her group novae. It is interesting to note that the GK Per light curve at the transition phase shows an increased scatter in brightness  estimates, which completely "fills" the possible depression. Apparently, this can be attributed to the appearance in the spectrum of strong emissions, which had a strong effect on instrumental photometric systems of observers. But for V368 Aql, the light curve reproduces well the average light curve of GK Per and does not show such a significant spread of light estimates at the transition phase of the outburst (this may be due to the proximity of instrumental observers systems in the case of V368 Aql). 

Dust condensation is not a rare phenomenon in the Puppis group, but it not as large-scale as the DQ Her group: it is possible they  are  compact dust condensations, since the relict shells of these novae have a fibrous and ragged structure, for example, GK Per \cite{Liimets2012}, against the clearly ellipsoidal shell with three (equatorial and two tropical) belts and polar caps at DQ Her \citep{Williams1978}. 

There is an assorted shapes of the light curve at the transition phase of the outburst: from the presence of light depression to its absence, while maintaining the general trend of light drop; characteristic brightness variations. In addition to this, we note that novae with higher outburst amplitudes have a brightness depression in the transition phase (V476 Cyg), and ones with low amplitudes (V603 Aql) have brightness fluctuations. 

V1668 Cyg has a unique set of photometry: the broadband V, or visual, and narrow-band "y" Strömgren systems. The first gives the average value of the energy of the radiated nova in the spectrum range of at least 800 A, including emission lines appearing during nebular phase  that emit energy an order of magnitude and more higher than the radiation in the continuum in the region of 20 A or more. The second specialized photometric band gives the average brightness in a narrow range (230 A), where there may be no bright emissions. Narrow-band "y" photometry by \cite{Gallagher1980, Kaler1986} gives a modified light curve (Fig.14), which, up to the moment of maximum dust concentration on the line of sight \citep{Gehrz1980}, coincides with the visual light curve, and from that moment it sharply increases the rate of a decline in brightness and by the "standard" beginning of the final decline in visual brightness, the nova almost returns to a level of quiescence brilliance \cite{Kaler1986}. Consequently, the visual, as other broadband, brightness of V1668 Cyg by the beginning of the final decline in brightness can be 2.5$^{m}$ higher than that in the continuum, and after another 70 days, the brightness of a nova will mainly determine by radiation only in emission lines. Up to return the visual brightness of V1668 Cyg to a quiescence level, several hundred more days remained in addtion to a time of the "y" photometry return.   

\begin{figure}[t]
	\centerline{\includegraphics[width=78mm]{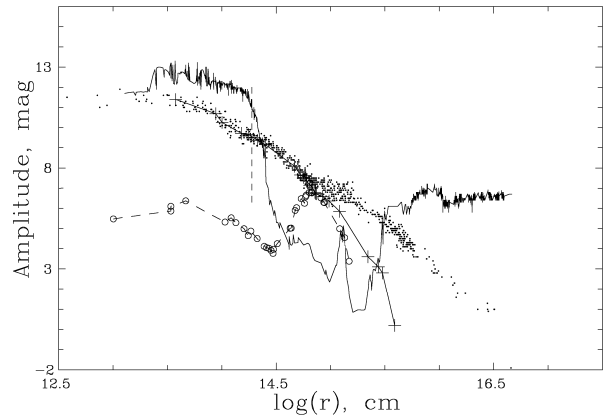}}
	\caption{Modified light curves of V1668 Cyg: visual \citep{Kafka2019} (dots), narrow-band "y" \citep{Gallagher1980, Kaler1986} (line with pluses) and infrared \citep{Gehrz1980} (dashed line with circles). The solid broken line is the light curve of the V1280 Sco. The vertical dashed line corresponds to the 15th day after the maximum brightness. \label{Fig14}}
\end{figure}

\subsection{Group V1974 Cyg}

Novae were allocated in the Cygni group through a comparison with V1974 Cyg as the prototype, which has a dense photometry, which gave an idea of all phases of the development of the outburst from the final brightness rise to returning of nova to a quiescence state (Fig.15). On the graph of the light curve in its middle part, the presence of "branches" of the light curve is clearly visible, which reflects the difference between the instrumental visual photometric systems of various AAVSO observers. 

The shape of the light curve was similar in shape to V603 Aql (Fig.15). But V603 Aql had a slightly larger slope, and the duration of the linear initial brightness decline was in the interval log(r)=13.6-15.8. Tthe initial brightness decline phase of V1974 Cyg started at log(r)$\approx$13.6 against 13.1 in V603 Aql. The difference between the outburst amplitudes reaches 1.5$^{m}$: the V1974 Cyg amplitude is 12.4$^{m}$ versus 10.9$^{m}$ for V603 Aql (the actual observed outburst amplitude is shown, and not the adopted light curve of the Puppis group). The absence of clear differences between the two novae leads us to the assumption that novae of the Cyg group with their smooth light curves can form the upper boundary of the Pup group.   

\begin{figure}[t]
	\centerline{\includegraphics[width=78mm]{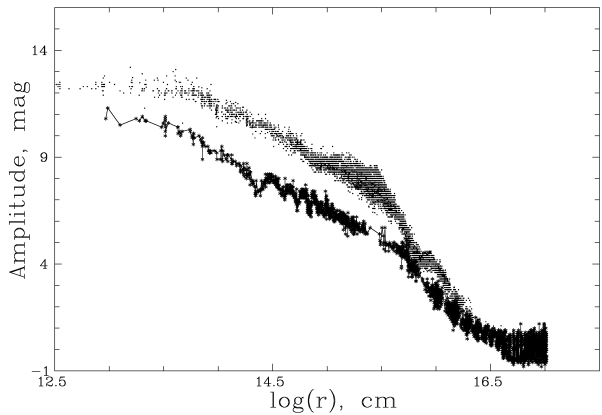}}
	\caption{Light curves of V1974 Cyg (dots) and V603 Aql (line with dotted line).\label{Fig15}}
\end{figure}

\begin{center}
	\begin{table}[t]%
		\centering
		\caption{List of novae of the V1974 Cyg group.\label{tab15}}%
		\tabcolsep=0pt%
		\begin{tabular*}{240pt}{@{\extracolsep\fill}rccc@{\extracolsep\fill}}
			\toprule
			\textbf{Nova} & \textbf{m${_q}$/t${_0}$} & \textbf{m1/m2 [ref]} & \textbf{Ref}  \\
			\midrule
			V1974 Cyg*&	16.9/2448676&	>21B/16.63 [1]&	2\\
			BY Cir&	18.5/2449732&	/17.8 [3]&	4\\
			DN Gem&	16.3/2419474&	/15.81 [5]&	6\\
			\textit{IL Nor}	&\textit{17.5p/2412639}&	\textit{/19.32B [7]}& \textit{6}\\
			V351 Pup&	19.5/2448617&	/19.55 [8]&	9\\
			V574 Pup&	20.1/2453330&	18.6&	10\\
			V5116 Sgr&	21/2453555&	/>18.1U [11]&	12\\
			RW UMi&	17p/2435730	&/18.64B  [13]	&14\\
			LU Vul&	22/2440054& >20.5/ [15]& 16\\							
			\bottomrule
		\end{tabular*}
		\begin{tablenotes}
			\item Note: Possible members of the subgroup are in italics. References: 1 - \citep{Collazzi2009}, 2 - \citep{Collins1992}, 3 - \citep{Woudt2003}, 4 - \citep{Liller1995a}, 5 - \citep{Szkody1994a}, 6 - \citep{Walker1933}, 7 - \citep{Tappert2012}, 8 - \citep{Downes2000}, 9 - \citep{Camilleri1992c}, 10 - \citep{Tago2004}, 11 - \citep{Sala2010}, 12 - \citep{Liller2005b}, 13 - \citep{Kaluzny1989}, 14 - \citep{Kukarkin1962}, 15 - \citep{Rosino1969}, 16 - \citep{Kohoutek1968}.
		\end{tablenotes}
	\end{table}
\end{center}

For confident inclusion a nova in the Cygni group, a complete and detailed light curve is needed, which will allow you to notice a number of distinctive features. First, the state of maximum brightness has a certain duration and ends near log(r)$\approx$13.6. Secondly, the final brightness drop begins at log(r)$\approx$15.5 and it occurs faster for the Cygni group than for novae of the Puppis group. Thirdly, a nova of the Cygni group has a short plateau at the phase of final brightness decline: $\Delta$log(r)$\approx$0.2 by log(r)$\approx$15.8. 

The belonging of the BY Cir light curve to the V1974 Cyg group with the parameters of Table 15 means that the outburst maximum occurred 2 days after the first but without-resulted image of \cite{Liller1995a} of the object area after the season of conjunction with the Sun and the visual brightness of the nova reached a maximum of 6$^{m}$. 

The DN Gem light curve \citep{Walker1933} is a hybrid of the Cygni group with the parameters of Table 9 and the Mus group (see below), where it could be included if the outburst amplitude could was increased by 2$^{m}$. But the presence of a plateau in the middle of the final decline in brightness has led us to favour the first option. 

The belonging of the BY Cir light curve only to this group is possible. In this case, the first observations \citep{Walker1933} relate to 1-2 days before a maximum brightness, which reached mpg=5-6$^{m}$. 

\subsection{GQ Mus group}

The prototype of the group was GQ Mus - the first nova in which X-flux was detected of \cite{Ogelman1984} and which had the longest phase of the initial brightness drop. The light curves of similar novae in modified scales had differences that distinguished them from other novae \citep{Rosenbush1999a}. 

The group was based on the light curves of two prototypes GQ Mus, Nova Muscae 1983, and EL Aql, Nova Aquilae 1927, (Fig.16, Table 16) and reduced to the average amplitude of the outbursts of the two prototypes. 

From the EL Aql light curve, we can conclude that the nova, apparently, only now (2012 is the date of the last UBVR photometry of \cite{Tappert2016}) has approached a quiescence state. GQ Mus should still weaken by 2$^{m}$ to the pre-nova brightness of m$_{v}$>21$^{m}$ \cite{Krautter1984}. The main difference of the light curves of this group is a noticeably smaller slope of the linear initial decline in brightness and its largest duration: it was about 500$^{d}$ for EL Aql, GQ Mus and V4633 Sgr. The transition from the initial brightness decline to the final decline in the logarithmic scale occurs quickly: the light curve can be represented by two straight lines with a transition point from one state to another with log(r)$\approx$15.8.

\begin{figure}[t]
	\centerline{\includegraphics[width=78mm]{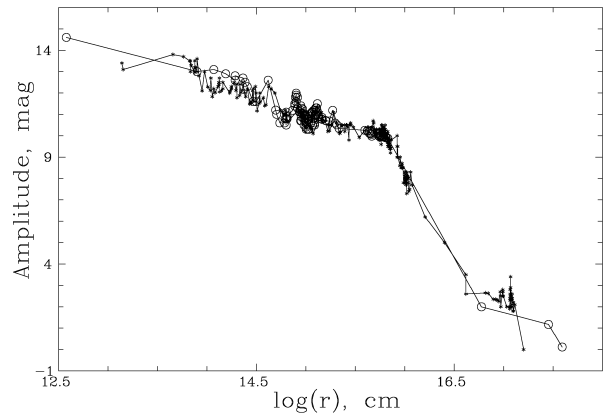}}
	\caption{Light curve of GQ Mus (line with dots) and EL Aql (line with circles).\label{Fig16}}
\end{figure}

\begin{center}
	\begin{table}[t]%
		\centering
		\caption{List of novae of the GQ Mus group.\label{tab16}}%
		\tabcolsep=0pt%
		\begin{tabular*}{240pt}{@{\extracolsep\fill}rccc@{\extracolsep\fill}}
			\toprule
			\textbf{Nova} & \textbf{m${_q}$/t${_0}$} & \textbf{m1/m2 [ref]} & \textbf{Ref}  \\
			\midrule
			GQ Mus&	21/45352&	>21/19.0 [1,2]&	3, 4\\
			EL Aql&	21p/25047	&/20.88 [5]&	6, 7\\
			V888 Cen&	20/49771&	>15/&	8\\
			V1039 Cen&	23/52183&	20.8B/ [9]&	10\\
			V1819 Cyg&	22.5/46647&	/20.33 [11]&	12\\
			V555 Nor&	25/57464&	>22/&	13\\
			\textit{V841 Oph}&	\textit{18/396145}&	\textit{/13.5 [14]}&	\textit{15}\\
			V2670 Oph&	25/54612&	>19.9/&	16\\
			V Per&	19.5p/10200&	/18.45B [17]&	18\\
			V4633 Sgr&	22/50896&	>21/>18.7 [19,20]&	21\\
			V5582 Sgr&	24/54881	&19.43I/ [22]&	23\\
			V744 Sco&	26.5p/27843&	/>15.6p	&24\\
			V1186 Sco	&23.5/53194&	>18/&	25\\
			V1310 Sco&	24.5/55245&	>18r/  [26]&	26\\
			V1657 Sco&	26.5/57779&	>17.3/&	27\\
			QU Vul&	20/46062&	/18.9V&	28	\\							
			\bottomrule
		\end{tabular*}
		\begin{tablenotes}
			\item Note: Possible members of the subgroup are in italics. References: 1 - \citep{Krautter1984}, 2 - \citep{Narloch2014}, 3 - \citep{Liller1983}, 4 - \citep{Bateson1983}, 5 - \citep{Tappert2016}, 6 - \citep{Wolf1927}, 7 - \citep{Beyer1929}, 8 - \citep{Liller1995b}, 9 - \citep{Holvorcem2001}, 10 - \citep{Liller2001b}, 11 - \citep{Downes2000}, 12 - \citep{Wakuda1986}, 13 - \citep{Jayasinghe2017}, 14 - \citep{Duerbeck1987a}, 15 - \citep{Parenago1949}, 16 - \citep{Nishiyama2008d}, 17 - \citep{Szkody1994a}, 18 - \citep{Walker1933}, 19 - \citep{Skiff1998}, 20 - \citep{Lipkin2008}, 21 - \citep{Liller1998a}, 22 - \citep{Mroz2015}, 23 - \citep{Sun2009}, 24 - \citep{Plaut1958}, 25 - \citep{Pojmanski2004}, 26 - \citep{Nishiyama2010a}, 27 - \citep{Nishimura2017}, 28 - \citep{Collins1984}.
		\end{tablenotes}
	\end{table}
\end{center}

V888 Cen is interesting due to the temporal weakening of the D4 V1280 Cyg type, followed by a plateau in the interval log(r)=15.7-16.0 and the brightness starts to decline again like in the V1280 Sco, but the observations ceased. An argument in favour of the GQ Mus group is the absence of a maximum state typical of the DQ Her group, and the slope of the initial brightness drop section was typical of the Mus group. At the initial decline phase, the brightness of the nova showed typical fluctuations around the general trend (Fig.16). We corrected the first brightness estimate of \cite{Liller1995b} by the extrapolated colour index v-r=0.8$^{m}$ \citep{Tse2001}. 

Outburst of ASASSN-16ra (V555 Nor) was detected by \cite{Jayasinghe2017} after a review of archival material. The spectrum 18 months after the maximum showed the properties of the nebular phase \cite{Chomiuk2017} and refers to the moment of the transition from the initial brightness decline to the final decline log(r)=15.84. Data on the X-ray flux, which is possibly typical of the novae in this group, is no. Unfortunately, the only source of photometry is the ASAS-SN sky survey with a limiting magnitude of 15-16$^{m}$ and the FWHM is $\approx$2 pixels or 16", which does not allow us to trace the behaviour of the nova at an interesting phase of the final brightness decline, since its brightness weakened below the limit equipment.

One of the old nova V841 Oph, whose outburst occurred in 1848, may belong to the Mus group. The light curve coincides well with the prototypes with the exception of the outburst amplitude. Our estimate of the brightness of a quiescence state differs from the observed brightness by 4.5$^{m}$, which is the biggest difference in our review. But we have no other options. We can only make the assumption that the maximum brightness could was 1m or more higher than that recorded.  

The classification of V Per, Nova Persei 1887, made full use of the property of the logarithmic representation of the abscissa scale: significant time shifts of the maximum moment to a much lesser extent shift the remote parts of the light curve in the logarithmic representation. If we take the first observation date as the zeroth approximation at the time of maximum, we see that the last points of the light curve exactly correspond to the final brightness decline of the Mus group prototypes with a slight shift. Therefore, by varying the maximum moment, we achieved a good coincidence of the light curves with the parameters of Table 16. 

Nova Sagittarii 2009, V5582 Sgr, was discovered by looking at observation archives three months after the outburst. This led to the scarcity of data for constructing a qualitative light curve: the spread of data reaches 1$^{m}$. High-precision photometry by the OGLE program \citep{Mroz2015} is available before the outburst and for the next season of object visibility and has not an overlap with the visual part of the light curve. Therefore, our interpretation of the light curve may raise some doubts (Table 16). But the visual part of the light curve shows a slope typical of the Mus group, and OGLE photometry satisfies the final brightness decline of this group prototypes. The inclusion of the V5582 Sgr in the Mus group means the large outburst amplitude and a post-nova brightness of about 24$^{m}$ (Table 16). \cite{Mroz2015} give the brightnenewss of the pre-nova I$\approx$19.43$^{m}$, but \cite{Sun2009} noted the absence of the pre-nova in the DSS surveys with a limit of m$_{r}$>20.8$^{m}$. This can be partially attributed to the large colour indices of the post-nova and to the orientation of the binary system relative to the line of sight. 

V744 Sco has only 6 photographic brightness estimates \cite{Plaut1958}, which admit membership in the Mus group (Table 16). 

The well-presented light curve of V1186 Sco is quite unusual due to the duration of the initial decline larger on $\Delta$log(r)>0.2 in comparison with other novae of this group. There is absent further photometry, but the photometry of the ASAS-SN review in 2016-2019 continues the initial linear decline in brightness, which is very unusual for the novae of this group: for V1186 Sco as a member of the Mus group, here would be typical at the current phase of the outburst to have a visual brightness of about 21$^{m}$. The Gaia mission database for the object with the coordinates of the nova gives the average brightness g=17.75$^{m}$. Therefore, there is good reason to exclude from consideration the latest photometry according to the ASAS-SN survey, but the length of the initial brightness decline section increased by $\Delta$log(r)>0.2 is possible. There are other peculiarities. \cite{Schwarz2007} noted the bright lines of [Ne II] (12.81 $\mu$m) and [Ne III] (15.56 $\mu$m). \cite{Tanaka2011} noted that O I and [O I] lines are stronger compared to the other lines.

Important details of the light curve were missed for the V1310 Sco: there are no data on the initial time of the outburst due to the season of the conjunction of the object with the Sun and on the moment of transition from the initial brightness decline to the final decline. But otherwise, the V1310 Sco's light curve matches the prototypes of the group. 

The maximum brightness of V1657 Sco was missed for no more than 5 days (Table 16). Photometry of the ASAS-SN sky survey, together with the AAVSO data, covered an almost 90-day time interval sufficient to determine membership in the Mus group. 

\subsection{RR Pic group}

Slow novae with the parameter t$_{3}$>100$^{d}$ and low amplitudes of outbursts 9-11$^{m}$ constitute the group under discussion. Light curves of RR Pic and DO Aql were the basis of group (Fig.17). 

The typical light curve of nova of the RR Pic group in the modified scales can be represented by two lines: almost horizontal, representing the state of the maximum, and inclined - the final decline in brightness. The almost horizontal part has slope variations of about 0: from positive to negative values.  Brightness variations can be superimposed on it with the amplitude of up to 2$^{m}$. The variations of RR Pic itself were unit and regular: rise from "zero" to a maximum, down to a minimum, to second maximum and transition to the final decline in brightness. 

The transition to the second trend occurs in the range of log(r) from 14.7 to 15.4. If we "controlled" the brightness range for the horizontal trend through the adopting outburst amplitude, then the slopes of the final brightness decline were approximately equal but the light curves itself in this part occupied the band of significant width, that is real because of large remoteness from the beginning of the maximum state t$_{0}$. 

The slope of the initial decline has become for us a formal criterion for a small systematization of novae in this group. The "P(ositive)" subgroup included novae with a slope equal to or greater than zero (RR Pic, DO Aql). The remaining novae made up the "N(egative)" subgroup (V2540 Oph), which are also characterized by several brightness variations throughout the state of maximum.

The light curves of three novae specify the range of the brightness amplitude and the logarithm of the effective radius, within which the light curves of the other novae are parallel to the borders (Fig.17, Table 17, 18, and 19).

\begin{figure}[t]
	\centerline{\includegraphics[width=78mm]{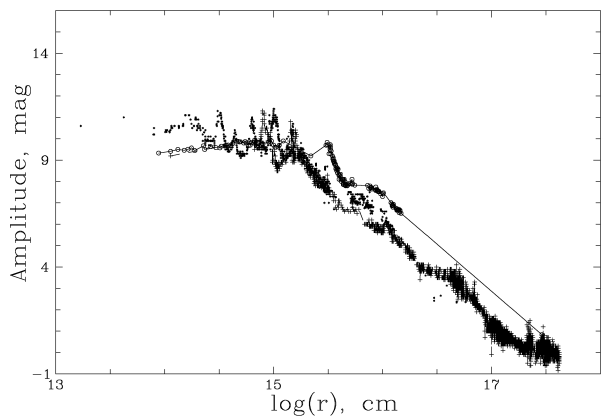}}
	\caption{Light curves RR Pic (dashed line with pluses) and DO Aql (solid line with circles) – P subgroup, V2540 Oph (points) – N subgroup. We draw attention to the location of the light curves at higher abscissas than in previous similar figures.\label{Fig17}}
\end{figure}

\begin{center}
	\begin{table}[t]%
		\centering
		\caption{Novae of the positive RR Pic subgroup.\label{tab17}}%
		\tabcolsep=0pt%
		\begin{tabular*}{240pt}{@{\extracolsep\fill}rccc@{\extracolsep\fill}}
			\toprule
			\textbf{Nova} & \textbf{m${_q}$/t${_0}$} & \textbf{m1/m2 [ref]} & \textbf{Ref}  \\
			\midrule
			DO Aql*	 & 18.5p/24415 &  	/17.9 [1] & 	2\\
			RR Pic* & 	12.2p/24245 & 	/12.7 [3] & 	\\
			V868 Cen & 	21.5/48300 & 	>22.5B/ [4]	 & 5\\
			AR Cir & 	21p/17230 & 	/19.59B [6] & 	7\\
			V2659 Cyg & 	20.5/56748 & 	/22 & 	\\
			V794 Oph & 	22.5/29420 & 	/18 [8] & 	9\\
			V849 Oph & 	19/22178 & 	/18.8 [10] & 	11\\
			V972 Oph & 	19p/36073	 & /16.627 [12]&	13\\
			V2944 Oph** & 	21.5/57110	 & 18.8B:/ & 	\\
			V2949 Oph & 	23/57283 & 	>21/ & 	14,15\\
			V3701 Oph  & 	21.5I/55470 & 	 & 	12, 16\\
			V400 Per & 	21/42300	 & 19.5B/19.5B [17]	 & 18\\
			V999 Sgr & 	19.3/18745 & 	/16.6 [19] & 	18\\
			V4077 Sgr & 	19/45225 & 	>20/ & 	20\\
			V4362 Sgr**	 & 20/49490 & 	>15/ & 	21\\
			V6345 Sgr  & 	23I/55270 & 	>20p/ 22 & 	12, 22\\
			V612 Sct & 	20.5/57927.5 & 	19.1g/ & 	23\\
			NR TrA	 & 20.5/54490	 & >19.2r   [24] & 	24\\
			CN Vel&	20p/17169&	/17.54 [25]	 & 7\\
			LMCN 1948-12a & 	24.5p/32890 &  & 		\\
			LMCN 2017-11a & 	24/58073	 & 	 & 26,27\\
			SMCN 2008-10a & 	24I/54777	 & 		 & 		\\
			\bottomrule
		\end{tabular*}
		\begin{tablenotes}
			\item Note: ** - novae with a dust condensation. References: 1 - \citep{Shafter1993}, 2 - \citep{Beyer1929}, 3 - \citep{Fuentes-Morales2018}, 4 - \citep{McNaught1991a}, 5 - \citep{Liller1991a}, 6 -\citep{Duerbeck1993}, 7 - \citep{Walker1933}, 8 - \citep{Duerbeck1987a}, 9 - \citep{Burwell1943}, 10 - \citep{Szkody1994a}, 11 - \citep{Strope2010}, 12 - \citep{Mroz2015}, 13 - \citep{delaRosa1959}, 14 - \citep{Nishiyama2015b}, 15 - \citep{Munari2017}, 16 - \citep{Mroz2014}, 17 - \citep{Collazzi2009}, 18 - \citep{Sanduleak1974}, 19 - \citep{Duerbeck1987b}, 20 - \citep{Honda1982b}, 21 - \citep{Sakurai1994}, 22 - \citep{Scholz2012}, 23 - \citep{Stanek2017b}, 24 - \citep{Brown2008},  25 - \citep{Tappert2013}, 26 - \citep{Shappee2014}, 27 - \citep{Kochanek2017}.
		\end{tablenotes}
	\end{table}
\end{center}

\begin{center}
	\begin{table}[t]%
		\centering
		\caption{Novae of the negative RR Pic subgroup.\label{tab18}}%
		\tabcolsep=0pt%
		\begin{tabular*}{240pt}{@{\extracolsep\fill}rccc@{\extracolsep\fill}}
			\toprule
			\textbf{Nova} & \textbf{m${_q}$/t${_0}$} & \textbf{m1/m2 [ref]} & \textbf{Ref}  \\
			\midrule
			V1369 Cen & 	14.5/56629 & 	15/ [28] & 	\\
			V1405 Cen	 & 22/57869 & 	18.8g: & 	29\\
			V465 Cyg & 	20p/32700	 & /20.3B & 	\\
			V408 Lup & 	22.5/58267  & 	19.4G:/  & 	30\\
			V390 Nor & 	20.8/54245 &   	>20/ & 	31\\
			V2214 Oph & 	20/47260 &  	20.5/ [32] & 	33\\
			V2540 Oph* & 	19.5/52293 & 	>21/ [34] & 	35\\
			V441 Sgr	 & 19.5p/26218 & 	/19.935 [12] & 	36\\
			V1310 Sgr & 	22.2p/27902 & 	>20p & 	37\\
			V5587 Sgr & 	22/55575	 & 18.4/<19.5[12] & 	38\\
			V5593 Sgr & 	22/56123 & 	>20/ & 	39, 40\\
			V5666 Sgr	 & 
			21/56668 & 	20.0 & 	40, 41\\
			V5667 Sgr & 	21/57064 & 	>16.6r	 & 40, 42\\
			V992 Sco** & 	19.3/48764 & 	18.2B/17.8 [17] & 	43\\
			V496 Sct & 	18/55143 & 	>19/ & 	44\\
			V378 Ser & 	22.5/53448 & 	>18r & 	45\\
			V549 Vel & 	20.5/58029 & 	16.9 & 	46\\
			PW Vul & 	18/45908 & 	16.9 & 	47	\\						
			\bottomrule
		\end{tabular*}
		\begin{tablenotes}
			\item Note: ** - novae with a dust condensation. References: 28 - \citep{Hornoch2013}, 29 - \citep{Stanek2017a}, 30 - \citep{Kaufman2018a}, 31 - \citep{Liller2007b} 32 - \citep{McNaught1988}, 33 - \citep{Wakuda1988}, 34 - \citep{Kato2002a}, 35 - \citep{Haseda2002}, 36 - \citep{Hoffleit1932}, 37 - \citep{Fokker1951}, 38 - \citep{Nishimura2011}, 39 - \citep{Kojima2012}, 40 - \citep{Walter2012}, 41 - \citep{Furuyama2014}, 42 - \citep{Nishiyama2015a}, 43 - \citep{Camilleri1992a}, 44 - \citep{Nishimura2009}, 45 - \citep{Pojmanski2005d}, 46 - \citep{Stanek2017c}, 47 - \citep{Wakuda1984}.
		\end{tablenotes}
	\end{table}
\end{center}

\begin{center}
	\begin{table}[t]%
		\centering
		\caption{Novae of the HR Del subgroup.\label{tab19}}%
		\tabcolsep=0pt%
		\begin{tabular*}{240pt}{@{\extracolsep\fill}rccc@{\extracolsep\fill}}
			\toprule
			\textbf{Nova} & \textbf{m${_q}$/t${_0}$} & \textbf{m1/m2 [ref]} & \textbf{Ref}  \\
			\midrule
			HR Del*	 & 12/39665 & 	/12.4 & 	\\
			\textit{V1548 Aql} & 	\textit{20.5/51940} & 	\textit{19.6B: [48]} & 	\\
			V365 Car & 	17.5p/32720 & 	/18.3 [49]	 & 50\\
			V723 Cas & 	16.5/49927 & 	<18/15.5 & 	51\\
			\textit{FM Cir} & 	\textit{14/58138} & 	\textit{17.3}g & 	52\\
			\textit{BS Sgr} & 	\textit{17p/21060} & 	\textit{/17.9 [53]} & 	\\
			\textit{V5558 Sgr} & 	\textit{15.5/54203} & \textit{	<20/ [54]} & 	\textit{55}\\
			FV Sct & 	20.5/36620 & 	21/18.7 [56] & 	\\
			\textit{RR Tel}	 & \textit{13/31450} & 	\textit{14B/16.5} [57] & 		\\				
			\bottomrule
		\end{tabular*}
		\begin{tablenotes}
			\item Note: Possible members of the subgroup are in italics. References: 48 - \citep{Uemura2001}, 49 - \citep{Woudt2002}, 50 - \citep{Henize1975}, 51 - \citep{Yamamoto1995}, 52 - \citep{Seach2018}, 53 - \citep{Tappert2015}, 54 - \citep{Henden2007b}, 55 - \citep{Sakurai2007}, 56 - \citep{Ringwald1996}, 57 - \citep{Mayall1949}.
		\end{tablenotes}
	\end{table}
\end{center}

The light curve of the V868 Cen was built taking into account the unique observation of a nova obtained 28 days before the first report of the outburst detection by \cite{Liller1991a} and given in the AAVSO database. It fits well with the typical light curve of the Pic group (Table 17). The uniqueness of the nova is also associated with possible dust condensation, as indicated by the appearance of IR excess according to the results of IR photometry near JD 2448429 of \cite{Harrison1991}; but for this IR excess it was not possible to select the appropriate temperature. This event in time coincided with a sharp decrease in brightness, as in V2944 Oph, and both corresponded to the D3 DQ Her type. Earlier, JD 2448407, the IR spectrometry of V868 Cen did not detect the presence of dust \cite{Smith1995}. The explanation for this contradiction can be the development of the dust component in the ejected matter during this 24-day time interval, or the formation of IR excess according to the results of broadband IR photometry was due to the development of intense emission IR lines. 

The light curve of the AR Cir outburst in 1906 in modified scales allows for almost the only possibility for inclusion in the RR Pic group. The difference of 1.5m between the measured brightness of the post-nova  by \cite{Duerbeck1993} and our estimation of the brightness of a quiescence state falls within the acceptable range due to the spatial orientation of the binary system with its bright circumstellar and near-system structures. Our reconstruction of the light curve as a linear continuation of the initial light decline according to the RR Pic as the template is in good agreement with the reconstruction of \cite{Duerbeck1993}.  The light curve of the AR Cir outburst in 1906 in modified scales allows for almost the only possibility for inclusion in the RR Pic group. The difference of 1.5m between the measured brightness of the post-nova by \cite{Duerbeck1993} and our estimation of the brightness of a quiescence state falls within the acceptable range due to the spatial orientation of the binary system with its bright circumstellar and near-system structures. Our reconstruction of the light curve as a linear continuation of the initial light decline according to the RR Pic as the template is in good agreement with the reconstruction of \cite{Duerbeck1993}.  

For V794 Oph \cite{Duerbeck1988} indicated the possibility of belonging to recurrent novae. But the modified light curve (Table 17) in this case is significantly shifted to the right of T Pyx with its very similar light curve (see further general discussion).  

V972 Oph was reached its maximum brightness state between JD 2436072, the last date when the object did not be presented on the photographic plate, and JD 2436091, the date the object was detected, but already at the maximum brightness state. The amplitude of the V972 Oph outburst satisfies the observed brightness of the quiescence state: V=16.627$^{m}$, V-I=1.551$^{m}$ \cite{Mroz2015}, R=15.8$^{m}$ \cite{Tappert2013}. 

There were no problems with the V2944 Oph classification (Table 17). Beside a similarity to the V400 Per in the state of maximum brightness, the V2944 Oph has its own peculiarity. At the beginning of the phase of the final decline at log(r)$\approx$15.4, a sharp decrease in brightness in an amplitude and in a duration was observed, very similar to dust condensation of the D4 DQ Her type. An almost stable weakened level of brightness remained for some time until the recovery of the trend typical for the final phase. 

V2949 Oph was detected at the limit of search program capabilities of \cite{Nishiyama2015b} and before the start of the invisibility season due to conjunction with the Sun, so the light curve is very incomplete. A typical local maximum in the state of maximum brightness made it possible to find the moment t$_{0}$, fitting the width of this local maximum to other novae of this group. Since the beginning of the new observation season, the nova has been significantly weakened in comparison with prototypes, which may be due to dust condensation of the D3 DQ Her type. 

The V400 Per outburst was detected 2 months after the maximal brightness, therefore the state of the maximum is poorly represented by observations: less than a dozen measurements of archived images. But the transition to the final brightness decline has a dense series of visual observations, which allows you to confidently perform the classification of the light curve (Table 17). V400 Per and V2944 Oph have very similar light curves: during the state of maximum brightness there was a local maximum by the same value for log(r)=14.25, or 15 days after reaching a nova the state of maximum brightness. V400 Per demonstrates a stable brightness of a quiescence state before and after an outburst over an interval of more than 30 years: to two equal values in Table 11 we add measurement of \cite{Szkody1994a} B=19.61$^{m}$. 

V4362 Sgr is interesting in the condensation of D3 DQ Her type dust in a ejected shell, as indicated by a temporary decrease in brightness. In 2002, 8 years after the outburst, a nebula with a diameter of about 4 arc seconds, classified as planetary, was discovered near the remnant of a nova \cite{Boumis2006}.

Nova WISE, V6345 Sgr, \citep{Mroz2015}, one of the weakest novae whose outburst was discovered as a result of reviewing the archives. The recorded outburst duration of more than 1000 days indicates that this is a classic nova of the Pic group (Table 17). The first brightness estimates date back to the time soon after the invisibility season, so the outburst moment should be extrapolated, which we did by fitting the slope of the final brightness decline. 

With the observational data of NR TrA, Nova Triangulae Australea 2008, there was arose the situation about the same as with V2573 Oph (see Section 3.7). After \cite{Brown2008} reported the outburst, an AAVSO database was supplemented the results of CCD observations of this object over the previous 3 months with slightly variable brightness, which is in good agreement with subsequent observations. As a result, this data went unnoticed, since the nova was classified in the AAVSO database as the fast Na nova. In fact, this is a slow nova of given group with an almost stable level of brightness until it was detected by \cite{Brown2008} during a local flare of small amplitude, about 2$^{m}$, after which the brightness returned to its previous level. The second equivalent flare through 10$^{d}$ ended by a monotonous decline in brightness, in which a third flare at 2.5$^{m}$ occurred at 25$^{d}$ after discovery. The final brightness decline was typical of the novae in this group. For practical use, the data of archival CCD observations were corrected by +0.6$^{m}$ to take into account differences in instrumental systems when combining the common area outside the flares. The amplitude of the flares in these CCD data is lower than in visual observations, which is explained by the difference in the position and width of the instrumental photometric systems and the presence of other spectral lines in each instrumental photometric band. In this case, the light curve before the detection of the outburst goes at the level of the detection limit of other observers. As a result, there is agreement with the RR Pic regarding the position of the local flares and the "stable" brightness after reaching the state of maximum brightness. At the phase of the final brightness decline, there was a "deviation" of the NR TrA light curve until the DO Aql light curve, but then the brightness returned to the RR Pic trend.

The CN Vel outburst was detected after the invisibility season, so the selection of the parameter t$_{0}$ for the modified abscissa scale was possible within 145 days before the first successful observation. The result of combining the CN Vel light curve with the prototype is given in Table 17. 

At least 100 days of the state of brightness maximum and the formed trend for the final decline of LMCN 1948-12a \citep{Henize1954} leave little doubt about its belonging to the RR Pic group. 

SMCN 2008-10a was discovered from archival materials of the OGLE-IV sky survey \cite{Mroz2016b}. The light curve has the shape typical of the RR Pic group: the duration of the maximum state is 80–120 days, the brightness fluctuations with the amplitude of about 1$^{m}$, and the beginning of the final brightness decline. 

Light curve of V1405 Cen shown a rapid decline in brightness at a time point close to the time of dust condensation in the DQ Her group of the D3 V1280 Sco type. But we do not associate this behaviour of brightness with a dust condensation, since photometry of \cite{Walter2012} did not show the development of IR brightness corresponding to such a process.  

The data for V465 Cyg light curve were used from a single source \cite{Miller1959}, since the uniform longest, 8-year series of photographic observations is given here. It should be borne in mind that the state of maximum brightness had a 130-day trend to decline from the maximum value to the transition point to the final decline in brightness, so the nova was included in subgroup N. 

The light curve of V408 Lup has the largest slope of the initial light decline trend, which brings it closer to the D3 DQ Her group. Fluctuations at the initial brightness decline phase in amplitude and characteristic time are very typical of the Pic group, and these became the basis for our inclusion of the nova in Table 18. 

The classification of the Nova Sagittarii 1930, V441 Sgr, was made according to 10 dates of observations of \cite{Hoffleit1932}. In the commentary to data, it was added that 4 lines of the data table are 88 images of the 1931 season, which showed a monotonous decline in the brightness of the nova. This commentary added confidence that the coincidence of these 4 lines with the linear portion of the light curve of the prototype in the interval log(r)=15.0-15.2 is real, and not just the result of our selection of parameters for the light curve in modified scales. Our estimate of the photographic brightness of a quiescence state differs from the real brightness by no more than 1m: in the OGLE Atlas of \citep{Mroz2015}, the brightness and colour index of the possible post-nova V=19.935$^{m}$ and V I=1.221$^{m}$, and the distance 0.75" from the coordinates according to the Simbad base are given. 

The classification of the V1310 Sgr as a slow nova \citep{Duerbeck1987a} is based on a very incomplete light curve \citep{Fokker1951}, with a fast rise in brightness during 6 days on more than 4.3$^{m}$, followed by a slow rise from 12.2$^{m}$ to 11.7$^{m}$, and finally the decline to 15.3$^{m}$ in within 526 days. The lack of good coordinates and maps of identification created the basis for the proposal of \cite{Tappert2012} to exclude the object from the list of novae on the basis of identification of a possible remnant with a star whose spectrum has the characteristics of a myrid. We constructed the light curve in the modified scales and did not find a contradiction with the possibility of classification as a member of the group of slow novae (Tabl.18). Our estimate of the photographic magnitude of the post-nova at 22.2$^{m}$ means that the visual magnitude can be 20-21$^{m}$. On DSS maps in the limit of 15 arc seconds near a possible bright myrid, there are more than 8 faint stars, therefore, we need to look for post-nova among them. The AAVSO database provides arguments in favour of the object as a classic nova and gives the corrected coordinates of the object weaker than 20$^{m}$. 

The V5666 Sgr outburst was detected after the invisibility season due to conjunction with the Sun, so the onset of the state of maximum brightness was not fixed. The parameters of Table 18 were selected according to the shape of the brightness fluctuations at the state of maximum according to their typical duration and position on the abscissa axis. After 2–3 characteristic variations in the state of maximum brightness, the light curve of V5666 Sgr fits well on the light curve of RR Pic.

The well-presented V992 Sco light curve \cite{Camilleri1992a, Camilleri1993, Kafka2019} allowed us to include this nova in this group with the parameters of Table 18. The conclusion about the presence of dust was made based on spectrophotometry in the range of 8–13 $\mu$m \cite{Smith1995}). A brief note of \cite{Harrison1992} about the unusual behaviour of brightness in the IR region refers to the time long before the start of temporary weakening of brightness due to the appearance of dust: the process of dust formation appears in the IR range almost immediately with the appearance of the first dust particles and only after 10-20 days the radiation begins to respond in the optical range, if condensation occurs on the line of sight \cite{Rosenbush1996c}. As of 2009, the brightness of the nova was above the quiescence level by 1.5$^{m}$.

\subsubsection{HR Del subgroup}

Close affinities with the RR Pic group have the novae of non-numerous HR Del subgroup (Fig.18, Table 19). The outburst amplitudes of these novae are noticeably lower, the final brightness decline occurs later: the rapid brightness decline at the beginning (in the logarithmic abscissa scale) is replaced by a slower linear final brightness decline, but faster than the RR Pic group. The decrease in the amplitude of the outburst also resulted in an increase in doubts about the membership of a number of novae in this group, which was also a reflection of the presence of peculiarities in light curves.

\begin{figure}[t]
	\centerline{\includegraphics[width=78mm]{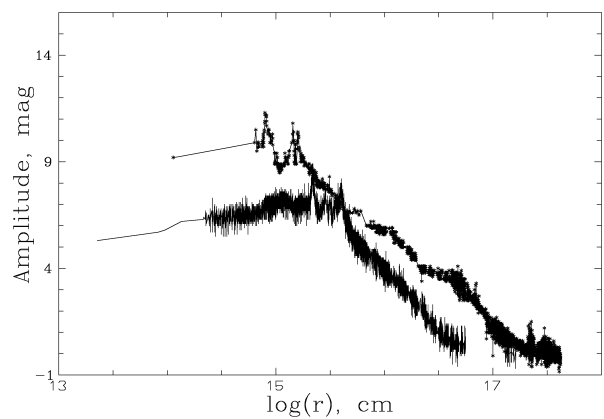}}
	\caption{Light curve of HR Del (solid broken line) compared to the light curve of RR Pic (solid line with dots).\label{Fig18}}
\end{figure}

Nova Aquilae 2001, V1548 Aql, was a faint object: its brightness had variations mainly around 13$^{m}$, therefore, observations are available only for the state of maximum brightness. The nova is included in the group as possible member (Table 19). 

The light curve of V723 Cas is similar to HR Del. The final part of the light curve gives reason to say that the star is almost back to a quiescence state. This means that the outburst amplitude in visual brightness is about 9$^{m}$, \cite{Munari1996} express the opinion that the amplitude in B band is of about 11.4$^{m}$. Such a outburst amplitude makes the V723 Cas an intermediate member of the RR Pic group and the HR Del subgroup, like the V5558 (see below). 

N Cir 2018, FM Cir, was included as a candidate for this subgroup: the light curve is in an intermediate position with the main RR Pic group. 

The V5558 Sgr classification is an interesting, hybrid case. The light curve in modified scales is more similar to HR Del. The parallel of V5558 Sgr with HR Del and another group member, V723 Cas, is conducted by \cite{Tanaka2011}. The outburst amplitude about of 13.4$^{m}$ \cite{Henden2007b} is noticeably larger than that typical of the HR Del subgroup (about 7$^{m}$). The pre-nova has not been identified: on the SIMBAD chart in the vicinity of the nova there are 3-4 objects, which creates certain difficulties for photometry. It remains to wait until the full finish of outbreak to confidently resolve the issue of classification.

For FV Sct, the slow Nova Scuti 1961, despite scarce data \cite{Savel'eva1960} (see also AFOEV), \cite{Nassau1961, The1967, Bertaud1967, Ringwald1996}, we were able to carry out the classification of the light curve. The choice was limited to two options: RR Pic or HR Del. Based on two local maxima and the slope of the final brightness decline, we selected parameters for the light curve, whose characteristics are similar to HR Del (Table 19). According to our reconstruction, a nova near the date of JD 2436620 reached the phase of the final brightness rise with a starting value of about 15m and with a further more slow increase to 13.5$^{m}$ within 880 days against 300 days for HR Del, in the final phase of which there were two brightness variations with an amplitude of 0.5-1.0$^{m}$. This was followed by a monotonous return of the nova to a quiescence state. During spectral observations of \cite{Ringwald1996} the nova has not yet completed the outburst and was 1.5$^{m}$ brighter than the quiescence state. 

\subsubsection{Novae of the LMC and the SMC}

2 novae in the Large and 1 in the Small Magellanic Clouds in the form of light curves were assigned by us to the RR Pic group (Fig.19, Table 17). 

 \begin{figure}[t]
 	\centerline{\includegraphics[width=78mm]{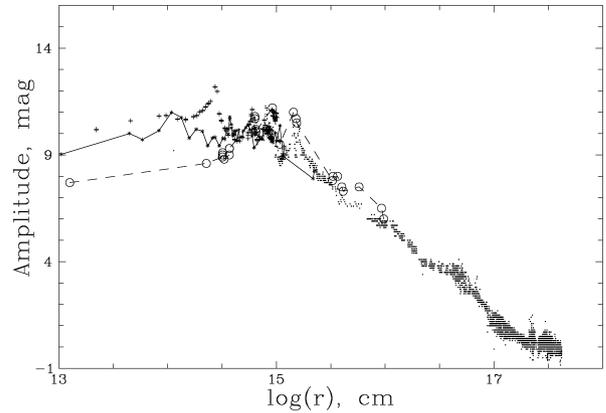}}
 	\caption{The light curves of novae of the RR Pic groups in the LMC and SMC: LMCN 1948-12a (dashed line with circles), LMCN 2017-11a (pluses), SMCN 2008-10a (solid line with dots) and RR Pic (dots) with the parameters Table 17.\label{Fig19}}
 \end{figure}
 
 Membership of 3 novae to the MCs (Table 17) makes it possible to estimate their absolute magnitude with minimal errors. 
 
The question of taking into account interstellar extinction for novae of the MOs was solved uniformly with the procedure described above. 

Then, having for LMCN 1948-12a and SMCN 2008-10a, respectively, the brightness at the maximum of mpg=13.3$^{m}$ and I=13.0$^{m}$ we obtain the absolute brightness MB=-6.8$^{m}$ and MI=-6.1$^{m}$. Unfortunately, these values are obtained for the photometric bands B and I; we cannot say anything confidently about the magnitudes in the V band because of the lack of multicolour photometry. It is available in the AAVSO database only for LMCN 2017-11a. In the interval JD 2458079-2458096, we have an average value of V-I=0.40$^{m}$ and a peak brightness of V=11.9$^{m}$. Since there is no data for the stars of the field near nova in the database of \cite{Haschke2011}, we use the average value of interstellar absorption A$_{V}$$\approx$0.2$^{m}$ for the LMC. Then the absolute brightness of LMCN 2017-11a in maximum will be M$_{V}$$\approx$-6.8$^{m}$.

\subsubsection{Preliminary discussion}

We can state that in 3 out of 38 novae of the Pic group in the substance ejected during the outburst, dust condensation only of the D3 or D4 DQ Her type occurred.

The HR Del subgroup is a heterogeneous association of low amplitude very slow novae. 

\subsection{V1493 Aql group}

Below is proposed for consideration a small problem group of novae with unique light curves. 

The reason for a creation the V1493 Aql group appeared in the last two decades, when outbursts of three novae with similar unique light curves occurred. The first example of light curve of this kind was presented in 1999 by Nova Aquilae, V1493 Aql. Three known at that time novae: V1493 Aql, V2362 Cyg, and V2491 Cyg, were combined in the catalogue of \cite{Strope2010} into the C class. To date, the number of such novae has increased thanks to the OGLE survey \citep{Mroz2015}. 

The prototypes for the novae of this group are V1493 Aql and V2362 Cyg (Fig.20, Table 20). They have well-presented light curves with the possibility of determining the maximum moment t$_{0}$. 

The search for the pre-nova of V1493 Aql ended without result, and the upper brightness limit was found to be no brighter than the DSS-2 sky survey limit \citep{Moro1999}. For definiteness, we accepted the brightness of the quiescence state of B$\geq$21$^{m}$. 

The pre-nova for V2362 Cyg was identified with a star of brightness r'=20.30$^{m}$ and i'=19.76$^{m}$ in the 2004 images of \citep{Steeghs2006}. A small colour index (r'-i') allows us to assume that the visual magnitude was of about 20$^{m}$. This sets the outburst amplitude for the second prototype. 

Such initial parameters (Table 20) for the two prototypes very well combine the initial sections of the light curves and allows almost to preserve the emerging initial trends of light curves also after secondary maxima that occurred at different shell radii (Fig.20). The light curves in the modified scales are "perfectly" smooth between these two linear sections. We will try to use this idealization further.  

A characteristic feature of the V1493 Aql groups is the secondary brightness maximum, the re-brightening, with an amplitude of 1-3$^{m}$ in the range log(r)=14.7-15.7. 

The peculiarly regular shape of the light curves of the two prototypes, which we obtained with minimal corrections to level the  outbursts amplitude, gives an idea of the main properties of the light curve of the V1493 Aql group in modified scales. In this we, as has been repeatedly was before, will be helped by the unique nova V1280 Sco. We use one of the properties of constructing light curves in modified scales: the position of the parts of the light curve at a distance of 100 days or more from the maximum light depends weakly on errors in determining the moment of maximum t$_{0}$. The V1280 Sco is notable for its three episodes of dust condensation and possible is four. As it became clear after studying the light curves of novae of this group and some novae of other groups (V1723 Aql from the D2 DQ Her group), a fourth episode occurs but rare. There is the depression on the V1280 Sco light curve at log(r)$\approx$15.5 that coincides with the secondary maximum of V2362 Cyg, after which dust condensation occurred \citep{Lynch2008, Arai2010}. As a result of very detailed IR observations, \cite{Arai2010} traced the temporary development of the condensation process and compared it with other details of the light curve. \cite{Arai2010} admitted that processes leading to secondary maximum are a trigger for dust condensation. In V1493 Aql, there are no IR observations after a secondary maximum, the only such observation was 8 days early \citep{Venturini2004}. At this time, dust condensation may not yet begin. Consequently, the position of the moments of the secondary maxima should be somewhat ahead of the onset of a dip connected with the dust condensation, a dust dip. This follows from the sequence of the dust condensation process, which in practice takes an interval of one to two dozen days \cite{Rozenbush1988b, Rosenbush1996c}. When conditions arise for the condensation of dust, dust nuclei manifest themselves primarily through the absorption in the UV region of the spectrum and its re-emission in the IR region of the spectrum. With an increase in radius, dust particles begin to affect the visual region of the spectrum, effectively absorbing here and re-emitting in the IR range. We define this sequence of processes as the first property of the light curves of novae of the V1493 Aql group: after a secondary maximum, re-brightening, a decline in the visual brightness due to condensation of dust on the line of sight can be observed. The second property: the slope of the linear initial decline of brightness is the same for the stars of this group. Third property: after the secondary maximum, there is a short linear section. For example, in V2362 Cyg, it occurs when log(r)$\approx$15.5-15.8. The slopes of this section are different, but it is close trending with the initial decline section.  

We also included in the V1493 Aql group the novae with less expressive secondary maxima in the middle part of the light curve, but these are also unique examples that distinguish their from the entire array of novae. 

To include a nova in the V1493 Aql group, the classical light curve is sometimes sufficient: the secondary maximum is a unique, inimitable detail. It remains from the available observations to estimate the brightness of a quiescence state or to assume its definite value if it is lower than the penetrating power of the observation complex.

\begin{figure}[t]
	\centerline{\includegraphics[width=78mm]{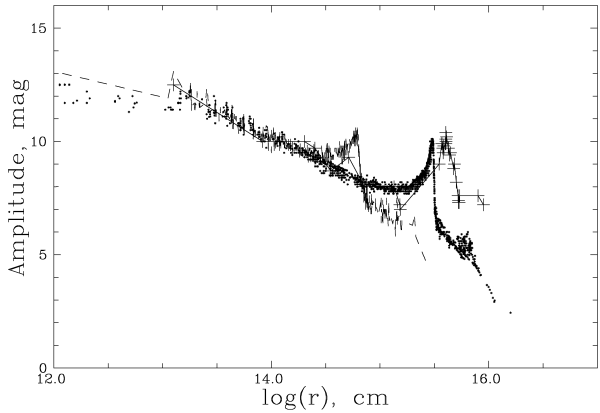}}
	\caption{Light curves of the prototypes of the Aquilae group: V1493 Aql - dashed line, V2362 Cyg - dots. The line with pluses is the light curve of the old nova CK Vul.\label{Fig20}}
\end{figure}

\begin{center}
	\begin{table}[t]%
		\centering
		\caption{List of novae of the V1493 Aql group.\label{tab20}}%
		\tabcolsep=0pt%
		\begin{tabular*}{240pt}{@{\extracolsep\fill}rccc@{\extracolsep\fill}}
			\toprule
			\textbf{Nova} & \textbf{m${_q}$/t${_0}$} & \textbf{m1/m2 [ref]} & \textbf{Ref}  \\
			\midrule
			V1493 Aql* & 	22/51373 & 	>21/ [1] & 	2 \\
			V2362 Cyg* & 	20/53830.5 & 	20/ [3] & 	4 \\
			V2573 Oph & 	20.5/52620 & 	>20.5B/ [5] & 	6 \\
			HZ Pup & 	18p/38040	 & /17.0B [7] & 	8 \\
			\textit{V5914 Sgr} & 	\textit{26.5I/53750} &  & 		\textit{9} \\
			V463 Sct  & 	20/51510 & 	>20/ [10] & 	11 \\
			CK Vul** & 	13/2331180 & 	/20.7R [12] & 	12 \\
			\textit{LMCN 1971-03a} & 	\textit{22.5/40979}	 &  & 	\textit{13, 14} \\
			LMCN 2001-08a  & 	24I/52125	 &  & 	15\\					
			\bottomrule
		\end{tabular*}
		\begin{tablenotes}
			\item Note: ** - t${_0}$ for CK Vul is as the correct Julian Date. Possible members of the subgroup are in italics. References: 1 - \citep{Moro1999}, 2 - \citep{Tago1999}, 3 - \citep{Munari2008}, 4 - \citep{Nishimura2006}, 5 - \citep{Kimeswenger2003}, 6 - \citep{Takao2003}, 7 - \citep{Szkody1994a}, 8 - \citep{Hoffmeister1965}, 9 - \citep{Mroz2015}, 10 - \citep{Kato2002b}, 11 - \citep{Haseda2000a}, 12 - \citep{Shara1985}, 13 - \citep{Graham1971a}, 14 - \citep{Ardeberg1973}, 15 - \citep{Mroz2016b}.
		\end{tablenotes}
	\end{table}
\end{center}

The light curve of V2573 Oph, Nova Ophiuchi 2003, had for over 70 days an unusual state of slightly variable brightness before the outburst discovery by \cite{Takao2003}, which even caused a discussion about the reliability of these observations [see \cite{Kimeswenger2003} for details]. By this time, only one nova was known, the V1493 Aql, with a similar detail of the light curve. As of right now, we can confidently use Tabur's data and include V2573 Oph in this small group of novae with unique light curves (Table 20). As a result of targeted searches of \cite{Kimeswenger2003}, there is a brightness estimate for the pre-nova: B$\geq$20.5$^{m}$ and R>19.5$^{m}$. A sharp drop in brightness after secondary maximum could has been accompanied by dust condensation, but IR observations of \cite{Russell2003} were performed before the secondary maximum, when conditions for dust condensation had not yet been created. We did not correct at +1$^{m}$ the data of \cite{Takao2003}, as this was made by \cite{Kimeswenger2003}; otherwise, the light curve in this area would go below and a new uniqueness of the modified light curve in this area would arise. With this data set, you can reconstruct the general view of the first half of the light curve. We can immediately say that the beginning of the outburst is shifted at least several tens of days from the first observations of \cite{Takao2003}: so that the detail of the light curve possibly associated with dust condensation corresponds to the D4 V1280 Sco type. Property (2) of novae in this group: equality of the slope of this part of the light curve with prototypes, also required such a removal of the beginning of the outburst. Now it becomes clear that it is necessary to bring close the moments of dust condensation in V2573 Oph and V2362 Cyg, i.e. for the latter as this moment, we take the beginning of the brightness drop after the secondary maximum, and for the first - the moment of a sharp increase in the rate of brightness drop after the secondary maximum. Matching is also done to the equating of brightness levels in the secondary maximum. Now we with the parameters of Table 20 achieve good agreement of the light curve with prototypes and other group members. Our estimation of the visual brightness of a quiescence state of m$\approx$20.5$^{m}$ confirms the pre-nova identification by \cite{Kimeswenger2003}. The maximum of outburst took place in the season of the conjunction of the object with the Sun no earlier than JD 2452620 and with a visual brightness of about 8.5$^{m}$. 

The HZ Pup outburst was observed by only one author \citep{Hoffmeister1965}, which ensures data uniformity and reliability. Therefore, the inclusion of this nova in this unique group from our point of view is not in doubt. Extrapolation to the outburst maximum with the parameters of Table 20 gives m$_{p}$$\approx$5.5$^{m}$ for the date JD 2438040. An interpolated secondary maximum occurred on date JD 2438104 with m$_{p}$$\approx$7$^{m}$, which is slightly higher than the observation data. A possible dust condensation after the secondary maximum occurred by the D2 type. 

A very incomplete, with a large scatter of data, V463 Sct light curve shows a pre-maximum plateau for at least 10 days, then a 15-day gap in the observations follows and immediately the maximum of brightness with the amplitude of about 1.5$^{m}$ is fixed. It resembles the light curves of the novae of this group. Spectral observations of \cite{Kato2002b} confirmed this object is the Fe II class nova. The last image of the sky without the nova was obtained on JD 2451492 or 90 days before the outbreak was detected. With this data set and the parameters of Table 20, it is possible to restore the light curve in the modified scales, which is in good agreement with the prototypes. The outburst began in the season of object invisibility due to conjunction with the Sun. The principal maximum was near JD 2451510, the brightness may have reached 8$^{m}$. As it follows from the first property of the novae of this group that dust condensation could occurred according to the D3 type. 

The uniqueness of the light curves of this novae group allows us to confidently classify the light curve of one of the oldest known nova CK Vul (Fig.20, Table 20). The outburst occurred in 1670 and the light curve was perfectly reconstructed by \cite{Shara1985}. A comparison of even the classic light curves of CK Vul and V1493 Aql shows an undeniable similarity. The parameters of Table 20 only state the main parameters of the pre-nova and the outburst itself. The modified light curve makes it possible to extrapolate the brightness of the maximum CK Vul: m$_{vis}$$\geq$0.5$^{m}$, reached no more than 6 days before the discovery of a nova. In Fig.20, the CK Vul light curve after the maximum follows practically as the prototypes, the secondary maximum occurs somewhat later than the V2362 Cyg secondary maximum. On the modified light curve of CK Vul (Fig.20), a typical sharp drop in brightness from the level of 8.9$^{m}$ is clearly recorded at log(r)=15.7. Our estimation of the brightness of a quiescence state is much brighter than the known stars - candidates for the post-nova CK Vul: at least 7$^{m}$. This is one of the two central stars of the bipolar nebula \cite{Hajduk2013}, which over 27 years increased its brightness by about 1.8$^{m}$ to R=19.9$^{m}$, while the second candidate accordingly decreased its brightness, which creates the basis for some speculation about spatial motions of stars and clouds for 300 years at a distance of post-nova of about 700 ps). Near the geometric center of the bipolar nebula is a radio source \cite{Hajduk2007}, which has a peculiar atomic and molecular composition \cite{Kaminski2017}. One thing is certain, that the brightness of a post-nova is very much attenuated by the dense dust shell ejected during the outburst and in which, after re-brightening, carbon dust condensed by the D4 V1280 Sco type (the molecular abundance includes a myriad of C-bearing species typical for carbon stars and many others \citep{Kaminski2017}). 

An outburst of OGLE-2006-NOVA-01, V5914 Sgr, was observed only in the III phase of the OGLE project \citep{Mroz2015} and the maximum outburst was reached in the object's invisibility season.\cite{Mroz2015} drew attention to the secondary maximum of the object and assumed a possible classification of the light curve as "C" according to the scheme of \cite{Strope2010}. The light curve has a secondary maximum typical for this group, but the amplitude of the secondary maximum is noticeably lower than that of the two prototypes. The gap in the observations of the OGLE survey before the first observation of an object in an outburst is 110 days. The lack of data on such a period of time creates some uncertainty in the further reconstruction of the light curve. We settled on a variant of the parameters of Table 20. This allows us to conclude almost unambiguously from a comparison with the V1280 Sco that the secondary maximum corresponds to dust condensation of the D2 type. The secondary maximum is set equal to the prototypes, i.e. about 12.2$^{m}$, which accordingly raises the initial linear portion of the light curve to the level of prototypes. From the existing light curve, it can be assumed that the general trend of the light curve before the secondary maximum is also preserved after it (see property 3 of the light curves of the novae in this group). The secondary maximum with a brightness of I=13.0$^{m}$ occurred near JD 2453750. The reconstructed light curve of V5914 Sgr has some similarity to LMCN 1971-03a (see the next subsection), but with interesting differences. The secondary maximum had equal amplitudes, but it was at different moments. After the secondary peak of V5914 Sgr, the light curves began to diverge. The light curve of V5914 Sgr kept the initial trend, i.e. was linear, while the brightness of LMCN 1971-03a began to increase the rate of decline. Differences from a typical light curve leave some doubts about the correct inclusion of LMCN 1971-03a in the V1493 Aql group. 

\subsubsection{The LMC novae, as members of the V1493 Aql group}

Table 20 contents the parameters for the two novae of the LMC as members of this group. 

Nova Dor 1971, LMCN 1971-03a, the only nova that we originally included according to the light curve in two groups: this and the DQ Her group. The nova had a shallow light minimum of the D1 type, which was the basis for inclusion in the DQ Her group. In duration, this minimum did not gone beyond typical values. The subsequent light curve at first also follows the shape typical of the DQ Her groups, but then the brightness nevertheless declines faster. The doubt concerned a small increase in brightness before this above noted minimum (here is interesting to pay attention to the NQ Vul light curve (Fig.8)), which differed from typical low-amplitude light fluctuations for DQ Her near the average level; the amplitude of this low secondary maximum is comparable to V5914 Sgr (see the previous subsection). This small increase and a faster drop in brightness in the second half of the final brightness decline, more typical of the Aquilae group, served us arguments in favour of the Aquilae group (but doubts about this decision remain and in Table 20 noted by italics). This also required on 2$^{m}$ more lower outburst amplitude and a correspondingly brighter brightness in a quiescence state (Table 20). The outburst maximum was perhaps around JD 2440979 and the visual magnitude of about 10.3$^{m}$. Earlier observations of the sky region without the outburst of LMCN 1971-03a were made near JD 2440975 with a limit value of 11.9$^{m}$ \cite{Bateson1971} (when taken account the almost zero colour index (B-V)<0.1$^{m}$ \citep{Ardeberg1973}), which taken place 4 days before our estimate of the moment of maximum (Table 20) and 1.6$^{m}$ weaker than our estimate of maximum brightness. 

\cite{Mroz2016b} described the outburst of LMCN 2001-08a by the possible time interval for a beginning of outburst between 2001 July 9 and August 9 and two maxima and, on this basis, attributed it to the C class according to \cite{Strope2010}. The unusual shape of the light curve of LMCN 2001-08a indicates its belonging to the Aquilae group (Table 20). The uniform observational material for LMCN 2001-08a of \cite{Mroz2016b} provides an opportunity to propose reconstruction of the light curve up to a secondary maximum. A nova outburst occurred near JD 2452125 with a maximum brightness near I=11.7$^{m}$. A few additional extrapolation points are such: JD 2452126, I = 11.7$^{m}$; (..2145, 15); (..2150, 15.2); (..2155, 15.35); (..2160, 15.4). Here it is necessary to understand that the outburst amplitude in the I band will be less than in the visual region; i.e., the amplitude will not higher 12.3$^{m}$, but let's say 11m (no data on colour indicators). Then the visual brightness of a quiescence state will be m$_{vis}$$\approx$22.7$^{m}$ (if the colour index is about 0$^{m}$). The amplitude of the secondary maximum is even higher than prototypes by almost 1$^{m}$. 

Repeating the above procedure, we find the absolute brightness of LMCN 2001-08a at the outburst maximum of the M$_{I}$$\approx$-6.9$^{m}$. Similarly, for LMCN 1971-03a, we estimate the absolute visual brightness of M$_{V}$$\approx$-8.4$^{m}$. 

If the colour index at the maximum of outburst is about zero, then we get the range of possible absolute visual brightness of novae of the Aql groups: (-6.9$^{m}$)-(-8.4$^{m}$). 

\subsection{Novae with unique light curves}

In the course of our review, light curves were found for which it was difficult to determine membership in a particular group because of several options for belonging to any group. By the shape of a light curve, they could be partially combined with the prototypes at a certain offset, which for the ordinate scale is permissible within 1-2$^{m}$, but on the abscissa scale, any shift greater than 0.2 would contradict our fundamental principle: no any shifts along the shell radius scale. The influence of the error in the parameter t$_{0}$ on the position of the critical points of the light curve (Table 1) is weakened with distance from the maximum light, especially at a distance of 100 or more days, which we discussed when describing the methods for constructing light curves in modified scales. 

Here we will discuss 4 such objects that have appeared recently. It is interesting that among old novae with good light curves, we did not encounter such situations. It is possible that these novae are members of unknown groups of novae that are even less numerous than the V1493 Aql group.

\begin{center}
	\begin{table}[t]%
		\centering
		\caption{Novae with unique light curves.\label{tab21}}%
		\tabcolsep=0pt%
		\begin{tabular*}{240pt}{@{\extracolsep\fill}rccc@{\extracolsep\fill}}
			\toprule
			\textbf{Nova} & \textbf{m${_q}$/t${_0}$} & \textbf{m1/m2 [ref]} & \textbf{Ref}  \\
			\midrule
			V906 Car & 	18.5/58198.9 & 	19.9m/  & 	1 \\
			V2491 Cyg & 	18/54566 & 	/18.03 [2] & 	3, 4 \\
			V445 Pup & 	21/51870 & 	13.1/19 [5] & 	6 \\
			V4643 Sgr & 19.5/51963.6 & /16.6 [7]  & 8 \\					
			\bottomrule
		\end{tabular*}
		\begin{tablenotes}
			\item References: 1 - \citep{Stanek2018b}, 2 - \citep{Zemko2018}, 3 - \citep{Nishiyama2008c}, 4 - \citep{Hounsell2016}, 5 - \citep{Platais2001}, 6 - \citep{Goranskij2010}, 7 - \citep{Mroz2015}, 8 - \citep{Liller2001a}.
		\end{tablenotes}
	\end{table}
\end{center}

\begin{figure}[t]
	\centerline{\includegraphics[width=78mm]{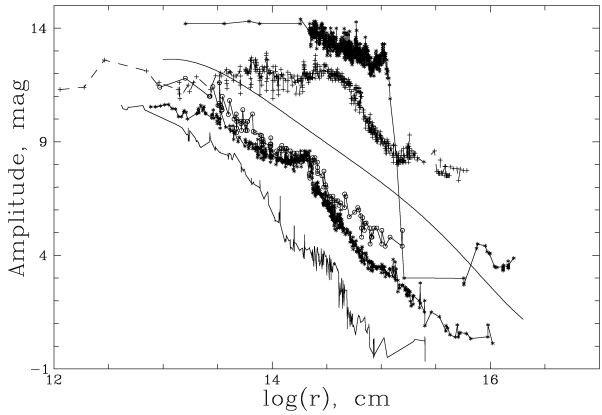}}
	\caption{Light curves of novae (from top to bottom): V445 Pup - line with dots (offset +2 along the ordinate axis); V906 Car - line  with pluses; polynomial representation of the light curve CP Lac - solid smooth line; V4643 Sgr - line with circles; V2491 Cyg - line with dots; the recurrent nova U Sco - the most bottom  broken line.\label{Fig21}}
\end{figure}

The light curve of the Nova Carinae 2018, V906 Car, can be characterized by the state of maximum brightness with slight differences similar to the V445 Pup (Table 21, Fig.21). We equated the amplitudes of the novae outbursts near log(r)$\approx$14.5 or approximately on the 25-30th day after the star reached the state of maximum brightness. Up to this point, the V445 Pup light curve has had rare observations with a stable brightness. The V906 Car in this state exhibits brightness variations with an amplitude of about 1$^{m}$. Following, both stars have a trend towards a decrease in brightness, which ends or with a shallow temporary weakening or slight re-brightening. The subsequent parts of the light curves are fundamentally different: if the V445 Pup experienced a rapid deep weakening of the light due to the condensation of dust on the line of sight \citep{Lynch2001}, then the brightness of the V906 Car shows a slow weakening. As of end 2019, the final phase of the brightness decline in V906 Car has not yet arrived, i.e. brightness can stabilize at a high level like the V1280 Sco. There is no data on dust condensation in the V906 Car, but the spectrum shows some peculiarities \citep{Izzo2018, Rabus2018}. 

After the "journey" of V445 Pup through the DQ Her and V1493 Aql groups, we settled on the current version of uniqueness for this nova. The V445 Pup \citep{Kanatsu2000, Haseda2000b} is an interesting case for a number of reasons. It is considered to be the nova, first and only helium nova \cite{Ashok2003}: emission lines of H I and He I were absent in the spectrum \cite{Wagner2001}. The light curve in our view (Table 21, Fig.21) can be considered as a outburst of nova with dust condensation of the D4 DQ Her type in the presence of a number of peculiarities. Firstly, the outburst amplitude, which, according to our estimate (Table 21) is about 12$^{m}$, should be counted from the brightness of the post-nova, which in 2009 had a brightness of V$\approx$19$^{m}$. Otherwise, according to the brightness of the pre-nova, B=14.3$^{m}$ \citep{Goranskij2010} or m$_{vis}$=13.1$^{m}$ \citep{Platais2001} we have an amplitude should be much lower (not higher than 5.9$^{m}$). Secondly, a very long and deep light minimum on 10$^{m}$, which according to the moment of the beginning of the brightness drop corresponds to the D4 V1280 Sco type. The nova outburst begins in the 54-day period of object invisibility and, in our estimation, the parameter t${_0}$ is most likely closer to the time of the outburst detection, since in the opposite case, the onset of light minimum shifts by +54$^{d}$ or $\Delta$log(r)$\approx$+(0.15-0.19). In this interpretation of the moment t${_0}$, a shallow temporal weakening near JD 2452025 may be associated with an small episode of dust condensation by the D3 type, and after the recovery of brightness, dust condensation of the D4 type followed. This recovery was interpreted by \citep{Goranskij2010} as the re-brightening. We mention one more interesting detail of the light curve that is also present in the light curve of the V906 Car. The continuous state of a slightly variable maximum brightness ended near log(r)$\approx$14.7, which corresponds to the beginning of dust condensation according to the D2 type, and gave way to a smooth trend towards a decline in brightness. On the other hand, if we accept the presence of re-brightening \citep{Goranskij2010}, then, since we included novae with small re-brightening amplitudes (for example, V5914 Sgr) in the Aquilae group, then V445 Pup can also be a member of the Aql groups. Another property of the V445 Pup light curve is related to the V1493 Aql groups: a prolonged state of little variable brightness at the beginning of the outburst. One more argument in favour of the V1493 Aql group may be the geometric shape of the shell ejected by the V445 Pup during an outburst \citep{Woudt2009}: it is almost identical to the CK Vul shell \citep{Hajduk2013}. 

The light curves of the other two novae of Table 21, V2491 Cyg and V4643 Sgr, are very similar (Fig.21). We used the presence of a small secondary maximum, or re-brightening, at log(r)$\approx$14.3 to combine the two light curves on the ordinate scale and estimate the outburst amplitude of V4643 Sgr. This estimation means a weaker brightness of the pre-nova than the brightness of the possible candidate in a pre-nova (see columns 2 and 3 of Table 21). If the amplitude is still lower, then this only strengthens the conclusion to include these two novae in the number of candidates for recurrent novae. For V2491 Cyg, such an assumption has already been made by \cite{Tomov2008}: when discussing the spectrum, an analogy drew the analogy to U Sco and V2487 Oph and based on the fact that the X-ray of V2491 Cyg was detected before the outburst similarly to V2487 Oph. We in own turn will add V4643 Sgr to the V2491 Cyg based on the similarity of the light curves and on the basis of the criterion of \cite{Duerbeck1987b}: positions in the diagram "outburst amplitude, light decline rate t$){m}$". In the next section, we generalize the criterion \cite{Duerbeck1987b} to the case of modified light curves: the light curves of repeated novae in modified scales are located lower than classical novae. Fig.21 shows a schematic light curve of the classic nova of CP Lac, which gives an idea of the relative position of these two types of novae: classic and recurrent (U Sco). It should be noted that \cite{Bruch2001} already noted similar details in the spectra of V4643 Sgr and V2487 Oph, whereas V2487 Oph only in 2009 received the status of a recurrent nova \citep{Pagnotta2009}. 

Of the interesting objects that we examined in our preliminary study \cite{Rosenbush1999a}, but which we left outside the scope of this study due to significant differences from typical examples, we mention PU Vul. A separate group was allocated for PU Vul. The modified light curve is characterized by a slightly variable state of maximum brightness, interrupted by dust condensation of the D4 DQ Her type, and lasting to log(r)$\approx$16.6, i.e. was the longest. The final decline in brightness began nevertheless earlier than that of the V1280 Sco. The late dust condensation of the V605 Aql and V4334 Sgr \citep{Rosenbush1999c} is very unusual in the light of the views of this study, but the duration of the entire outburst did not go beyond the known limits of the final phases of the novae that we met here.

Such solitary examples may indicate the existence of unique characteristics of binary systems, which can cause unique outbursts of novae.

\section{General discussion} 

Changing the methodology and setting the rules for constructing light curves of novae made it possible to confirm our preliminary conclusion \citep{Rosenbush1999a} about the possibility of grouping novae according to the shape of the light curve and to establish the existence of several groups of classic novae in which the light curves have a certain shape. This result was expressed in a distribution by 7 groups almost 250 novae from almost 450 known novae in addition to 3 groups among the known 10 recurrent novae (Paper I). We are inclined to attribute the rest novae to other types of eruptive stars, such as WZ Sge, in particular, to the dwarf novae known as the TOADs, which were not the subject of our review. About 70 remaining stars due to fragmented light curves were left without classification. Out of the scope of this study, there were very slow outbursts of V605 Aql, V4334 Sgr, etc., which are considered born-again objects \citep{Schonberner2008}. 

The total distribution of classic novae by group is presented in Table 22.

\begin{center}
	\begin{table}[t]%
		\centering
		\caption{The distribution of novae by groups and galactic affiliation.\label{tab22}}%
		\tabcolsep=0pt%
		\begin{tabular*}{240pt}{@{\extracolsep\fill}rccc@{\extracolsep\fill}}
			\toprule
			\textbf{Group, subgroup} & \textbf{Galaxy} & \textbf{LMC} & \textbf{SMC}  \\
			\midrule
			V1493 Aql & 	6 3\% & 	2 & 	0 \\
			V1974 Cyg & 	8  3\%	 & 0 & 	0 \\
			\midrule
			DQ Her= &  64 27\%  &  6 &  3 \\
			=D1+D2+D3  & =18+30+16 & =0+6+2 & =0+2+0 \\
			Her1 & 	28  14\% & 	1 & 	1 \\
			\midrule
			CP Lac	 & 21  9\% & 	1 & 	0 \\
			GQ Mus & 	15  7\%	 & 0 & 	0 \\
			RR Pic+HR Del & 	37+4  17\% & 	2+0	 & 1+0 \\
			CP Pup+Flash & 	42+9 21\% & 	1 & 	0 \\
			\midrule
			Total & 	235 & 	13 & 	4 \\					
			\bottomrule
		\end{tabular*}
		\begin{tablenotes}
			\item Note. Table reflects statistics of only novae with the confident determination.
		\end{tablenotes}
	\end{table}
\end{center}

The largest DQ Her group includes three subgroups according to the type of dust condensation. The Her1 subgroup consists of novae for which there is no data on the dust condensation, and by the shape of the light curve in the state of maximum brightness, they should probably be included in the D1 subgroup: the absence of a light minimum at the transition phase of the outburst is associated with the condensation of dust outside the line of sight to the central source radiation. In this case, the population of the subgroups decreases with increasing duration of the state of maximum brightness as 3:2:1. But this ratio does not take for account that after the D1 condensation could occurred the D2 and the D3, and may be also situation that after the D2 but without the D1, occurred the D3. Then the ratio will be near equilibrium. 

The V1974 Cyg group is one of the smallest groups that we have identified. This may raise some doubts about the legality of choosing such a group. But it is obvious that it occupies its own niche - an intermediate position between the GQ Mus and CP Pup groups: in terms of outburst amplitude it is closer to the first group, and in the short duration of the maximum brightness state it is closer to the Pup group. There are two features that distinguish it from these two groups: the light curve at the end of the initial decline phase gradually decreases its slope to almost zero, and there is a short plateau in the middle of the final brightness decline phase. 

Novae of the RR Pic group stand out from all novae with the longest state of maximum brightness, which immediately passes into the phase of the final decline.

The V1493 Aql group is also not numerous, but the shape of the light curves is very unique. 

Novae of the LMC and the SMC follow these statistics of the distribution of novae into groups. 

To simplify the comparison of the summary light curves of the selected groups, we presented them in a schematic form without taking into account the temporary dip (the DQ Her group) or the light depression (the CP Pup group) at the transition phase and the secondary brightness maximum at the end of the initial light decline phase (the V1493 Aql group) (Table 23). The second and fourth columns of Table 23 contain the coordinates of the starting and ending points of the linear sections of the light curves for the phases of initial and final declines: (log(r), A[mplitude]). The coordinates of the three points from which the average values are given in brackets are given in those cases where a typical light curve has a slight systematic deviation from linearity in this range. The third and fifth columns contain the slope coefficients of these sections. 

In Fig.22 we presented these schematic light curves (Table 23) for some groups of novae. A comparison allows us to note some similarity of the schematic light curves for the Aql, Cyg, Lac, and Pup groups. They occupy a small band and for their reduction into one common light curve, a shift on the amplitude scale is sufficient, which is permissible due to the dependence of the outburst amplitude on the spatial orientation of the binary system, the progenitor of a nova, with respect to the observer's line of sight \citep{Warner1987}. It is also permissible due to the fact that the amplitude was a fairly free parameter: we ascribed the amplitudes of 2-4 prototypes to the entire group. But the individualities of the light curve of novae of each group (depression, or fluctuations or a smooth trend) object to such a combination and, apparently, we can talk about a family consisting of these four groups of novae. 

Above this family are the light curves of the DQ Her and GQ Mus groups. The last group can be considered as a kind of extreme case for the first. The basis for this view is the close slopes of the light curves at the phases of initial and final light declines and close amplitudes of outbursts. 
 
 Below the family of four groups locates the RR Pic group.
 
 By this means we can speak of three families of light curves of novae with a population of 36\%, 47\%, and 17\%, respectively.

\begin{figure}[t]
	\centerline{\includegraphics[width=78mm]{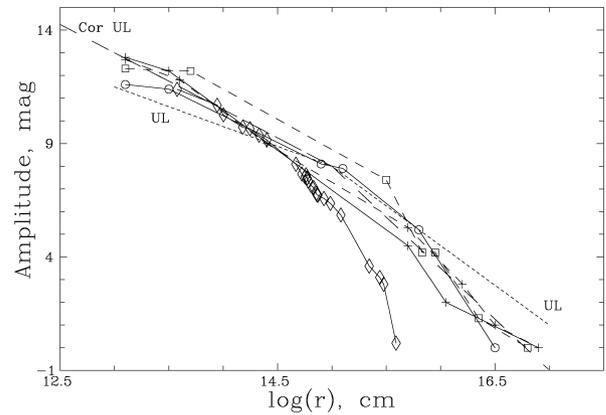}}
	\caption{Schematic light curves of some groups of novae (Table 23). Dashed line with squares - Cyg group; solid line with pluses - Lac group; solid line with circles - Aql group; dashed line with pluses – Pup group. The solid line with rhombus is the V1668 Cyg light curve in the "y" band of the Strömgren system. Short-dashed line (UL) - the universal law of \cite{Hachisu2006}; the long dashed line (Cor UL) is one of the possible corrections of the universal law.\label{Fig22}}
\end{figure}

\begin{center}
	\begin{table*}[t]%
		\caption{The main parameters of the schematic modified light curves of the prototypes of (sub)groups.\label{tab1}}
		\centering
		\begin{tabular*}{500pt}{@{\extracolsep\fill}lcccr@{\extracolsep\fill}}
			\toprule
			\textbf{Group,} & \textbf{First and last points} & \textbf{Slope coefficient}  & {\textbf{First and last points}}  & \textbf{Slope coefficient}   \\
			\textbf{subgroup} & \textbf{ of the initial decline,}  & \textbf{of the initial}  & {\textbf{of the final }}  & \textbf{of the final}   \\
			\textbf{} & \textbf{log(r)/A}  & \textbf{decline}  & {\textbf{log(r)/A}}  & \textbf{decline}   \\
			\midrule
			Aquilae & 13.5/11.4-14.9/8.1  & -2.4  & 15.8/5.2-16.5/0  & -7.4 \\
			Cygni & 13.7/12.2-15.5/7.4 & -2.7  & 15.83/4.2-(15.95/4.2)-16.35/1.3  & -8.3   \\
			DQ Herculis & 13.8/12.3-(15.1/10.5)-15.5/8  & -5.7  & 16.2/6-16.75/1.5  & -8.2   \\
			Her1 & 13.9/12.3-(14.5/11.0)-15.45/7.1  & -3.4  & 15.65/5.5-(15.9/4.1)-16.6/1.6  & -4.0   \\
			Lacertae & 13.5/12.2-15.7/4.5  & -3.5  & 15.7/4.5-16.05/2  & -7.1   \\
			Muscae & 13.1/14-15.8/9.9  & -1.5  & 15.80/9.9-16.73/2.4  & -8.1   \\
			Pictoris & 13.1/8.5-15/10  & +0.8  & 15/10-17.35/0	 & -4.2   \\
			Puppis & 13.6/11.8-(14.4/9.0)-15.7/5.3  & -3.1  & 16.2/5.2-16.5/3  & -5.3   \\
			\bottomrule
		\end{tabular*}
	\end{table*}
\end{center}

\subsection{Groups of novae and universal decline law}

\cite{Hachisu2006} proposed the universal law of a light curve. The light curve of nova according to this law has a slope of m$\propto$t$^{-1.75}$ in the first half of light curve, and a slope of m$\propto$t$^{-3.5}$ in the second one, where M is magnitudes and t is time in days from maximum magnitude. To test this law it was recommended to present a light curve at the y-band of the Strömgren photometric system, because it is free from strong emission lines. In Fig.22, this law is represented by two dashed lines. It can be noted that in our presentation of the light curves of novae, there is no agreement between the universal law and the observations both at the phase of an initial decline in brightness and at the final decline phase. The observed light curves have a significantly greater slope at both phases. Other groups also do not follow the universal law. The existence of groups of novae based on the shape of the light curve means that each group has its own kind of universal law. For a family of 4 groups (Fig.22), we can specify the adjusted slope coefficients in both parts of the universal law, the initial and final, and assume that novae in this band obey it with their own characteristics. The adjusted universal law for visual brightness can have coefficients close to the following values: -2.5 and -4.5, and a knot point with coordinates (log(r),A)$\approx$(15,8). The remaining groups (Mus, Her, Pic) have their own laws. It can be noted that only the initial brightness decline of the light curve for the Her and Mus groups has a slope close to the universal law of -1.75 (Table 23), while the Lac and Pup groups and the Her1 subgroup with smooth light curves have a slope of the initial brightness decline typical of the second part universal law -3.5 according to \cite{Hachisu2006}.

\subsection{Groups of novae and processes in a binary system}

The binning of modified light curves into groups also reflects the difference in physical processes of a bursting star and in its vicinity, which we already noted in our preliminary study \cite{Rosenbush1999b}.

Dust condensation in the ejected shell always occurs in novae of the DQ Her and V1493 Aql groups. Condensation sometimes occurs in the CP Pup and RR Pic groups. For V888 Cen, a nova of the Mus groups, we noted possible dust condensation of the D4 V1280 Sco type. Perhaps the following interesting circumstance has a real basis. In the DQ Her group, the dust condensation only by the D3 type occurs less frequently than the first two types. At the same time, in the RR Pic group, in 4 recorded cases, dust condensation occurred only according to types D3 and D4. Here we can once again mention V408 Lup, the Pic group, with a light curve very similar to the D3 and D4 DQ Her subgroups; this leads us to the conclusion that there is a possible close affinity for these groups or a transition between them due to the difference in the distribution of matter in the shells ejected during an outburst. Of course, there is necessary to understand that the formation of the shell occur initially on the stellar surface.

Dust condensation occurs at four values of the effective radius of the ejected shell: D1 at the radius of 3.2$\times$10$^{14}$ cm, D2 - 1$\times$10$^{15}$ cm, D3 - 1.8$\times$10$^{15}$ cm, D4 - 5$\times$10$^{15}$ cm. Consequently, conditions necessary for the condensation of dust already exist or only now create at these distances. [\cite{Rozenbush1996a} ascribed the highest activity of processes on the transition phase of an outburst (maximal optical thickness of the dust, disappearance of the absorption spectra and appearance of the nebular spectra) to a shell radius of about 5$\times$10$^{14}$ cm.] These values of the dust condensation radii do not take into account possible changes by +(100-300) km s$^{-1}$ the ejecta outflow velocity after each episode of dust condensation as a result of the acquisition by dust an additional acceleration from the radiation pressure and the transfer of momentum from the dust to the surrounding gas. It is possible that if such a change in speed is taken into account, then this will make it possible to bring close the phases of the final brightness decline in the light curves.

For example, in V5668 Sgr, the D3 DQ Her subgroup (Table 5), 120 days after the maximum brightness, the dust condensation on the line of sight reached its maximal optical density, which resulted in a maximum dip in brightness. Following, the general trend of the brightness decline has recovered, i.e. that is dust particles density was decreased nearly to null both due to evaporated a dust by the radiation of a hot central source and scattered away by the radiation pressure force. The expansion speed of Vexp=650 km s$^{-1}$ \citep{Banerjee2016} due to the additional speed from a dust increased, for example, by 250 km s$^{-1}$, and became about 900 km s$^{-1}$. Consequently the radius of the shell that was fixed by the ALMA observations \citep{Diaz2018} should be increased by 40\%, as well as the distance obtained with the expansion parallaxes should be correspondingly increased. The shape of the shell corresponding to this change in velocity is not deformed, which means that the change in velocity refers to the entire shell. It is not uninteresting to note that before the temporal weakening of brightness, V5668 Sgr showed "pulsations" of brightness with a multi-period of 12–20 days. Similar brightness pulsations in the case of variables with variability of the type R Cor Bor are responsible for creating conditions for the condensation of dust in the upper atmosphere of these stars. Light curve in the log-scale clearly shows that the drop in brightness due to dust condensation began after one of these "pulsations" of the brightness. On the ALMA image of the V5668 Sgr shell at 230 GHz \citep{Diaz2018}, an internal elongated structure is clearly visible, possible identical to the equatorial belt of DQ Her, in which dust condensation occurred (see for details Section 2.1). After the completion of the temporary dip, i.e. a vanishing dust, the X-ray flux increased sharply by 3-4 orders of magnitude: in Fig. 11 of the review by \cite{Page2019} this is clearly seen through comparing the detailed X-ray light-curve and the UV light-curve.

In connection with the condensation of carbon dust in the shells of novae, it will not be out of place in the future to pay attention to stars with variability of the R Coronae Borealis type. The main variability of these stars is associated with the condensation of carbon dust, and modelling of processes in their atmospheres has shown that a low hydrogen content, but not too low, is required for dust condensation (\cite{Jurcsik1996, Goeres1996}, for details see also review of \cite{Rosenbush1996b}).

The absence of temporary light weakening in some novae of the DQ Her group can be interpreted by the non-uniformed distribution of dust in the envelope ejected during the outburst \cite{Rozenbush1988a, Rozenbush1988b} and from this we can obtain an idea of the inclination of the orbit of the binary system in which the outburst erupted.

It is very interesting the distribution of matter in the shell of the very old nova CK Vul \citep{Hajduk2013, Eyres2018}: elongated in one direction and compressed in the perpendicular one. All this looks like two hourglass – like shells with a 1:4 aspect ratio and nested one with the other with mutually tilted axes: two shells were ejected by a nova  with different velocities at different times and rotated by a small angle due to the precession of the binary system or for another reason. 

\cite{Slavin1995}, as a result of a review of the shells of the remnants of the novae, concluded that "… these data indicate a possible correlation between nova speed class and the ellipticity of the resulting remnants - those of faster novae tend to contain randomly distributed clumps of ejecta superposed on spherically symmetric diffuse material, whilst slower novae produce more structured ellipsoidal remnants with at least one and sometimes several rings of enhanced emission".

\cite{Rosenbush1999b} also came to a similar conclusion, but with respect to the preliminary selection of novae groups based on the shapes of light curves. Novae of the DQ Her group possess ellipsoids with equatorial and, sometimes, tropical belts. For the RR Pic group, non-bright ellipsoids with a brighter equatorial belt are typical. Smooth light curves of novae of the CP Lac and V1974 Cyg groups correspond to homogeneous ellipsoids. The CP Pup groups have a very uneven distribution of matter in the shell, such as GK Per \cite{Liimets2012}. It is possible that heterogeneity in the shells may form as the uniform shell expands \cite{Lloyd1997}.

Novae are bright sources of radiation in almost the entire range of the spectrum. In our review, we examined the patterns of parameters in the optical and occasionally in the infrared, the UV and the radio ranges, and mentioned rare observations of gamma radiation. X-ray observations are systematic, as evidenced by a review of \cite{Page2019}. We used the rich material in this review to compare the behaviour of optical light curve and X-ray light-curves and make interesting conclusions. This discussion can be considered a test of our assumption that the novae of the same group have the same physical and geometric parameters.

Our main attention was paid to significant changes in X-ray flux, for the characteristics of which we determined several simple parameters: the moment of the beginning of a significant increase in X-ray flux (S), the order of value of the increase of flux r1 to a maximal level, and the moments of a reach and an end of the maximum level of X-ray flux (P1, P2), the order of value of the decline of flux after the maximal level r2, and the moment of slowing down the drop in flux (F). Unfortunately, X-ray observations before the transition phase of outburst are still scarce.

Rare X-ray observations of V2362 Cyg, the Aql group, with a monotonic decrease in visual brightness after the secondary maximum on the interval (P2,F)=(15.7,16.0) recorded a variation of X-ray flux, with rise and with decline, and with a change in intensity on  order of magnitude.

In V1974 Cyg, the prototype of the group of the same name, the amplification of weak X-ray was recorded at the point S=14.9, when there was a slowdown in the decline of optical radiation (Fig.15). It reached its maximum level by the finish of the slowed-down brightness drop in P1$\approx$15.6: the intensity increased by about 4 orders of magnitude, i.e. the source became brighter by 10m. Then the optical radiation sharply accelerated the fall, but the X-ray was kept at a maximum level to the point P2$\approx$15.8. Here (point P2) the optical radiation decline stopped near point F, but during this time the X-ray flux fell by 2-3 orders of magnitude.

For three novae of the Lac groups: V407 Lup, V959 Mon, and V597 Pup, a general conclusion can be made: during the second half of the initial decline phase, the intensity of the X-ray source was either at a maximum level or no more than one order of magnitude weaker than this level, and X-ray source attenuation occurred in a typical range of transition from the initial decline phase to the final decline (P2$\approx$15.4, F$\approx$15.9).

Novae of the CP Pup group have their own peculiarities, tied to the details of the visual light curve. The X-ray of V339 Del increased by two orders of magnitude between the moment of the onset of depression and the moment of reaching the minimum level of brightness in its. When the general trend of the visual light curve was restored, the X-ray flux began to weaken and it was "disappeared" when the rate of visual light decline slowed down near point F$\approx$15.7. The X-ray of the V1535 Sco has been high since the start of the observations. At the beginning of a shallow depression in the transition phase of the outburst, it began to weaken and with the end of the depression, before the general trend of the visual light curve was restored, the X-ray weakened by two orders of magnitude. The observations of V458 Vul, a prototype of the subgroup with flares in of the CP Pup group, started with log(r)$\approx$15.6 and recorded a low level of X-ray radiation, which sharply increased by two orders of magnitude near the point S=15.7 and was rapidly changing with the same high amplitude at this rise. During subsequent flashes typical for this subgroup, the X-ray flux weakened to a "quiet" level and recovered after the completion of the next typical optical flash, which is clearly seen in Fig. 3 of \cite{Page2019}. The X-ray flux remained at a high level until the completion of X-ray observations near log(r)=16.1. Until the completion of the nova outburst remained still about 3$^{m}$.

For novae of the DQ Her groups, we can be summarized that it is possible that the transition phase of outburst is characterized by a high level of the X-ray flux, which is constantly accessible to observations in the absence of dust on the line of sight or shortly after scattering of dust on the line of sight as a result of its removal to great distances from nova. V5668 Sgr between the moments corresponding to the points S=15.3 and P1=15.4 sharply increased the X-ray flux by two orders of magnitude after complete scattering of the dust medium by the line of sight. And immediately (P2=15.4) to the time corresponding to point F=15.6 returned to its original level (this is clearly seen already in Fig.11 of \cite{Page2019}). The X-ray flux of SMCN 2016-10a began to amplify before the transition phase (S<13.8), consisting of a series of shallow light minima D1, D2, etc. To the point P1=14.6 it increased by almost two orders of magnitude and, starting from P2=15.4, the X-ray returned to its initial level near the point F=15.6, which corresponds to a typical of time for the recovery of visual brightness after the dust scattering. SMCN 2012-06a, the Her1 subgroup, was detected in the X-ray range with a high radiation level on 136th day after a nova outburst. Between the points P2=15.6 and F=15.7 the X-ray radiation of the nova weakened by more than one order of magnitude.

Two novae of the RR Pic group, V1369 Sco and V549 Vul, behaved very similarly. Detection of the high-level X-ray flux coincides with a temporary decrease in the visual magnitude by $\approx$1$^{m}$. Some time after the restoration of the general trend of the visual light curve, the X-ray weakened by an order of magnitude. The corresponding points are S1=15.0, P1=15.1, P2=15.4, F=15.6.

The X-ray behaviour of the unique nova V2491 Cyg is shown in Fig.23. The amplitude of an outburst of the X-ray is comparable to the maximal amplitude of nova outburst in the optical spectral range, when the bolometric luminosity of a nova is mainly determined by radiation in the optical wavelength range. Visual brightness after local re-brightening at log(r)=14.3 accelerated the decline. At log(r)=14.9, the decline of visual brightness sharply slowed down, which was also accompanied by a short-term delay in the fall of X-ray radiation.

\begin{figure}[t]
	\centerline{\includegraphics[width=78mm]{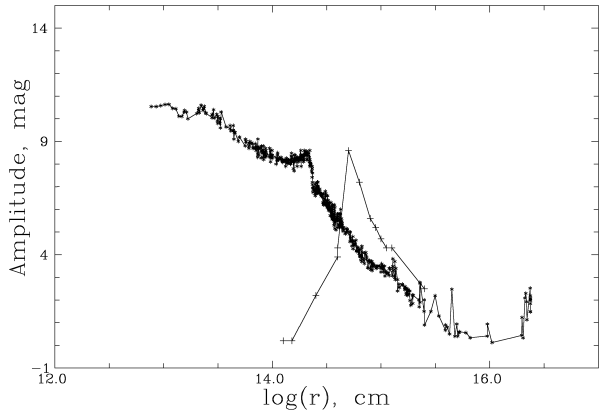}}
	\caption{V2491 Cyg: the visual light curve versus the X-ray light-curve.\label{Fig23}}
\end{figure}

Apparently, classical novae are becoming the X-ray sources mainly at the end of the initial decline phase or the beginning of the final decline. Recurrent novae in the X-ray region were often recorded from the first day of the outburst, as these are faster phenomena compared to classic novae.

Three candidates (V2672 Oph, V1534 Sco, and V1707 Sco) were included in the group of recurrent novae with a prototype T CrB based on the similarity of visual light curves. General conclusion: X-ray appears in the first moments of the nova outburst and remains at a high level during a time of the monotonic decrease in a visual brightness to a point located in the range P2=13.9-14.8, when it falls one-two orders of magnitude in the second half of the decline of optical flux near the point F=14.6-15.1. The earlier beginning of the fall of the X-ray in V1534 Sco coincided with a sharp acceleration of the fall in visual brightness near the abscissa 13.9; the X-ray returned to a original level before the start of a small secondary brightening of a nova.

The X-ray of the V745 Sco was recorded from the first moments of the optical outburst. At the end of the linear initial decline in the optics, the X-ray flux near point S$\approx$13.64 began to rapidly amplify by two orders of magnitude and was at the maximum level in the interval (P1,P2)$\approx$(13.80,13.91). To the point F1$\approx$14.2, the X-ray weakened by almost two orders of magnitude and, by the end of the optical outburst, weakening by another two orders of magnitude, returned to a quiescence level near F2$\approx$15.4.

The maximal level of the X-ray flux of U Sco refers to the interval between points P1$\approx$14.2 and P2$\approx$14.5, after which it fell by two orders of magnitude near the outburst end by F$\approx$14.9. It is noteworthy that a relatively stable level of maximum X-ray radiation corresponds to "pulsations" near the average level of UV radiation (Fig.6 of \cite{Page2019} or to the interval of slowed decline of visual brightness.

A violation of the monotonous decline in visual brightness of recurrent novae of the RS Oph group, of RS Oph itself and two candidates for this group (KT Eri and LMCN 2009-02a), in the second half of the outburst occurred simultaneously with the temporary appearance of the intense X-ray. At KT Eri the X-ray flux very quickly and with the largest amplitude by 4 orders of magnitude increased around S$\approx$P1$\approx$14.9, remained at this level longer than the other two novae of group: to P2$\approx$15.4, and fell by 4 orders of magnitude to F$\approx$15.6. The amplitude of the optical outburst of KT Eri was also larger by about 2.5m or one order of magnitude.

The increase in X-ray flux from the slow recurrent nova T Pyx coincided with the beginning of the plateau in the middle part of the brightness decline (see UV light curve in Fig.11 of \cite{Page2019}): point S$\approx$15.1. The X-ray reached its maximum level by the end of this plateau on the UV light curve (see the same Fig.11): point P1$\approx$15.2. After the second small plateau, near log(t-t$_{0}$)+C$\approx$P2$\approx$15.4, the attenuation of X-ray radiation began, which ended during yet another short plateau-like section of F$\approx$15.7.

Two candidates the slow recurrent novae of the CI Aql group, HV Cet and LMCN 1968-12a, have data about the X-ray flux. But we will not discuss HV Cet, so as not to be misled, as our classification of her membership in the group is still in question due to the large gap at the beginning of the outburst and the longer outburst duration, as well as the significant difference in the behaviour of the X-ray from the second candidate. The amplification of the X-ray flux by one order of magnitude in LMCN 1968-12a began in a state of maximum brightness in front of the first typical plateau (S$\approx$13.8), reached a maximum level at the beginning of this plateau (P1$\approx$14.2) and remained this level up to ending of plateau of about P2$\approx$14.6 . The fall of the X-ray flux by two orders of magnitude occurred before the deceleration rate of the brightness decline near F$\approx$14.8.

The entrance to the phase of the final brightness decline in almost all novae groups has the same form: a sharper drop in brightness begins. This occurs near the moment log(r)$\approx$15.5. Thus, the final decline or the nebular phase begins after the high-energy mechanism producing x-ray radiation is "turned off". The "turned on" of this mechanism refers to various earlier moments of the outburst, for which, unfortunately, there are not yet enough observational data. The maximal X-ray flux of the DQ Her group novae are observed either after the completion of a temporary light minimum at the transition phase or after complete scattering/destruction of the dust in the ejected shell \citep{Gehrz2018}. The X-rays as a review of \cite{Page2019} is found in novae in all 7 groups.

Already from the first radio observations of novae \citep{Hjellming1970} it is known that a nova is a bright variable radio source. The fact that the intensity of radio emission V1723 Aql \citep{Weston2016} increased during periods of time coinciding with the episodes of dust condensation by the D2, D3 and D4 type DQ Her was noted in our discussion of its light curve. But there is no need to speak about any regularity, since the radio emission of V959 Mon \citep{Healy2017} reached a peak in intensity before the transition of nova from the initial decline phase to the final decline phase and cannot be associated with dust condensation due to its absence or insignificance in novae of the CP Lac group.

In connection with the proposal to distinguish the families of novae, let us pay attention to the presence of intermediate polar (IP) and candidates for them. The Catalog of IPs and IP Candidates of K.Mukai (see footnote 4) contains a list of more than one hundred such objects with a confidence level from 1 to 5th high degree. The sample is very small to draw conclusions except for one: for 6 and 7 of such objects, among novae of the DQ Her and CP Pup families, respectively, there are no objects in the Pic family or we have a 1:1:0 ratio for the DQ Her, CP Pup, and RR Pic families. This is with a family population ratio of about 3:2:1, respectively (Table 22). Additionally, we note that with a second, low degree of confidence, two candidates for IPs are among the RNs. V1500 Cyg and V407 Lup from the CP Pup family are polar and intermediate polar, respectively \citep{Kaluzny1988, Aydi2018}.

\subsection{On absolute magnitudes}

To determine the absolute magnitude of novae, the empirical dependence proposed by \cite{Zwicky1936} and now known as the "maximum-magnitude-rate-of-decline" (MMRD) relation is often used. But \cite{Schaefer2018}, as a result of comparing methods for determining distances to novae, came to the conclusion that it was necessary to exclude the use of the MMRD relation for this purpose.

In a current form, the MMRD relation was obtained by \cite{Arp1956}, using observations of 30 novae of the galaxy M31, and was proposed two variants of the dependence of the absolute magnitude novae both on the rate brightness decline t$_{m}$, i.e. the known MMRD, and on the duration during which the nova is brighter than definite magnitude, in particular, m$_{pg}$=20.0$^{m}$.

Using \cite{Arp1956} photometry, unfortunately limited by a brightness range of up to 4.5$^{m}$ in the region of the peak magnitude, we evaluated the possible belonging of the novae of M31 galaxy to our groups.

The light curve of the nova N1 from the list of \cite{Arp1956}, due to the small range after the maximum brightness, allows two options: it may be or a recurrent nova, either the U Sco group or the T CrB group, or the second option it is the classic nova of the CP Pup group with a light curve similar to CP Pup itself. For the nova N9 we are agree with \cite{Shafter2015} that it is possibly recurrent nova and we include it in the T Pyx group. For the nova N3, the situation is similar to the nova N1, but either the U Sco group or the T CrB group with a maximum brightness of almost 0.5$^{m}$ brighter than \cite{Arp1956} is acceptable; thus, it rises above the dependence "maximum magnitude, duration" Fig.31 of \cite{Arp1956}. The light curve of the nova N4, which in the opinion of \cite{Arp1956} deviates more strongly from dependence due to strong interstellar absorption, according to our ideas, in addition to possible belonging to the D1 DQ Her subgroup, can also belong to the TOADs dwarf novae (in the lower left quarter of Fig.24).

The general conclusion for this set is as follows. Novae of the CP Lac  and CP Pup groups and D1 DQ Her subgroup are located at the top of the dependencies of Fig.31,32 of \cite{Arp1956}. They had a peak photographic magnitude of about 15.9$^{m}$. 6 novae of the RR Pic group definitely occupied only the lower part of the dependence, i.e. had the peak photographic magnitudes of about 17.8$^{m}$. 9 novae of the D2 and D3 DQ Her subgroups had peak magnitudes in the range of 15.9-17.8$^{m}$. It can be estimated that by the distance modulus m-M=24.4$^{m}$ and by the interstellar extinction A$_{B}$$\approx$0.5$^{m}$ for the galaxy M31, this corresponds to the absolute magnitude M$_{B}$$\approx$-9.0$^{m}$ for novae of the CP Lac and CP Pup groups and the D1 DQ Her subgroup and M$_{B}$$\approx$-7.1$^{m}$ for the RR Pic group (see also above Section 3.6). Novae of the DQ Her group have intermediate values sequentially from the D1 subgroup through the D2 subgroup to the D3.

Concerning the second parameter of \cite{Arp1956} - the duration d$_{m}$ of the state of a brightness of a nova higher than a certain level of magnitude m - we can say that it is equivalent to the duration of the state of maximum brightness in our definition (see Section 2), followed by the linear initial decline phase. Novae of the CP Lac and CP Pup groups have the highest t$_{m}$ values or the lowest duration d$_{m}$. Novae of the RR Pic group are the slowest novae. The DQ Her group contains novae with a wider range of t$_{m}$ or d$_{m}$ parameters. Therefore, along the MMRD relation, they will be located from left to right, following this order: CP Lac, CP Pup, GQ Mus, V1974 Cyg, (D1, Her1, D2, D3) DQ Her, RR Pic. It turns out that this sequence also tracks the change in the shape, or parameters, of the shells that are ejected by nova during outburst. From the pronounced disk shape in the slow novae of the RR Pic group, through ellipsoids with bright belts in novae of the DQ Her group to ellipsoids with heterogeneities in the V1974 Cyg, CP Pup, CP Lac groups.

Figure 23 shows the dependence of the duration d$_{m}$ on the peak magnitude of nova according to Table 1 from \cite{Arp1956} with the indication of a group, including the possible belonging. According to position of the novae groups, it would be more correct to name our three families as bright (groups CP Lac, CP Pup, V1974 Cyg), intermediate (DQ Her) and weak (RR Pic). We add that the possible distribution of stars from the list of \cite{Arp1956} among our groups corresponds to Table 22. But according to the signs of both the outburst amplitude and the duration of maintaining a high brightness level, or by the parameter t$_{m}$, the bright novae are the intermediate family DQ Her (+GQ Mus). A large spread in the amplitudes of outbursts: from the highest to low, is available for the family of bright novae with their highest parameters t$_{m}$. The faint family, RR Pic, has low and lowest amplitudes, but the largest values of the parameter t$_{m}$.

\begin{figure}[t]
	\centerline{\includegraphics[width=78mm]{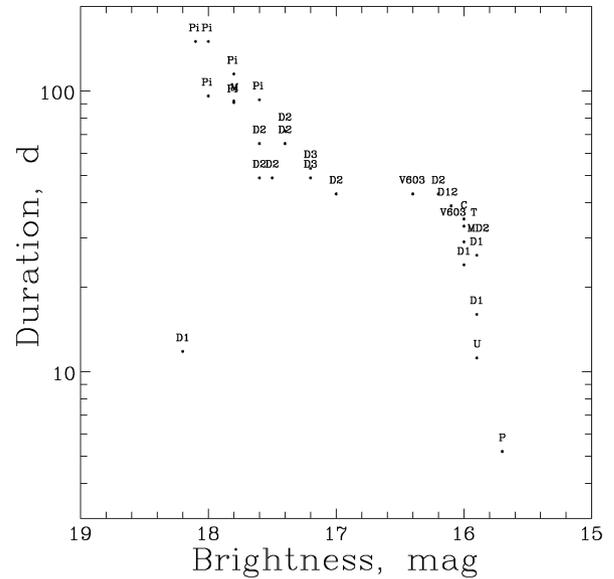}}
	\caption{The ratio of "maximal brightness, duration" for novae of the Arp’ list indicating possible membership in groups: P, U - CP Pup or U Sco; D1, D2, D3 - DQ Her, D12 - one of the subgroups D1 or D2 DQ Her; Pi - RR Pic; M - GQ Mus; MD2 - GQ Mus or DQ Her; V603 - similar to V603 Aql (the CP Pup group); C - V1974 Cyg.\label{Fig24}}
\end{figure}

This provides a basis in addition to the conclusion of \cite{Slavin1995} on the existence of "a possible correlation between nova speed class and the ellipticity of the resulting remnants" to say the following. The development of a nova outburst, which is recorded in the form of a light curve or other time sequence of changes in stellar parameters, is determined by the distribution of the matter and variations in the intensity of its radiation in the shell ejected during an outburst. With a lack of ejection mass, the formation a long time existing expanding shell does not occur, and the outburst is recorded as an outburst, for example, of a recurrent nova. A jump of +1m near 43d is noteworthy (Fig.24). If the upper part of the dependence to shift by this value, then we can obtain either a linear dependence of the maximum brightness Max on d$_{m}$: Max$\approx$15.9+0.01$\times$d$_{m}$, or power-law one: Max=15.60-0.0163$\times$d$_{m}$-(0.21$\times$d$_{m}$)$^{2}$  corrected by +1$^{m}$ at d$_{m}$>43$^{d}$. This transition zone is formed by novae of the DQ Her group from the D1 subgroup to the D3 subgroup. 43d correspond to log(r)$\approx$14.7 or novae of the D2 subgroup. We can say that here, within the group, occurs a sharp change from the almost uniform distribution of the matter over the ejected shell to the belt-like distribution. It will be of place here to recall a dust condensation in novae of the RR Pic group: only the D3 or D4 type (see the beginning of Section 4.2), or the depression of novae of the CP Pup group. In short, this refers to the transition phase of the schematic light curve of \cite{Payne-Gaposchkin1957}. This dependence reflects the state of the shell during its formation at the maximum of outburst. Therefore, it is also valid for recurrent novae: from the "bright" U Sco to the "faint" T Pyx.

From our presentation of the light curves, it was clearly seen that the shape of the light curves of novae in a limits of one group stabilizes by 15 days after the maximum. This coincides with another approach in determining the absolute magnitude of novae, in which \cite{Buscombe1955} substantiated the equality of absolute magnitude of any nova on the 15th day after the maximum. \cite{Shara2018} confirm this proposal with observations of novae in other galaxies. But this approach may not work for novae of the RR Pic group that have very long states of maximum.

Further discussion is required here.

We performed a classification of the light curves of some novae of the MCs and estimated their absolute magnitudes; but, unfortunately, the photometric bands in the observations were different: photographic or the B band, visual or V, R, I and non-system. Comparison of our estimates and review data of \cite{Shafter2013} shows close values with a difference of no more than 0.5$^{m}$. The most numerous is the D2 DQ Her subgroup in the MCs. Table 8 allows us to present the light curves of this subgroup in the absolute scale. Since the initial data are presented in different photometric bands in Fig.25 we did not highlight concrete novae except for SMCN 1994-06a. For the latter, a correction was made to the value in Table 8 by $\Delta$M$_{B}$=-1$^{m}$ to reduce the variance of the points of all light curves before the start of a light minimum; this correction may be due to inaccurate accounting for interstellar extinction A$_{B}$ for SMCN 1994-06a. It is clearly seen from Fig.25 that the maximal magnitude has a scatter of about 2$^{m}$, which is comparable with the difference in the average maximum brightness levels of fast and slow novae in the set of \cite{Arp1956}. At the same time, the brightness values at 15$^{d}$ after the maximum have the smaller scatter. Consequently it is preferable to use the proposal of \cite{Buscombe1955, Shara2018} about the brightness of novae on the 15th day after the maximum. But this approach may face difficulties for novae of the RR Pic group with a prolonged state of maximal brightness and several variations of brightness in this state. Apparently, it will be necessary to look for an individual approach to each group.

\begin{figure}[t]
	\centerline{\includegraphics[width=78mm]{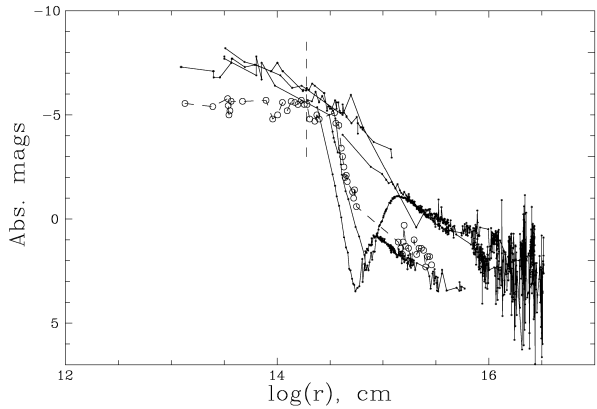}}
	\caption{Light curves of the D2 DQ Her subgroup of the MCs in the absolute magnitude scale. The vertical dashed line corresponds to the moment t=15d. Dashed line with circles - SMCN 1994-06a.\label{Fig25}}
\end{figure}

From Table 4 for 6 novae of the MCs, members of DQ Her group,  with absolute visual or V magnitude, we can estimate the average absolute brightness of about -7.7$^{m}$ at maximum, which is close to the above average absolute value of novae in the upper left quarter of Fig.24. The average outburst amplitude of novae of the DQ Her group is about 13$^{m}$. Therefore, the ordinate scale of light curves (Fig.3-6) can be calibrated in absolute magnitudes by simply equalizing a brightness level of 13$^{m}$ of amplitude scale to an absolute magnitude of -7.7$^{m}$.

\subsection{Relationship of classic and recurrent novae}

When reviewing the modified light curves of recurrent novae, we paid attention to some similarity their characteristics to classical novae.

So, in the classical nova V1974 Cyg and the recurrent nova RS Oph coincided the slopes of light curves for the initial decline. The distinction was related to the large difference in the amplitudes of the outbursts. There is another common detail on the light curves of these two novae: the plateau-like portion before the final decline. In V1974 Cyg, it is in the interval log(r)=14.9-15.4; in RS Oph, it is shorter and lies in the interval log(r)=14.8-15.0. It is interesting to compare the general details of the RS Oph and V1974 Cyg light curves with the behaviour of the X-rays \cite{Hachisu2007, Bode2008, Krautter1996, Page2019} (Fig.26) (definition key points see also in Section 4.2). The X-ray flux of RS Oph began to amplify in the middle of the linear portion of the light curve (point "s" in Fig.26), reached a maximum towards the beginning of a short plateau (point p1). At the end of this plateau (point p2) at a point with coordinates (log(r)$\approx$14.98, A$\approx$7.5$^{m}$ (in the scale of Fig. 26), a rapid decline of the X-ray flux by two orders of magnitude, and at point "f" the decline sharply slowed down. The X-ray flux of V1974 Cyg changed with noticeable differences. The growth of the X-ray was fixed at point S when the decline of the optical radiation was slowed down. The X-ray reached its maximum level to the ending of the slowed-down brightness decline at log(r)$\approx$15.6 (point P1). Then the optical radiation sharply accelerated its decline, but the X-ray was kept at a maximum level to the point P2. Here (point P2), the decline of the optical flux stopped to point F, but during this time the X-ray declined by 2-3 orders of magnitude.

We add to this the general features of the light curves of the classical Pic group and the recurrent T Pyx group. This is a slow final rise to a maximum, several brightness fluctuations in a state of maximum and a quick transition to an almost linear slow final decline. The differences lie in a shorter final rise in the latter, and, accordingly, an earlier and faster final decline. The difference in outburst amplitudes of about 2$^{m}$ does not look like a significant difference. T Pyx has a shell from one of the previous outbursts \cite{Duerbeck1979}. The candidate for the recurrent nova LMCN 2005-11a (Paper I) has a plateau on the light curve that coincides with a position of the D3 type dust condensation episode of the V1280 Sco. According to the SMARTS, the brightness of this nova in the K band had a corresponding local maximum, which can be interpreted as an indication of dust condensation. The amplitude of the LMCN outburst 2005-11a was similar to a typical recurrent nova.

In a joint graph, the light curves of recurrent novae are in the inner region with respect to classic novae. The existence of such zone of recurrent novae is a generalization of the proposal of \cite{Duerbeck1987b, Duerbeck1988}: for the search for recurrent novae among novae to use the "amplitude-t$_{3}$-time" diagram. For example, V2491 Cyg and V4643 Sgr fall into this inner zone, but the light curves of them differ from the known recurrent novae, so we presented them in a separate group of novae with unique light curves (see Section 3.8).

In some cases, it is necessary to know the amplitude of the outburst in order to make an informed choice between classifying the nova as recurrent or classical. An example of such a situation is LMCN 2019-11a\footnote{http://www.cbat.eps.harvard.edu/unconf/followups/J05145365-7009486.html}. \cite{Walter2019} classified her as He-N nova. The modified light curve according to AAVSO data satisfies the light curve both the recurrent V745 Sco group, especially V394 CrA, and the classic V838 Her subgroup. The difference in the amplitudes of the outbursts of these two interpretations reaches 3$^{m}$. Two possible progenitors proposed by \cite{Pessev2019} have a brightness in the middle of this range, which leaves the question unanswered: is it a classic or recurrent nova.

Due to the existence of a "zone of recurrent novae", attention can be paid to some objects at the opposite end of the abscissa scale, which were initially classified as nova, but were then identified as "born-again objects" \cite{Schonberner2008}. This is a very slow final thermal pulse of the helium-burning shell and the light curve of such flash lead us to the interesting question of the limit light curve. The duration of the nova outburst apparently has a limit of 70-100 years. Born-again objects remain in a state of high luminosity for a long time, but near this duration limit quickly calms down. The final decline phase of modified light curve of the symbiotic nova PU Vul can also be proposed as a limit example. Obviously, the V1500 Cyg light curve with the highest amplitude among the classic novae: 18-20$^{m}$ can offer a limitation in amplitude; \cite{Pavlenko2018} report variability in modern brightness in the range of 19-21$^{m}$.

\begin{figure}[t]
	\centerline{\includegraphics[width=78mm]{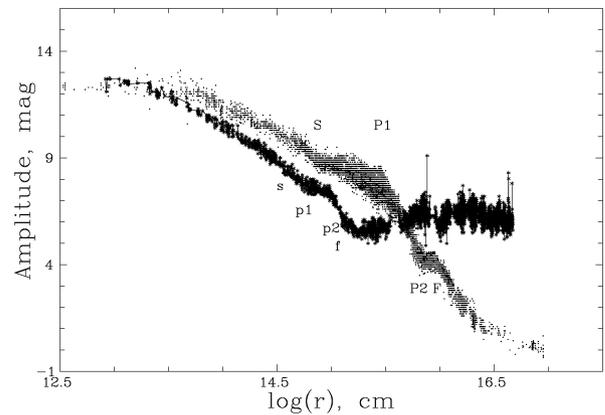}}
	\caption{Light curves of V1974 Cyg (higher points) and RS Oph (lower line with dots). For convenience of comparison, the RS Oph light curve is shifted only along the ordinate axis by +5.5$^{m}$. For designations: S, P1, P2, F, s, p1, p2 and f see text.\label{Fig26}}
\end{figure}

\section{Conclusion}

A modification of the light curve of a nova, proposed by \cite{Rozenbush1996a, Rosenbush1999a, Rosenbush1999b, Rosenbush1999c, Rosenbush1999d, Rosenbush1999e}, met expectations of its applicability to the classification of novae. Nova Scorpii 2007, V1280 Sco, became a key object in formulating a uniform construction rule of modified light curves, which precluded the arbitrary interpretation of the normalization of coordinate axes for light curves. The complex shape of the visual light curve of the V1280 Sco during the transition phase of the outburst should be understood as a sequence of three, or even 4, episodes of dust condensation in the shell ejected during the outburst, when it was occurring on the line of sight.

Our review of the light curves of a known recurrent novae and several hundred classic novae confirmed the current view that the shape of the light curve can be the basis for the distribution of novae by groups. At this phase, the criterion for belonging of unknown object to a certain group of novae is a visual estimation of the coincidence of the light curves of the prototype of the certain group with this object, otherwise it is made conclusion on a possible belonging. The classification uses a complete light curve.

Replacing the linear scale of the argument with a logarithmic scale made possible to contrastingly highlight and to display as direct lines (linearise) the main sections of the light curve corresponding to the phases of the initial and final decline, and also to fix key points of the transition between the phases of the outburst. At this phase, it is proposed that we go from the presentation of the light curve in the time scale to the scale of the radius of the shell, which is ejected during a nova outburst. The scale of radius is normalized by the radius of the star in the moment of maximum brightness in the case of fast novae or in the time the star reaches the state of maximum brightness in the case of slow novae, when actually the shell ejects. In fact, the well-known classical schematic light curve after the maximum can be represented by two straight lines, while the initial decline is divided into two parts by the transition phase. Each group had 1-5 prototypes, which also set a scale for the amplitude of the outburst, to which the amplitudes of the novae with unknown levels of quiescence brightness were equated.

About 400 novae are currently known. For less than 100 of them, it was not possible to classify the light curve mainly due to the lack of a sufficiently densely presented and prolonged light curve. The oldest nova in our review is Nova Vulpeculae 1670 (CK Vul), for which it was established that it belongs to the unique V1493 Aql group. It is proposed to consider a small part of novae as possibly recurrent novae or dwarf novae. The remaining almost 250 novae were distributed by 7 groups with subgroups (there are slight characteristic differences between the light curves of the prototypes of the subgroup): from the largest DQ Her group with subgroups (more than 60 novae) to the smallest unique V1493 Aql group (7 novae). The GQ Mus, V1974 Cyg, CP Pup, CP Lac, and RR Pic groups were also highlighted. Some groups overlap or pass one into another, forming a wide band in modified scales: the Aql, Cyg, Lac, Pup groups. Perhaps this is due to the existence of three families of novae. It is possible the existence of unknown unique, small groups of novae. The phases of the schematic light curve (\cite{McLaughlin1942, Payne-Gaposchkin1957} obtained the calibration in the scales by both axes.

As a result of a review of light curves of more than 70 novae of the LMC and the SMC, respectively 17 and 5 of these novae  was distributed by 7 groups.

The relationship of novae groups, or forms of the light curve, with processes in the circumstellar and near-system environment is discussed. With rare exceptions (V1280 Sco), binary systems after an outburst return to their original state.

The material ejected during an outburst forms expanding shell of a ellipsoidal shape.  Shells of novae with dust condensation (the DQ Her group) has the structure with a belt. The CK Vul shell gives an idea of the possible shell shape of other members of the V1493 Aql group. Novae of the RR Pic group characterized by a bright belt. In the Lac, Pup, and Cyg groups, the ellipsoidal shells do not show a pronounced regular structure. Our review of light curves of the novae from the study of novae in the galaxy M31 of \cite{Arp1956} allowed us to evaluate the type of light curves of some of them and compare it with the position on the MMRD relation: novae of the RR Pic group with bright belts are located at the faint end of the relation, and fast novae with shells without a pronounced regular structure are at the opposite end. This confirms the conclusion of \cite{Slavin1995} that the main factor determining the speed class of a nova is the geometric shape of a ejected shell. Or, more simply put, the main factor shaping the form of an empirical MMRD relation is the distribution of matter in the shell ejectd by a nova during its outburst. The existence of the empirical MMRD relation allows us to say that the difference in the amplitudes of novae outbursts within the same group, for example, CP Lac and V1500 Cyg from the CP Lac group, may be due to the different brightness of the quiescence state of binary systems in which these novae broke out.

Novae of the DQ Her group always form a dust structures. And in a quiescence state, they can be intermediate polars. Intermediate polars have not yet been found among novae of the RR Pic group. Polars and intermediate polars are found among novae of the Lac group. 

The existence of novae groups (by forms of light curves) allows us to confidently distinguish the recurrent nova from the classic nova. There are rare examples of hybrids of these two types of novae that occupy an intermediate, unique position, such as the V838 Her. This scheme of grouping light curves makes it possible to quickly determine whether a newly erupted nova belongs to a well-known group or it is something unique.
 
Unique nova V1280 Sco maintains its "unique" status: the only nova that has been at an unusually high and stable level of brightness for 10-12 years after the outburst. V1280 Sco in 2019 has not yet entered the phase of the final decline, but it began to show the semi-regular light variability after a period of comparative quiet.

Further development of the application of the shell radius scale can consist in taking into account changes in the expansion velocity of the ejected shell in the presence of dust condensation processes.  As a result of the radiation pressure force, dust particles acquire a momentum that, when particles drift through the gaseous medium, is transferred to atoms, and the matter of shell has an add speeds up to 200-300 km s$^{-1}$ and more, as is demonstrated by observations of stars with dust condensation (see review of \cite{Rosenbush1996b}. This will provide an opportunity to create several new points for calibrating the radius scale and go to the real radius scale of the nova and its shell. Another direction in the development of the ideas of this study may be the transition from the scale of the amplitude to the scale of absolute brightness. The accumulation of photometric data of novae at the phase of rose the brightness from a quiet state will allow reconstructing the unified light curve of novae of a certain group at this phase too.

The existence of the proposed classification scheme for novae by form of light curves may mean that in certain binary systems, when certain conditions are met, the outburst of certain nova with certain characteristics may take place.

For practical reproduction of the proposed classification scheme, it is sufficient to construct the light curves of the prototypes from the AAVSO database or another enough full database with the parameters of Tables 2,3 and others, and compare the studied object with prototypes.


\section*{Acknowledgements}

We thank the AAVSO observers who made the observations on which this project is based, the AAVSO staff who archived them and made them publicly available. The BAAVSS database is acknowledged as the (part) source of data on which this article was based. This research has made use of the AFOEV database, operated at CDS, France. This research has made use of the NASA's Astrophysics Data System and the SIMBAD database, operated at CDS, Strasbourg, France. This work has made use of data from the European Space Agency (ESA) mission Gaia (https://www.cosmos.esa.int/gaia), processed by the Gaia Data Processing and Analysis Consortium (DPAC, https://www.cosmos.esa.int/web/gaia/dpac/consortium) and the Photometric Science Alerts Team (http://gsaweb.ast.cam.ac.uk/alerts). Funding for the DPAC has been provided by national institutions, in particular the institutions participating in the Gaia Multilateral Agreement. At the most early phase of our study the state of novae was traced by the VSNET data and we are grateful all observers for this possibility. The author is grateful to the administration of the MAO NAS of Ukraine for the opportunity to maintain my staff status without any mutual financial obligations. Author is thankful L.Laurits for useful discussion, and to the Valga vallavalitsus, Estonia, for the financial support, which allowed us to carry out this interesting investigation.









\nocite{*}
\bibliography{Rosenbush2}%



\end{document}